\documentclass[3p,preprint,hyperref]{elsarticle}
\input{NCFF_aux}

\begin{document}

\begin{frontmatter}
\title{Numerical computation of Fox functions}

\author[torino]{Giampiero Passarino}
\ead{giampiero@to.infn.it}

\address[torino]{\csuma}


\begin{abstract}
 \noindent
In this work we discuss techniques for the numerical computation of Fox functions that
represent Feynman integrals. Illustrative examples based on Sinc numerical methods and
Quasi{-}Monte Carlo methods are given.
\end{abstract}

\end{frontmatter}
\vspace{-1.cm}
{\footnotesize{
\tableofcontents
}}

\newpage
%
%
%
\section{Introduction \label{intro}}
Feynman integrals are indispensable for precision calculations in quantum field theory; generically they
are characterized by a branch cut structure.
The connection between Feynman integrals and Fox functions~\cite{oFox,compH,HS,Hus,BSid,More} 
has been systematized in \Bref{Passarino:2024ugq}.
The Fox $\mrH$ functions are alternative representations for Mellin{-}Barnes (hereafter MB) contour 
integrals~\cite{Boos:1990rg}; they provide a general framework and a concise notation.

Techniques for two{-}fold MB integrals have been developed in \Bref{Friot:2011ic}.
Investigation of multiple MB integrals by means of multidimensional residues is also dicussed in \Bref{ZT}.
In a second approach MB integrals are computed by introducing conic hulls or triangulations of point configurations.  
Both the resonant and non{-}resonant cases can be handled by these methods which are automatized in the
{\tt{MBConicHulls.w1}}
\footnote{MBConicHulls webpage \url{https://github.com/SumitBanikGit/MBConicHulls}}
package~\cite{banik2024geometrical}.
Numerical evaluation of MB integrals and related problems are also described in \Bref{Freitas:2010nx}.
Mellin{-}Barnes representations for $\mrL${-}loop  diagrams are discussed in 
\Brefs{Kalmykov:2012rr,Kalmykov:2016lxx}.

The object to be computed is
\bq
\mrH = \Bigl[ \prod_{\mrj=1}^{\mrr}\,\int_{\mrL_\mrj}\,\frac{\mrd \mrs_\mrj}{\tip} \Bigr]\,
\mrF\lpar \mrs_1\,\dots\,\mrs_\mrr \rpar\,\prod_{\mrj=1}^{\mrr}\,\mrz_\mrj^{\mrs_\mrj} \spp
\label{MBinta}
\eq
The function $\mrF$ is a quotient of products of Euler Gamma functions;
in the most general case the paths of integration are straight lines parallel to the imaginary axis 
avoiding the poles of the integrand~\cite{HTF}. 
Therefore, introducing $\mrs_\mrj = \sigma_\mrj + \mri\,\mrt_\mrj$, we can rewrite \eqn{MBinta} as
\bq
\mrH = \Bigl[ \prod_{\mrj=1}^{\mrr}\,\int_{- \infty}^{+ \infty}\,\frac{\mrd \mrt_\mrj}{2\pi} \Bigr]\,
\mrF\lpar \mrs_1\,\dots\,\mrs_\mrr \rpar\,\prod_{\mrj=1}^{\mrr}\,\mrz_\mrj^{\mrs_\mrj} \spp
\label{MBintb}
\eq
In this work we present the set of conditions controlling the convergence of the integral in \eqn{MBintb} 
and define the Sinc 
approximation~\cite{Sbook,Sinc,SUGIHARA2004673,doi:10.1137/1.9781611971637} of $\mrH$. Special cases of 
Sinc approximations have been discussed in \Brefs{PhysRevD.61.125001,Petrov_2001,osti_40205105,Freitas:2010nx,SSinc}.

Numerical integration in \eqns{MBinta}{MBintb} (or numerical approximation) is relatively simple when the
integrand converges exponentially. Having in mind the connection with Feynman integrals~\cite{Passarino:2024ugq}
we see that such is the case in the Euclidean region; working in the physical region requires extensions of the
method used.

There are many papers dealing with Fox functions~\cite{oFox,compH,HS,Hus} and different authors use different
notations, making necessary to create a comparison of the vocabularies of different notations.

The outline of the paper is as follows:
after a general introduction we introduce univariate Fox functions in \sect{UFF},
including criterions for convergence and contiguity relations;
bivariate Fox functions in \sect{BFF};
multivariate Fox functions in \sect{MFF};
modified Fox function and their connection with infrared divergent Feynman integrals in \sect{ModFF};
analytic continuation in \sect{AC};
singularities of the Fox function in \sect{Hls};
Sinc numerical methods in \sect{SEXP};
hypergeometric methods in \sect{hmb};
factorization/decomposition techniques in \sect{FD};
technical details are given in \sect{AFUN};
numerical examples are given in \sect{exa}.
\section{Univariate Fox function \label{UFF}}
In this work the univariate Fox function is defined as follows:
\bq
\mrH\,\Bigl[ \mrz\,;\,\lpar {\mathbf \alpha}\,,\,{\mathbf a} \rpar\,;\,
\lpar {\mathbf \beta}\,,\,{\mathbf b} \rpar \Bigr] =
\int_{\mrL} \frac{\mrd \mrs}{\tip}\,
\frac{
 \prod_{\mrj=1}^{\mrm}\,\eG{\alpha_{\mrj} + \mra_{\mrj}\,\mrs}
     }
     {
 \prod_{\mrj=1}^{\mrn}\,\eG{\beta_{\mrj} + \mrb_{\mrj}\,\mrs}
     }\,\mrz^{\mrs} \spc
\label{UFFone}
\eq
where ${\mathbf \alpha}$ and ${\mathbf \beta}$ denote vectors of complex numbers while
${\mathbf a}$ and ${\mathbf b}$ denote vectors of real numbers. Furthermore, $\eG{\mrz}$ is the Euler Gamma 
function~\cite{HTF}.

Alternatively we can use the definition of \Bref{HTF}:
\bq
\mrH\,\Bigl[ \mrz\,;\,
 \lpar {\mathbf a}\,,\,{\mathbf A} \rpar\,;\,
 \lpar {\mathbf b}\,,\,{\mathbf B} \rpar\,;\,
 \lpar {\mathbf c}\,,\,{\mathbf C} \rpar\,;\,
 \lpar {\mathbf d}\,,\,{\mathbf D} \rpar \Bigr] =
\int_{\mrL} \frac{\mrd \mrs}{\tip}
\frac{
 \prod_{\mrj=1}^{\mrm}\,\eG{\mra_\mrj + \mrA_\mrj\,\mrs}\,
 \prod_{\mrj=1}^{\mrn}\,\eG{\mrb_\mrj - \mrB_\mrj\,\mrs}
    } 
    {
 \prod_{\mrj=1}^{\mrp}\,\eG{\mrc_\mrj + \mrC_\mrj\,\mrs}\,
 \prod_{\mrj=1}^{\mrq}\,\eG{\mrd_\mrj - \mrD_\mrj\,\mrs}
     }\,\mrz^{\mrs} \spp
\label{UFFtwo}
\eq
It is assumed that all the $\mrA_\mrj\,\dots\,\mrD_\mrj$ are real and positive, and that the path of integration 
is a straight line parallel to the imaginary axis  (with indentations, if necessary) to avoid the poles of
the integrand. 

There is a third definition given in \Bref{compH}:
\bq
\mrH\,\Bigl[ \mrz\,;\,(\mra_1,\mrA_1)\,\dots\,(\mra_\mrp,\mrA_\mrp)\,;\,
(\mrb_1,\mrB_1)\,\dots\,(\mrb_\mrq,\mrB_\mrq) \Bigr] =
\int_{\mrL} \frac{\mrd \mrs}{\tip}
\frac{
\prod_{\mrj=1}^{\mrm}\,\eG{\mrb_\mrj + \mrB_\mrj\,\mrs}\,\prod_{\mrj=1}^{\mrn}\,\eG{1 - \mra_\mrj - \mrA_\mrj\,\mrs}
     }
     {
\prod_{\mrj=\mrm+1}^{\mrq}\,\eG{1 - \mrb_\mrj - \mrB_\mrj\,\mrs}\,
\prod_{\mrj=\mrn+1}^{\mrp}\,\eG{\mra_\mrj + \mrA_\mrj\,\mrs}
     }\,\mrz^{ - \mrs} \spp
\label{UFFthree}
\eq
It is important to observe that in \Bref{compH} the integration contour separates the poles of 
$\eG{\mrb_\mrj + \mrB_\mrj\,\mrs}$ from the poles of $\eG{1 - \mra_\mrj - \mrA_\mrj\,\mrs}$.

Using the definition of \eqn{UFFtwo} we introduce
\bqa
\ovA = \sum_{\mri=1}^{\mrm}\,\mrA_\mrj &\qquad \dots \qquad& \ovD= \sum_{\mrj=1}^{\mrq}\,\mrD_\mrj \spc
\nl  
\ova = \Re \sum_{\mri=1}^{\mrm}\,\mra_\mrj &\qquad \dots \qquad& \ovd= \Re \sum_{\mrj=1}^{\mrq}\,\mrd_\mrj \spc
\eqa
and define the following parameters:
\bqa
\alpha = \ovA + \ovB - \ovC - \ovD \spc &\qquad&
\beta  = \ovA - \ovB - \ovC + \ovD \spc 
\nl
\lambda = \frac{1}{2}\,\lpar \mrp + \mrq - \mrm - \mrn \rpar + \ova + \ovb - \ovc - \ovd \spc &\qquad&
\rho =  
\prod_{\mrj=1}^{\mrm}\,\mrA_{\mrj}^{\mrA_\mrj}\, 
\prod_{\mrj=1}^{\mrn}\,\mrB_{\mrj}^{ - \mrB_\mrj}\, 
\prod_{\mrj=1}^{\mrp}\,\mrC_{\mrj}^{ - \mrC_\mrj}\, 
\prod_{\mrj=1}^{\mrq}\,\mrD_{\mrj}^{\mrD_\mrj} \spp
\label{cpar}
\eqa 
If we use the definition of \eqn{UFFthree} the parameters are given by
\bqa
\alpha = \sum_{\mrj=1}^{\mrn}\,\mrA_\mrj - \sum_{\mrj=\mrn+1}^{\mrp}\,\mrA_\mrj +
\sum_{\mrj=1}^{\mrm}\,\mrB_\mrj - \sum_{\mrj=\mrm+1}^{\mrq}\,\mrB_\mrj \spc &\qquad&
\beta = \sum_{\mrj=1}^{\mrq}\,\mrB_\mrj - \sum_{\mrj=1}^{\mrp}\,\mrA_\mrj \spc
\nl
\lambda = \sum_{\mrj=1}^{\mrq}\,\mrb_\mrj - \sum_{\mrj=1}^{\mrp}\,\mra_\mrj +
\frac{1}{2}\,(\mrp - \mrq) \spc &\qquad&
\rho = \prod_{\mrj=1}^{\mrp}\,\mrA_{\mrj}^{ - \mrA_\mrj}\,
\prod_{\mrj=1}^{\mrq}\,\mrB_{\mrj}^{\mrB_\mrj} \spp
\eqa 
\paragraph{Convergence of the MB integral} \hspace{0pt} \\
Using the notations of \eqn{UFFtwo} with $\phi = \marg(\mrz)$ and $\mrs = \sigma + \mri\,\mrt$ we have the four different 
cases described in \Bref{HTF} and repeated here for the benefit of the reader:
\bei

\item[\ovalbox{I}] $\alpha > 0$. The integral converges absolutely for $\mid \phi \mid < \alpha \pi/2$ (the
point z = 0 is tacitly excluded.).

\item[\ovalbox{II}] 
$\alpha = 0, \beta \not= 0$. The integral does not converge for
complex $\mrz$. For $\mrz > 0$ it converges absolutely if $\sigma$ is so chosen that
$\beta\,\sigma + \lambda < - 1$
and there exists an analytic function of $\mrz$, defined over $\mid \phi \mid < \pi$ 
whose values for positive $\mrz$ are given by $\mrH$ defined in \eqn{UFFtwo}.

\item[\ovalbox{III}] $\alpha = \beta = 0$, $\lambda < -1$
The integral converges absolutely for all positive $\mrz$ (but not for complex $\mrz$)
and represents a continuous function of $\mrz$ ($0 < \mrz < \infty$ ). There are now two
analytic functions, one regular in any domain contained in $\mid \phi \mid < \pi$, $\mid \mrz \mid > \rho$
whose values for $\mrz > \rho$ are represented by $\mrH$, and another regular
in any domain contained in $\mid \phi \mid < \pi$, $0 < \mid \mrz \mid < \rho$ whose values for
$0 < \mrz < \rho$ are represented by $\mrH$. The two functions are in general distinct.

\item[\ovalbox{IV}]
$\alpha = \beta = 0$, $- 1 < \lambda < 0$.
The integral converges (although not absolutely) for $0 < \mrz < \rho$ and for $\mrz > \rho$. There are two analytic 
functions of the same nature as in the preceding case. There is a discontinuity
at $\mrz = \rho$ and the integral does not exist there, though it may have a principal
value. 

\eei

A special case of interest is $\alpha > 0$ and $\phi = \alpha \pi/2$. Introducing $\mrs = \sigma + \mri\,\mrt$ the 
large $\mid \mrt \mid$ behavior of the integrand is controlled by
\bq
\mid \mrt \mid^{\beta\,\sigma + \lambda} \spc
\eq
requiring $\beta\,\sigma + \lambda < - 1$; however, convergence in the numerical treatment can be very slow.

As a matter of fact, when computing Feynman integrals in the physical region this happens for $\alpha = 2$ and
$\phi = \pi$. Following the notations of \eqn{UFFtwo} the most general form of $\mrH$ corresponding to
$\alpha = 2$ and $\beta = 0$ requires $\ovC = \ovA - 1$ and $\ovD= \ovB - 1$.
A possible solution consists in writing the original integral as $\mrH_\sigma$ and in using instead
$\mrH_{\sigma^{\prime}}$ with $\sigma^{\prime}$ such that $\beta\,\sigma^{\prime} + \lambda \muchless - 1$. The original integral
is obtained by subtracting the poles contained in the strip $\sigma^{\prime} < \Re \mrs < \sigma$. 

However, in many cases
$\alpha = 2$ is associated with $\beta = 0$. Here we can improve the convergence only if we can modify $\lambda$ and
making $\lambda$ more negative requires changing the value of the parameters. Using the notations of \eqn{UFFthree} this
requires decreasing the value of the $\mrb$ parameters and/or increasing the values of the $\mra$ parameters; 
this can be done by using contiguity relations. Although we can find useful contiguity relations for
(generalized) hypergeometric functions, \eg
\bq
\mrc\,(\mrc - 1)\,(\mrz - 1)\,\ghyp{2}{1}(\mrc - 1) + 
\mrc\,\Bigl[ \mrc - 1 - (2\,\mrc - \mra - \mrb- 1)\,\mrz \Bigr]\,\ghyp{2}{1}(\mrc) +
(\mrc - \mra)\,(\mrc - \mrb)\,\ghyp{2}{1}(\mrc + 1) = 0 \spc
\label{F21cont}
\eq
these results cannot be easily generalized to Fox functions. In \eqn{F21cont} $\ghyp{2}{1}(\mrc - 1)$ stands
for $\hyp{\mra}{\mrb}{\mrc - 1}{\mrz}$. 
\paragraph{Caveats} \hspace{0pt} \\
Different notations for the Fox function hide substantial differences connected to the path of integration.
For instance,
\bei

\item[1)] In \Bref{HTF} and in \Bref{HS} the path of integration is a straight line parallel to the
imaginary axis avoiding the poles of the integrand.

\item[2)] In \Bref{compH} the integration contour separates the poles of $\eG{\mrb_\mrj + \mrB_\mrj\,\mrs}$
from the poles of $\eG{1 - \mra_\mrj - \mrA_\mrj\,\mrs}$.

\eei
Therefore Fox functions, as defined by different authors, although having the same integrand, are not necessarily the 
same; the difference depends on $\sigma$. There are simple examples illustrating this fact. Given
\bq
\mrH = \int_{\sigma - \mri\,\infty}^{\sigma + \mri\,\infty}\,\frac{\mrd \mrs}{\tip}\,
\eG{ - \mrs}\,
\frac{
      \eG{\mra + \mrs}\,\eG{\mrb + \mrs}
     }
     {
      \eG{\mrc + \mrs}
     }\,\mrz^{\mrs} \spp
\label{Hexa}
\eq
If $\sigma$ separates the poles at $\mrs = - \mra - \mrj, \mrs = - \mrb - \mrk$ from the poles at
$\mrs = \mrn$, where $\mrj, \mrk, \mrn \in \Zf^{+}_{0}$, then 
\bq
\mrH = \frac{\eG{\mra}\,\eG{\mrb}}{\eG{\mrc}}\,\hyp{\mra}{\mrb}{\mrc}{ - \mrz} \spp
\label{HeF}
\eq
Furthermore, \eqn{HeF} requires $\mid \marg \,\mrz\, \mid < \pi$. The same $\mrH$ with arbitrary $\sigma$ is not
always given in terms of a Gauss hypergeometric function.
The integral in \eqn{Hexa} corresponds to $\alpha = 2$ and $\beta = 0$, therefore it converges for
$\mid \marg \,\mrz\, \mid < \pi$. However it is easily seen that for $\mrz< 0$ it converges if 
$\Re\,(\mra + \mrb - \mrc) < 0$, although it is not a Gauss hypergeometric function.
\paragraph{Contiguity} \hspace{0pt} \\
The first step in discussing contiguity relations is to understand the reduction of Fox functions to 
Meijer functions~\cite{HTF}.
A simple reduction can be obtained when we have $\eG{\alpha + \mri/\mrk\,\mrs}$ where $\mri, \mrk$ are integers. For
instance we start from
\bqa
\mrH^{\mrm\,,\,\mrn}_{\mrp\,,\,\mrq}( \mrz ) &=&
\int_{\mrL}\,\frac{\mrd \mrs}{\tip}\,\frac{\mrN}{\mrD}\,\mrz^{ - \mrs} \spc
\nl
\mrN = \prod_{\mrj=1}^{\mrm}\,\eG{\mrb_{\mrj} - \mrs}\,
         \prod_{\mrj=1}^{\mrn - 1}\,\eG{1 - \mra_{\mrj} + \mrs}\,\eG{1 - \mra_{\mrn} + \frac{\mri}{\mrk}\,\mrs} \spc
&\qquad&
\mrD = \prod_{\mrj=\mrm+1}^{\mrq}\,\eG{1 - \mrb_{\mrj} + \mrs}\,\prod_{\mrj=\mrn+1}^{\mrp}\,\eG{\mra_j - \mrs} \spc
\eqa
and, using a well{-}known propery of the Fox function, derive
\bqa
\mrH^{\mrm\,,\,\mrn}_{\mrp\,,\,\mrq}( \mrz ) &=&
\mrk\,\int_{\mrL}\,\frac{\mrd \mrs}{\tip}\,\frac{\mrN^{\prime}}{\mrD^{\prime}}\,\mrz^{ - \mrk\,\mrs} \spc
\nl
\mrN^{\prime} = \prod_{\mrj=1}^{\mrm}\,\eG{\mrb_{\mrj} - \mrk\,\mrs}\,
         \prod_{\mrj=1}^{\mrn - 1}\,\eG{1 - \mra_{\mrj} + \mrk\,\mrs}\,\eG{1 - \mra_{\mrn} + \mri\,\mrs} \spc
&\qquad&
\mrD^{\prime} = \prod_{\mrj=\mrm+1}^{\mrq}\,\eG{1 - \mrb_{\mrj} + \mrk\,\mrs}\,
         \prod_{\mrj=\mrn+1}^{\mrp}\,\eG{\mra_j - \mrk\,\mrs} \spp
\eqa
Next we use 
\bq
\eG{1 - \mrc_\mrj + \mrk_\mrj\,s} = \eG{\mrk_\mrj\,\lpar s + \frac{1 - \mrc_\mrj}{\mrk_\mrj} \rpar} \spc
\quad
\eG{\mrk\,\mrz} = (2\,\pi)^{1/2 - 1/2\,\mrk}\,\mrk^{\mrk\,\mrz - 1/2}\,
\prod_{i=0}^{\mrk-1}\,\eG{\mrz + \frac{i}{\mrk}} \spp
\label{mth}
\eq
and obtain a Meijer function~\cite{HTF}. A simple example is as follows: given 
\bq
\mrH^{1,2}_{2,2}(\mrz) = \int_{\mrL}\,\frac{\mrd \mrs}{\tip}\,
\frac{\eG{\mrb_1 - \mrs}\,\eG{1 - \mra_1 + \mrs}\,\eG{1 - \mra_2 + \frac{2}{3}\,\mrs}}
     {\eG{1 - \mrb_2 + \mrs}}\,\mrz^{ - \mrs} \spc
\eq
we derive a $\mrG^{3,5}_{5,6}$ Meijer G function,
\bqa
\mrH^{1,2}_{2,2}(\mrz) &=& 3\,\lpar 2\,\pi \rpar^{ - 3/2}\,2^{1/2 - \mra_2}\,3^{\mrb_1 + \mrb_2 - \mra_1 - 1/2}\,
\mrG^{3,5}_{5,6}\lpar \frac{9}{4}\,\mrz^3 \rpar \spc
\qquad
\mrG^{3,5}_{5,6} = \int_{\mrL}\,\frac{\mrd \mrs}{\tip}\,\frac{\mrN}{\mrD}\,
\lpar \frac{9}{4}\,\mrz^3 \rpar^{- \mrs} \spc
\nl
\mrN &=&
\eG{1 - \frac{1}{3}\,\mra_1 + \mrs}\,
\eG{1 - \frac{1}{2}\,\mra_2 + \mrs}\,
\eG{\frac{1}{2} - \frac{1}{2}\,\mra_2 + \mrs}\,
\eG{\frac{1}{3} - \frac{1}{3}\,\mra_1 + \mrs}\,
\eG{\frac{2}{3} - \frac{1}{3}\,\mra_1 + \mrs}
\nl
{}&\times&
\eG{\frac{1}{3}\,\mrb_1 - \mrs}\,
\eG{\frac{1}{3} + \frac{1}{3}\,\mrb_1 - \mrs}\,
\eG{\frac{2}{3} + \frac{1}{3}\,\mrb_1 - \mrs} \spc
\nl
\mrD &=&
\eG{1 - \frac{1}{3}\,\mrb_2 + \mrs}\,
\eG{\frac{1}{3} - \frac{1}{3}\,\mrb_2 + \mrs}\,
\eG{\frac{2}{3} - \frac{1}{3}\,\mrb_2 + \mrs} \spp
\eqa
The advantage of using Meijer $\mrG$ functions is based on the fact that $\mrz\,\mrG(\mrz)$ (using the notations of 
\eqn{UFFthree}) is the same function where all the $\mra, \mrb$ parameters go to $\mra +1, \mrb + 1$.

In the general case the integrand contains Euler Gamma functions of the form $\eG{\mrd + \mrD\,\mrs}$ where
$\mrD$ is irrational. The corresponding Meijer $\mrG\,${-}function has 
$\mrD = 1$. If one or more parameters are small we can use the following 
identity~\cite{compH}:
\bqa
{}&{}& \mrH\,\Bigl[ \mrz\,;\,(\mra_1,\mrA_1)\,\dots\,(\mra_\mrp,\mrA_\mrp)\,;\,
                             (\mrb_1,\mrB_1)\,\dots\,(\mrb_\mrq,\mrB_\mrq) \Bigr] = \mrz^{ - \eta}
\nl
{}&\times& \mrH\,\Bigl[ \mrz\,;\,(\mra_1 + \eta\,\mrA_1,\mrA_1)\,\dots\,(\mra_\mrp + \eta\,\mrA_\mrp,\mrA_\mrp)\,;\,
                                 (\mrb_1 + \eta\,\mrB_1,\mrB_1)\,\dots\,(\mrb_\mrq + \eta\,\mrB_\mrq,\mrB_\mrq) \Bigr] \spp 
\label{malab}
\eqa
Therefore, we will be in the situation where the function contains $\eG{\mrd + \mrD\,\mrs}$ with
$\mid \mrd \mid > 1$,
and we can use~\cite{HTF,HTgam}
\bq
\eG{\mrd + \mrD\,\mrs} \sim \eG{\mrd + \mrs}\,\sum_{\mrn=0}^{\infty}\,\mrc_{\mrn}\,\mrd^{\mrD_{-}\,\mrs - \mrn} \spc
\eq
where 
\bq
\mrD_{\pm} = \mrD \pm 1 \spc \qquad
\mrc_0 = 1 \spc \quad
\mrc_1 = \frac{1}{2}\,\mrD_{-}\,\lpar \mrD_{+}\,\mrs - 1 \rpar\,\mrs \spp 
\label{Gratio}
\eq
For the determination of higher order coefficients in \eqn{Gratio} ($\mrc_2 , \dots$) see Eq.~(12) of \Bref{HTgam}.
Approximations for more general quotients of Gamma functions can be found in \Bref{BRaa}.

In special cases contiguity relations for generalized hypergeometric functions are available. If the MB
integral is a Meijer functions
\bq
\mrG^{\mrm,\mrn}_{\mrp,\mrq}\lpar \mrz\,;\,\mathbf{a}\,;\,\mathbf{b} \rpar \spc
\eq
and the conditions given in Section~(5.3) of \Bref{HTF} are satisfied then the Meijer function can be written
as a combination of $\ghyp{\mrp}{\mrq}$ functions; when $\mrp = \mrq + 1$ we can use 
\bq
\Bigl[ \mra_1 + \mrz\,\sum_{\mrj=1}^{\mrp}\,(\mra_{\mrj+1} - \mrb_\mrj) \Bigr]\,
\ghyp{\mrp+1}{\mrp}\lpar \mrz\,;\,\mathbf{a}\,;\,\mathbf{b} \rpar +
\mrz\,\sum_{\mrj=1}^{\mrp}\,
\frac{
\prod_{\mrk=1}^{\mrp}\,(\mrb_\mrj - \mra_{\mrk+1})
     }
     {
\prod_{\mrk=1\,,\,\mrk\not= \mrj}^{\mrp}\,(\mrb_\mrj - \mrb_\mrk)
     }\,(1 - \frac{\mra_1}{\mrb_\mrj})\,\ghyp{\mrp+1}{\mrp}(\mrb_\mrj + 1)
 = \mra_1\,(1 - \mrz)\,\ghyp{\mrp+1}{\mrp}(\mra_1 + 1) \spp
\eq
Relevant contiguity relations can be obtained by using a simple procedure. Let us consider an example,
\bqa
\mrH = \int_{\mrL}\,\frac{\mrd \mrs}{\tip}\,\mrR\lpar \{\mra\}\,,\,\{\mrb\} \rpar 
\qquad & \qquad \mrR= \frac{\mrN}{\mrD} \spc
\nl
\mrN = \prod_{\mrj=1}^{2}\,\eG{1 - \mra_\mrj - \mrs}\,\prod_{\mrj=1}^{3}\,\eG{\mrb_\mrj + \mrs}
\quad & \quad
\mrD = \eG{1 - \mrb_4 - \mrs}\,\prod_{\mrj=3}^{4}\,\eG{\mra_\mrj + \mrs} \spp
\eqa
We employ the notation $\mrR(\mrb_1 - 1)$ to denote the contiguous function in which $\mrb_1$ is replaced by
$\mrb_1 - 1$ but with all other parameters left unchanged. 
We need to use the following relation:
\bq
\mrz\,\mrR(\{\mra\}\,,\,\{\mrb\}) = \mrR(\{\mra + 1\}\,,\,\{\mrb + 1\}) \spc \qquad
\mrz\,\eG{\mrz} = \eG{\mrz + 1} \spc
\label{Rbo}
\eq
However, this relation has to be properly understood. What it is meant is:
given $\mrR(\mrs\,,\,\mrz) = \mrf(\mrs)\,\mrz^{-\mrs}$ we have
\bq
\mrz\,\int_{\sigma - \mri\,\infty}^{\sigma + \mri\,\infty}\,\mrd \mrs\,\mrR =
\mrz\,\int_{\sigma - \mri\,\infty}^{\sigma + \mri\,\infty}\,\mrd \mrs\,\mrf(\mrs)\,\mrz^{ - \mrs} =
\int_{\sigma - 1 - \mri\,\infty}^{\sigma + 1 + \mri\,\infty}\,\mrd \mrs\,\mrf(\mrs + 1)\,\mrz^{ - \mrs} =
\int_{\sigma - \mri\,\infty}^{\sigma + \mri\,\infty}\,\mrd \mrs\,\mrf(\mrs + 1)\,\mrz^{ - \mrs} + \Sigma \spc
\label{Rbt}
\eq
where $\Sigma$ is the sum of the residues in the strip $\sigma - 1 < \Re \mrs < \sigma$. We use the following notation:
$\mrR^{\prime} = \mrf(\mrs + 1)\,\mrz^{-\mrs}$. Given the combination $\mrC = (\mrx + \mry\,\mrz)\,\mrR$ we obtain
\bq
\int_{\sigma - \mri\,\infty}^{\sigma + \mri\,\infty}\,\mrd \mrs\,\,\mrC =
\mrx\,\int_{\sigma - \mri\,\infty}^{\sigma + \mri\,\infty}\,\mrd \mrs\,\,\mrR  +
\mry\,\int_{\sigmap - \mri\,\infty}^{\sigmap + \mri\,\infty}\,\mrd \mrs\,\,\mrR^{\prime} =
\int_{\sigma - \mri\,\infty}^{\sigma + \mri\,\infty}\,\mrd \mrs\,\,(\mrx\,\mrR + \mry\,\mrR^{\prime}) +
\Sigma \spp
\eq
Therefore, if we can find a non trivial solution for $\mrx$ and $\mry$ such that $\mrx\,\mrR + \mry\,\mrR^{\prime} = 0$,
we have a non homogeneous recurrence relation.
Going back to our example we introduce $\mrp_\mrn(\mrz) = \mrx_\mrn + \mry_\mrn\,\mrz$ and write the following 
combination,
\bq
\mrC =
\mrp_0\,\,\mrR +
\mrp_1\,\mrR( - \mra_1) +
\mrp_2\,\mrR( - \mra_2) +
\mrp_3\,\mrR(\mrb_1 - 1) +
\mrp_4\,\mrR(\mrb_2 - 1) +
\mrp_5\,\mrR(\mrb_3 - 1) \spc
\eq
and use
\bqa
\eG{1 - \mra_\mrj - \mrs} &=&  - (\mra_\mrj + \mrs)\,\eG{ - \mra_\mrj - \mrs} \spc 
\nl
\eG{ - 1 - \mra_\mrj - \mrs} &=&  - \eG{ - \mra_\mrj - \mrs}/(1+\mra_\mrj+\mrs) \spc \quad
\mrj= 1,2
\nl
\eG{\mrb_\mrj - 1 + \mrs} &=&  \eG{\mrb_\mrj + \mrs}/(\mrb_\mrj-1+\mrs) \spc \quad
\mrj=1,2,3
\nl
\eG{1 + \mrb_\mrj + \mrs} &=&  (\mrb_\mrj + \mrs)\,\eG{\mrb_\mrj + \mrs} \spc \quad
\mrj= 1,2,3
\nl
\ieG{\mra_\mrj + 1 + \mrs} &=&  \ieG{\mra_\mrj + \mrs}/(\mra_\mrj+\mrs) \spc \quad
\mrj= 3,4
\nl
\ieG{1 - \mrb_4 - \mrs} &=&  - \ieG{ - \mrb_4 - \mrs}/(\mrb_4 + \mrs) \spp
\eqa
After integrating $\mrC$ we obtain an integral of the following form:
\bq
\int_{\sigma - \mri\,\infty}^{\sigma + \mri\,\infty}\,\mrd \mrs\,
\mrQ(\mrs)\,\mrf(\mrs\,,\,\mathbf{x}\,,\,\mathbf{y}) \spp
\eq
\bq
\mrQ =
\frac{
\eG{ - \mra_1 - \mrs}\,\eG{ - \mra_2 - \mrs}\,\eG{\mrb_1 + \mrs}\,\eG{\mrb_2 + \mrs}\,\eG{\mrb_3 + \mrs}
     }
     {
\eG{\mra_3+ \mrs}\,\eG{\mra_4 + \mrs}\,\eG{ - \mrb_4 - \mrs}
     } \spp
\eq
Imposing $\mrf = 0$ we obtain a solution for $\mrx_\mrj$ and $\mry_\mrj$ valid for all value of $\mrs$. 
The analytic result is rather long and not particularly interesting, therefore we will present the numerical result 
for a particular choice of the cofficients, \ie
\[
\begin{array}{llll}
\mra_1 = 0 \quad & \quad
\mra_2 = 0.4 \quad & \quad
\mra_3 = 1.8 \quad & \quad
\mra_4= 1.25 \\
\mrb_1 = 0.41 \quad & \quad
\mrb_2 = 0.295 \quad & \quad
\mrb_3 = 0.8 \quad & \quad
\mrb_4 = 0.2 \\
\end{array}
\]
Setting $\mrx_0 = \mrx_1 = 1$ we obtain
\bqa
\mrx_2 &=& 1.2 \spc \qquad
\mrx_3 = \mrx_4 = 0 \spc
\nl
\mry_0 &=& 1 \spc \quad
\mry_1 = \mry_2 = 0 \spc \quad
\mry_3= 62.740 \spc \quad
\mry_4 = - 56.798 \spc \quad
\mry_5 = - 6.398 \spp
\eqa
As a result we can compute the original $\mrH$ in terms of a linear combination of $\mrH$ functions where 
$\lambda \to \lambda - 1$ and the procedure can be repeated.

Contiguity relations for the Fox function have been analyzed by Buschman and summarized in \Bref{compH} but
none of them is of any use for improving the power behavior.

The most difficult part in the numerical computation of Fox functions is to extend the algorithms derived for
univariate functions to multivariate functions; the multivariate integral may be divergent, although single,
iterated integrals converge.

\section{Bivariate Fox function \label{BFF}}
Our definition of the bivariate Fox function will be the following:
\bq
\mrH\,\Bigl[ {\mathbf z}\,;\,\lpar {\mathbf a}\,,\,{\mathbf A} \rpar\,;\,
\lpar {\mathbf b}\,,\,{\mathbf B} \rpar \Bigr] = 
\Bigl[ \prod_{\mrj=1}^{2}\,\int_{\mrL_\mrj}\,\frac{\mrd \mrs_\mrj}{\tip} \Bigr]\,
\frac{\mrN}{\mrD}\,\mrz_1^{\mrs_1}\,\mrz_2^{\mrs_2} \spc
\label{defBFF}
\eq
\bqa
\mrN &= &
\prod_{\mrj=1}^{\mrM_\mra}\,\eG{\mra_{1,\mrj,0} + \mrA_{1,\mrj,1}\,\mrs_1 + \mrA_{1,\mrj,2}\,\mrs_2}\,
\prod_{\mrj=1}^{\mrN_\mra}\,\eG{\mra_{2,\mrj,0} + \mrA_{2\,\mrj,1}\,\mrs_1 - \mrA_{2\,\mrj,2}\,\mrs_2}
\nl
{}&\times&
\prod_{\mrj=1}^{\mrP_\mra}\,\eG{\mra_{3,\mrj,0} - \mrA_{3,\mrj,1}\,\mrs_1 + \mrA_{3,\mrj,2}\,\mrs_2}\,
\prod_{\mrj=1}^{\mrQ_\mra}\,\eG{\mra_{4,\mrj,0} - \mrA_{4,\mrj,1}\,\mrs_1 - \mrA_{4,\mrj,2}\,\mrs_2} \spc
\nl
\mrD &=&
\prod_{\mrj=1}^{\mrM_\mrb}\,\eG{\mrb_{1,\mrj,0} + \mrB_{1,\mrj,1}\,\mrs_1 + \mrB_{1,\mrj,2}\,\mrs_2}\,
\prod_{\mrj=1}^{\mrN_\mrb}\,\eG{\mrb_{2,\mrj,0} + \mrB_{2,\mrj,1}\,\mrs_1 - \mrB_{2,\mrj,2}\,\mrs_2}
\nl
{}&\times&
\prod_{\mrj=1}^{\mrP_\mrb}\,\eG{\mrb_{3,\mrj,0} - \mrB_{3,\mrj,1}\,\mrs_1 + \mrB_{3,\mrj,2}\,\mrs_2}\,
\prod_{\mrj=1}^{\mrQ_\mrb}\,\eG{\mrb_{4,\mrj,0} - \mrB_{4,\mrj,1}\,\mrs_1 - \mrB_{4,\mrj,2}\,\mrs_2} \spp
\label{BFFND}
\eqa
The tensors $\mra\,,\,\mrb$ are complex while the tensors $\mrA\,,\,\mrB$ are real and positive.
The contours $\mrL_\mrj$ are straight lines defined by $\mrs_\mrj = \sigma_\mrj + \mri\,\mrt_\mrj$ such that the
integrand in \eqn{BFFND} has no singularity for $\mathbf{s}_\mrj \in \mathbf{L}_\mrj$.
\paragraph{Convergence} \hspace{0pt} \\
We start by considering the iterated integrals, \ie consider, for instance, the $\mrs_1$ integral.
By comparing \eqn{UFFtwo} and \eqn{BFF} we obtain
\bq
\mrm_1 = \mrM_\mra + \mrN_\mra \spc \quad
\mrn_1 = \mrP_\mra + \mrQ_\mra \spc \quad
\mrp_1 = \mrM_\mrb + \mrN_\mrb \spc \quad
\mrq_1 = \mrP_\mrb + \mrQ_\mrb \spc
\eq
\bqa
\mra^{(1)}_\mrj = \mra_{1,\mrj,0} + \mrA_{1,\mrj,2}\,\mrs_2 \spc &\qquad& 1 \le \mrj \le \mrM_\mra \spc
\nl
\mra^{(1)}_\mrj = \mra_{2,\mrj-\mrM_\mra,0} - \mrA_{2,\mrj-\mrM_\mra,2}\,\mrs_2 \spc &\qquad&
\mrM_\mra + 1 \le \mrj \le \mrM_\mra + \mrN_\mra \spc
\nl
\mrb^{(1)}_\mrj = \mra_{3,\mrj,0} + \mrA_{3,\mrj,2}\,\mrs_2 \spc &\qquad& 1 \le \mrj \le \mrP_\mra \spc
\nl
\mrb^{(1)}_\mrj = \mra_{4,\mrj-\mrP_\mra,0} - \mrA_{4,\mrj-\mrP_\mra,2}\,\mrs_2 \spc &\qquad&
\mrP_\mra \le \mrj \le \mrP_\mra + \mrQ_\mra \spc
\eqa
with similar results for $\mrc^{(1)}_\mrj$ and $\mrd^{(1)}_\mrj$. The remaining coefficient are simple,
\bqa
\mrA^{(1)}_\mrj = \mrA_{1,\mrj,1} \spc &\qquad& 1 \le \mrj \le \mrM_\mra \spc
\nl
\mrA^{(1)}_\mrj = \mrA_{2,\mrj-\mrM_\mra,1} \spc &\qquad& \mrM_\mra + 1 \le \mrj \le \mrM_\mra + \mrN_\mra \spc
\eqa
with similar result for $\mrB^{(1)}_\mrj\,\dots\,\mrD^{(1)}_\mrj$. We can now determine $\alpha_1, \beta_1, \lambda_1$
and $\rho_1$ and use the conditions given in \sect{UFF}. The same procedure determines the convergence of
the $\mrs_2$ integral. 
The overall convergence can be discussed by rewriting \eqn{BFF} as follows:
\bq
\mrN = \prod_{\mrj=1}^{\mrm}\,\eG{ \alpha^{\prime}_\mrj + \sum_{\mrk=1}^{2}\,\mra^{\prime}_{\mrj,\mrk}\,\mrs_{\mrk}}
\spc \quad
\mrD = \prod_{\mrj=1}^{\mrn}\,\eG{ \beta^{\prime}_\mrj + \sum_{\mrk=1}^{2}\,\mrb^{\prime}_{\mrj,\mrk}\,\mrs_{\mrk}} \spp
\label{ppar}
\eq
If we introduce
\bqa
\mrF_{1,\mrj} &= &
\prod_{\mrj=1}^{\mrM_\mra}\,\eG{\mra_{1,\mrj,0} + \mrA_{1,\mrj,1}\,\mrs_1 + \mrA_{1,\mrj,2}\,\mrs_2}
\spc \quad
\mrF_{2,\mrj} = \prod_{\mrj=1}^{\mrN_\mra}\,\eG{\mra_{2,\mrj,0} + \mrA_{2,\mrj,1}\,\mrs_1 - \mrA_{2,\mrj,2}\,\mrs_2}
\spc \nl
\mrF_{3,\mrj} &=& \prod_{\mrj=1}^{\mrP_\mra}\,\eG{\mra_{3,\mrj,0} - \mrA_{3,\mrj,1}\,\mrs_1 + \mrA_{3,\mrj,2}\,\mrs_2}\,
\spc \quad
\mrF_{4,\mrj} = \prod_{\mrj=1}^{\mrQ_\mra}\,\eG{\mra_{4,\mrj,0} - \mrA_{4,\mrj,1}\,\mrs_1 - \mrA_{4,\mrj,2}\,\mrs_2} \spc
\eqa
it follows that
\bq
\mrN =
\prod_{\mrj=1}^{\mrM_\mra}\,\mrF_{1,\mrj}\,
\prod_{\mrj=\mrM_\mra+1}^{\mrM_\mra + \mrN_\mra}\,\mrF_{2,\mrj-\mrM_\mra}\,
\prod_{\mrj=\mrM_\mra+\mrN_a+1}^{\mrM_\mra + \mrN_\mra + \mrP_\mra}\,\mrF_{3,\mrj-\mrM_\mra-\mrN_\mra}\,
\prod_{\mrj=\mrM_\mra+\mrN_\mra+\mrP_\mra+1}^{\mrM_\mra + \mrN_\mra + \mrP_\mra+\mrQ_\mra}\,
\mrF_{4,\mrj-\mrM_\mra-\mrN_\mra-\mrP_\mra} \spc
\eq
with a similar result for $\mrD$; all the parameters in \eqn{ppar} can be determined. Overall convergence will
be discussed in \sect{MFF}.
\paragraph{Contiguity} \hspace{0pt} \\
We also have contiguity relations; consider the following example:
\bq
\mrH = \Bigl[ \prod_{\mrj=1}^{2}\,\int_{\mrL_\mrj}\,\frac{\mrd \mrs_\mrj}{\tip} \Bigr]\,
\frac{
\eG{ - \mrs_1}\,\eG{1.52 + \mrs_1}\,\eG{1.41 + \mrs_1 + \mrs_2}\,\eG{ - \mrs_2}\,\eG{1.63 + \mrs_2}
     }
     {
\eG{5.95 + \mrs_1 + \mrs_2}
     } \,\mrz_1^{ - \mrs_1}\,\mrz_2^{ - \mrs_2}
\label{H2cr}
\eq
After introducing $\mrp_\mrn = \mrx_\mrn + \mry_\mrn\,\mrz$ we write the following combination,
\bqa
{}&{}& \mrp_0\,\eG{1 - \mra_1 - \mrs}\,\eG{\mrb_1 + \mrs}\,\eG{\mrb_2 + \mrs}\,\ieG{\mra_2 + \mrs} 
\nl
{}&+&
       \mrp_1\,\eG{ - \mra_1 - \mrs}\,\eG{\mrb_1 + \mrs}\,\eG{\mrb_2 + \mrs}\,\ieG{\mra_2 + \mrs} 
\nl
{}&+&  \mrp_2\,\eG{1 - \mra_1 - \mrs}\,\eG{\mrb_1 - 1 + \mrs}\,\eG{\mrb_2 + \mrs}\,\ieG{\mra_2 + \mrs} 
\nl
{}&+&
       \mrp_3\,\eG{1 - \mra_1 - \mrs}\,\eG{\mrb_1 + \mrs}\,\eG{\mrb_2 - 1 + \mrs}\,\ieG{\mra_2 + \mrs} \spc
\eqa
which, applied to the $\mrs_1$ integral in \eqn{H2cr}, gives the following solution:
\bqa
\mrx_0 &=& 1 \spc \qquad
\mrx_2 = 0 \spc \qquad
\mrx_3 = 0 \spc \qquad
\mrx_1 = - \frac{13}{25} \spc
\nl
\mry_0 &=& 1 \spc \qquad
\mry_1 = 0 \spc \qquad
\mry_2 = 0 \spc \qquad
\mry_3 = \frac{227}{50} \spp
\eqa
There are relations not involving contiguity which can be used to improve the rate of convergence, For instance,
given
\bq
\mrH = \mrF^{(2)}_{\mrD}\lpar \mrb\,;\,\mra_1\,,\,\mra_2\,;\,\mrc\,;\, - \mrz_1\,,\, - \mrz_2 \rpar \spc
\eq
we derive the following equation:
\bqa
\mrH &=&
         \mrJ_1 + (\mrz_1 + 1)^{-1}\,\Bigl\{
         \Bigl[ (1 + 2\,\mrc) + (2 - \mra_1 - \mrb + 3\,\mrc)\,\mrz_1 \Bigr]\,\mrJ_2 -
         (1 - \mrb + \mrc)\,\mrz_1\,\mrJ_3 
\nl
{}&+&
         \Bigl[ \mrc\,(\mrc + 1) + (1 - \mra_1 + 3\,\mrc - \mrc\,\mra_1 + 2\,\mrc^2 - \mrb - 
            \mrb\,\mrc)\,\mrz_1 \Bigr]\,\mrJ_4
\nl
{}&-&
         (1 - \mra_1 + 2\,\mrc - \mrc\,\mra_1 + \mrc^2 - \mrb + \mrb\,\mra_1 - \mrb\,\mrc)\,\mrz_1\,\mrJ_5 \Bigr\} \spc
\eqa
\bq
\mrJ_\mrj= \Bigl[ \prod_{\mrj=1}^{2}\,\int_{\mrL_\mrj}\,\frac{\mrd \mrs_\mrj}{\tip} \Bigr]\,
\mrR_\mrj\,\mrz_1^{\mrs_1}\,\mrz_2^{\mrs_2} \spc \quad
\mrR_\mrj = \eG{ - \mrs_1}\,\eG{ - \mrs_2}\,\mrT_{\mrj} \spc
\eq
\bqa
\mrT_1 &=&
\frac{
\eGs{1 + \mrs_2}\,\eG{\mra_1 + \mrs_1}\,\eG{\mra_2 + \mrs_2}\,
\eG{1 + \mrc + \mrs_2}\,\eG{\mrb + \mrs_{12}}
}{
\eGs{\mrs_2}\,\eG{2 + \mrc + \mrs_2}\,\eG{1 + \mrc + \mrs_{12}} 
} \spc
\nl
\mrT_2 &=&
\frac{
\eG{1 + \mrs_2}\,\eG{\mra_1 + \mrs_1}\,
\eG{\mra_2 + \mrs_2}\,\eG{1 + \mrc + \mrs_2}\,\eG{\mrb + \mrs_{12}}
}{
\eG{\mrs_2}\,\eG{2 + \mrc + \mrs_2}\,\eG{1 + \mrc + \mrs_{12}}
} \spc
\nl
\mrT_3 &=&
\frac{
\eG{1 + \mrs_2}\,\eG{\mra_1 + \mrs_1}\,
\eG{\mra_2 + \mrs_2}\,\eG{\mrb + \mrs_{12}}
}{
\eG{\mrs_2}\,\eG{2 + \mrc + \mrs_{12}}
} \spc
\nl
\mrT_4 &=&
\frac{
\eG{\mra_1 + \mrs_1}\,\eG{\mra_2 + \mrs_2}\,
\eG{1 + \mrc + \mrs_2}\,\eG{\mrb + \mrs_{12}}
}{
\eG{2 + \mrc + \mrs_2}\,\eG{1 + \mrc + \mrs_{12}} 
} \spc
\nl
\mrT_5 &=&
\frac{
\eG{\mra_1 + \mrs_1}\,\eG{\mra_2 + \mrs_2}\,\eG{\mrb + \mrs_{12}}\,
}{
\eG{2 + \mrc + \mrs_{12}}
} \spc
\eqa
where $\mrs_{12} = \mrs_1 + \mrs_2$. The $\mrJ_\mrj$ MB integrals are not Lauricella functions~\cite{Ext,FDMB} but w.r.t. the
$\mrs_1$ integrals they correspond to a shift $\lambda_1 \to \lambda_1 - 1$.
\section{Multivariate Fox function \label{MFF}}
The results of \sect{BFF} can be generalized to the case of $\mrr > 2$ variables; we introduce the following notation:
\bqa
\Gamma_1\lpar \mra_{1,\mrj,0}\,;\,\mrA_{1,\mrj,1}\,,\dots\,,\mrA_{1,\mrj,\mrr} \rpar &=&
\eG{\mra_{1,\mrj,0} + \mrA_{1,\mrj,1}\,\mrs_1 + \dots + \mrA_{1,\mrj,\mrr}\,\mrs_\mrr} \spc
\nl
\Gamma_2\lpar \mra_{2,\mrj,0}\,;\,\mrA_{2,\mrj,1}\,,\dots\,,\mrA_{2,\mrj,\mrr} \rpar &=&
\eG{\mra_{2,\mrj,0} + \mrA_{2,\mrj,1}\,\mrs_1 + \dots + \mrA_{2,\mrj,\mrr-1}\,\mrs_{\mrr-1} -
\mrA_{2,\mrj,\mrr}\,\mrs_{\mrr}} \spc
\nl
{} &\dots\dots\dots& 
\nl
\Gamma_\mrR\lpar \mra_{\mrR,\mrj,0}\,;\,\mrA_{\mrR,\mrj,1}\,,\dots\,,\mrA_{\mrR,\mrj,\mrr} \rpar &=&
\eG{\mra_{\mrR,\mrj,0} - \mrA_{\mrR,\mrj,1}\,\mrs_1 - \dots - \mrA_{\mrR,\mrj,\mrr}\,\mrs_\mrr} \spc
\eqa
with $\mrR = 2^{\mrr}$. The numerator of the integrand becomes
\bq
\mrN = 
\prod_{\mrj=1}^{\mrM_{\mra\,1}}\,\Gamma_1\lpar \mra_{1,\mrj,0}\,;\,\bigl\{\mrA_{1,\mrj,\mrk}\bigr\} \rpar\,
\prod_{\mrj=2}^{\mrM_{\mra\,2}}\,\Gamma_2\lpar \mra_{2,\mrj,0}\,;\,\bigl\{\mrA_{2,\mrj,\mrk}\bigr\} \rpar\,
\dots\,
\prod_{\mrj=1}^{\mrM_{\mra\,\mrR}}\,\Gamma_\mrR\lpar \mra_{\mrR,\mrj,0}\,;\,\bigl\{\mrA_{\mrR,\mrj,\mrk}\bigr\} \rpar
\spc
\eq
where 
\bq
\bigl\{\mrA_{\mri,\mrj,\mrk}\bigr\} = \mrA_{\mri,\mrj,1}\,,\,\dots\,,\mrA_{\mri,\mrj,\mrr} \spp
\eq
We can generalize \eqn{ppar} by rewriting $\mrH$ as follows:
\bq
\mrH = \Bigl[ \prod_{\mrj=1}^{\mrr}\,\int_{\mrL_\mri}\,\frac{\mrd \mrs_\mrj}{\tip} \Bigr]\,
\frac{\mrN}{\mrD}\,\prod_{\mrj=1}^{\mrr}\,\mrz_\mrj^{ - \mrs_\mrj} \spc
\label{convcomp}
\eq
\bq
\mrN = \prod_{\mrj=1}^{\mrm}\,\eG{\alpha_\mrj + \sum_{\mrk=1}^{\mrr}\,\mra_{\mrj\,\mrk}\,\mrs_\mrk} \spc \quad
\mrD = \prod_{\mrj=1}^{\mrn}\,\eG{\beta_\mrj + \sum_{\mrk=1}^{\mrr}\,\mrb_{\mrj\,\mrk}\,\mrs_\mrk} \spp
\eq
We introduce the following matrix:
\bqa
{\mathbf A}_{\mrj} &=& \lpar \mra_{\mrj , 1}\,\,\dots\,,\mra_{\mrj , \mrr} \rpar \quad \hbox{for} \quad
\mrj = 1\,,\,\dots\,,\mrm \spc
\nl
{\mathbf A}_{\mrm + \mrj} &=& \lpar \mrb_{\mrj , 1}\,\,\dots\,,\mrb_{\mrj , \mrr} \rpar \quad \hbox{for} \quad
\mrj = 1\,,\,\dots\,,\mrn \spp
\label{Amat}
\eqa
It follows that the integral in \eqn{convcomp} converges if
\bq
\mrF = \frac{\pi}{2}\,\sum_{\mrj=1}^{\mrm+\mrn}\,{\mathrm{sgn}}(\mrm + \frac{1}{2} - \mrj)\,
\mid {\mathbf A}^{\mrt}_\mrj\,\mrt_\mrj \mid - \mid {\mathrm arg}({\mathbf z})^{\mrt}\,{\mathbf t} \mid 
\to + \infty \spc
\label{Fcond}
\eq
for ${\mathbf t} \in \Rf^{\mrr}$ and $\mid\mid\, {\mathbf t}\,\mid\mid \to + \infty$, with $\mrt_\mrj = \Im \mrs_\mrj$.
We can interpret \eqn{Fcond} by defining the following matrices
\[
{\mathbf R}_{\mrj , \mrj_1 , \dots , \mrj_{\mrr - 1}} =
\left(
\begin{array}{c}
{\mathbf A}_{\mrj} \\
{\mathbf A}_{\mrj_1} \\
\dots \\
{\mathbf A}_{\mrj_{\mrr - 1}} \\
\end{array}
\right)
\qquad\qquad
{\mathbf P}_{\mrj_1 , \dots , \mrj_{\mrr - 1}} =
\left(
\begin{array}{c}
\marg({\mathbf z}) \\
{\mathbf A}_{\mrj_1} \\
\dots \\
{\mathbf A}_{\mrj_{\mrr - 1}} \\
\end{array}
\right)
\]
The convergenge problem is then interpreted by using the following theorem~\cite{HS}:
\begin{itemize}

\item[$\bullet$]
Define a sequence
\bq
1 \le \mrj_1 < \,\dots\, < \mrj_{\mrr-1} \le \mrm + \mrn \spc \quad
\mathrm{rank}\,{\mathbf R}_{\mrj_1 , \dots , \mrj_{\mrr-1}} = \mrr - 1
\label{Hseq}
\eq
then the generalized, multivariate, Fox function is given by a convergent integral if for any sequence
given in \eqn{Hseq} we have
\bqa
{}&1)& \quad  \rho\lpar \mrj_1\,,\,\dots\,,\,\mrj_{\mrr-1} \rpar =
\sum_{\mrj=1}^{\mrm + \mrn}\,\mathrm{sgn}\lpar \mrm + \frac{1}{2} - \mrj \rpar\,
\bmid\, \mathrm{det}\,{\mathbf R}_{\mrj , \mrj_1 , \dots , \mrj_{\mrr-1}} \bmid > 0 \spc
\nl
{}&2)& \quad
\frac{\pi}{2}\,\rho\lpar \mrj_1\,,\,\dots\,,\,\mrj_{\mrr-1} \rpar >
\bmid\,\mathrm{det}\,{\mathbf P}_{\mrj_1 , \dots , \mrj_{\mrr-1}} \bmid \spp
\label{ccond}            
\eqa

\end{itemize}
From \Bref{HS} we also obtain the following result: if there exists at least one sequence such that
the corrseponding $\rho$ is zero then the integral diverges for all 
${\mathbf z} \in \Cf^{\mrr}\,/\,\Rf^{\mrr}_{+}$.
As stated in \Bref{HS} the theorem does not give us any information about the convergence
(or divergence) of integral for $\mathbf{z} \in \Rf^{\mrN}_{+}$; the corresponding
multiple integral converges if there are some additional conditions~\cite{HY}.
Let us consider the following example:
\bq
\mrH = \Bigl[ \prod_{\mrj=1}^{2}\,\int_{\mrL_\mrj}\,\frac{\mrd \mrs_\mrj}{\tip} \Bigr]\,
\frac{
\eG{-\mrs_1}\,\eG{-\mrs_2}\,\eG{\mra_1 + \mrs_1 + \mrs_2}\,\eG{\mra_2 + \mrs_1}\,\eG{\mra_3 + \mrs_2}
     }
     {
\eG{\mrb_1 + \mrs_1}\,\eG{\mrb_2 + \mrs_2}
     }\,\mrz_1^{\mrs_1}\,\mrz_2^{\mrs_2} \spp
\label{sexa}
\eq
Given the matrix of \eqn{Amat} we start by considering the case $\mrz_\mrj \in \Rf_{+}$.  It follows
\bq
\mrF = \pi\,\Bigl[ \mid \mrt_1 \mid + \mid \mrt_2 \mid + \mid \mrt_1 + \mrt_2 \mid \Bigr] > 0 \spc
\eq
where $\mrs_\mrj = \sigma_\mrj + \mri\,\mrt_\mrj$.
The convergence of the iterated integrals is controlled by
\bqa
\alpha_1 &=& 2 \spc \quad \beta_1 = 0 \spc \quad \lambda_1 = \mra_1 + \mra_2 - \mrb_1 + \sigma_2 - 1 \spc
\nl
\alpha_2 &=& 2 \spc \quad \beta_2 = 0 \spc \quad \lambda_2 = \mra_1 + \mra_3 - \mrb_2 + \sigma_1 - 1 \spp
\eqa
To discuss the double integral we need to examine the large $\mid \mrt_\mrj \mid$ behavior, separating 
exponential from power behavior of the integrand. Let us consider the $\mrt_1 - \mrt_2$ plane and the
asymptotic behavior in different regions.
\bei

\item[\ovalbox{R1}] $\mid \mrt_2 \mid < \mrd$, $\mrt_1 \to + \infty$. The exponential behavior is
\bq
\exp\{ - ( \phi_1 + \pi)\,\mrt_1 \} \spc
\eq
where we used $ \mrt_1 > 0$ and $\mid \mrt_1 + \mrt_2 \mid \sim \mid \mrt_1 \mid$. If $\phi_1 = - \pi$ we have
to consider the power behavior, given by
\bq
\mrt_1^{\mrp_1} \spc \qquad \mrp_1 = \mra_2 - \mrb_1 + \sigma_2 - 1 \spc
\eq
requiring $\mra_2 - \mrb_1 + \sigma_2 < 0$.

\item[\ovalbox{R2}] $\mid \mrt_2 \mid < \mrd$, $\mrt_1 \to - \infty$. The exponential behavior is controlled by
$(\phi_1 - \pi)\,\mid \mrt_1 \mid$. Therefore, if $\phi_1 = \pi$ we must require $\mra_2 - \mrb_1 + \sigma_2 < 0$.

\item[\ovalbox{R3}] $\mid \mrt_1 \mid < \mrd$, $\mrt_2 \to + \infty$. If $\phi_2 = - \pi$ we must require
$\mra_1 + \mra_3 - \mrb_2 + \sigma_1 < 0$.

\item[\ovalbox{R4}] $\mid \mrt_1 \mid < \mrd$, $\mrt_2 \to - \infty$. If $\phi_2 = \pi$ we must require
$\mra_1 + \mra_3 - \mrb_2 + \sigma_1 < 0$.

\item[\ovalbox{R5}] $\mrt_1, \mrt_2 \to + \infty$. The exponential behavior is
\bq
\exp\{ - (\phi_1 + \pi)\,\mrt_1 - (\phi_2 + \pi)\,\mrt_2 \} \spc
\eq
with a power behavior given by
\bq
\mrt^{\mrp_1}\,\mrt_2^{\mrp_2}\,(\mrt_1 + \mrt_2)^{\mrp_{12}} \spc \qquad
\mrp_1 = \mra_2 - \mrb_1 - \sigma_1 - \frac{1}{2} \spc \quad
\mrp_2 = \mra_3 - \mrb_2 - \sigma_2 - \frac{1}{2} \spc \quad
\mrp_{12} = \mra_1 + \sigma_1 + \sigma_2 - \frac{1}{2} \spp
\eq
If $\mid \phi_2 \mid < \pi$ and $\phi_1 = - \pi$ we require $\mrp_1 < - 1$ and $\mrp_{12} < - 1$. Clearly
if $\mid \phi_1 \mid < \pi$ and $\phi_2 = - \pi$ we require $\mrp_2 < - 1$ and $\mrp_{12} < - 1$. Finally,
if $\phi_1 = \phi_2 = - \pi$ we require $\mrp_1 < - 1, \mrp_2 < - 1$ and $\mrp_{12} < - 1$.

\item[\ovalbox{R6}] $\mrt_1, \mrt_2 \to - \infty$. The exponential behavior is
\bq
\exp\{ (\phi_1 - \pi)\,\mid \mrt_1 \mid + (\phi_2 - \pi)\,\mid \mrt_2 \mid \} \spc
\eq
with a power behavior given by
\bq
\mid \mrt \mid^{\mrp_1}\,\mid \mrt_2 \mid^{\mrp_2}\,\mid \mrt_1 + \mrt_2 \mid^{\mrp_{12}} \spp
\eq
We have power behavior when $\phi_1$ and/or $\phi_2$ are equal to $\pi$.

\item[\ovalbox{R7a}] $\mrt_1 \to + \infty$, $\mrt_2 \to - \infty$, $\mid \mrt_1 \mid > \mid \mrt_2 \mid$. In this region
we have $\mid \mrt_1 + \mrt_2 \mid = \mrt_1 + \mrt_2$ and the exponential behavior is controlled by
\bq
(\phi_1 - \pi)\,\mid \mrt_1 \mid + \phi_2\,\mid \mrt_2 \mid \spp
\eq
\item[\ovalbox{R7b}] $\mrt_1 \to + \infty$, $\mrt_2 \to - \infty$, $\mid \mrt_1 \mid < \mid \mrt_2 \mid$. In this region
we have $\mid \mrt_1 + \mrt_2 \mid = - \mrt_1 - \mrt_2$ and the exponential behavior is controlled by
\bq
(\phi_2 - \pi)\,\mid \mrt_2 \mid + \phi_1\,\mid \mrt_1 \mid \spp
\eq
\item[\ovalbox{R8}] $\mrt_1 \to - \infty$, $\mrt_2 \to + \infty$ and  $\mid \mrt_1 \mid > \mid \mrt_2 \mid$ or
$\mid \mrt_1 \mid < \mid \mrt_2 \mid$. The exponential behavior gives the same results as in region [R7].

\eei

As far as the convergence of $\mrH$ is concerned we can accomodate
\bq
\phi_1 = - \pi \spc \quad - \pi \le \phi_2 \le 0
\qquad \hbox{or} \qquad
\phi_2 = - \pi \spc \quad - \pi \le \phi_1 \le 0 \spp
\eq
In regions $7$ and $8$ this corresponds to $\mrt_1$ exponential, $\mrt_2$ exponential unless $\phi_2 = 0$ and
$\mrt_2$ exponential, $\mrt_1$ exponential unless $\phi_1 = 0$.

Special cases are given by Lauricella functions~\cite{Ext,FDMB,SKL}; furthermore,
when $\mrn = 2$, the Lauricella functions correspond to the Appell hypergeometric series of two variables~\cite{Asur}. 
Generalized Lauricella functions have been introduced in \Bref{TBL}.
\section{Modified Fox function \label{ModFF}}
There is a new function needed in computing infrared divergent Feynman integrals~\cite{Passarino:2006gv},
\bq
\mrH_{\uppsi}\Bigl[ \mathbf{z}\,;\,(\mathbf{\alpha}\,\mathbf{a})\,;\,(\mathbf{\beta}\,,\,\mathbf{b})\,;\,
(\gamma\,,\,\mathbf{c}) \Bigr] =
\Bigl[ \prod_{\mrj=1}^{\mrr}\,\int_{\mrL_\mri}\,\frac{\mrd \mrs_\mrj}{\tip} \Bigr]\,
\frac{\mrN}{\mrD}\,\prod_{\mrj=1}^{\mrr}\,( - \mrz_\mrj)^{\mrs_\mrj} \spc
\label{Hpsi}
\eq
\bq
\mrN = \prod_{\mrj=1}^{\mrm}\,\eG{\alpha_\mrj + \sum_{\mrk=1}^{\mrr}\,\mra_{\mrj\,\mrk}\,\mrs_\mrk}\,
\uppsi(\gamma + \sum_{\mrk=1}^{\mrr}\,\mrc_\mrk\,\mrs_\mrk) \spc \quad
\mrD = \prod_{\mrj=1}^{\mrn}\,\eG{\beta_\mrj + \sum_{\mrk=1}^{\mrr}\,\mrb_{\mrj\,\mrk}\,\mrs_\mrk} \spc
\eq
where $\uppsi(\mrz)$ is the digamma function~\cite{HTF}.
We give a simple example:
\bq
\mrI = \pi^{\ep/2}\,\eG{1 - \frac{\ep}{2}}\,\mu^{ - \ep}\,\int_0^1 \mrd \mrx_1\,\int_0^{\mrx_1} \mrd \mrx_2\,\Bigl[
\mrs\,\mrx_1\,\mrx_2 + \mrm^2\,\mrx_1 - ( \mrs + \mrm^2)\,\mrx_2 \Bigr]^{ - 1 + \ep/2} \spp
\eq
The limit $\ep \to 0$ is given by
\bq
\mrm^2\,\mrI = \frac{\mrI_{-1}}{\epb} + \mrI_{0} + \ord{\ep} \spc \qquad
\epb^{\,-1}= \ep^{-1} + \frac{1}{2}\,\Bigl[ \uppsi(1) + \ln \pi + \ln\frac{\mrm^2}{\mu^2} \Bigr] \spc
\eq
\bqa
\mrI_{-1} &=& \mrF^{(2)}_{\mrD}\lpar 1\,;\,1\,,\,1\,;\,2\,;\,\mrx_{-}^{-1}\,,\,\mrx_{+}^{-1} \rpar \spc
\nl
\mrI_{0} &=& - \frac{1}{2}\,\Bigl[ \prod_{\mrj=1}^{2}\,\int_{\mrL_\mrj}\,\frac{\mrd \mrs_{\mrj}}{\tip} \Bigr]\,
\eG{ - \mrs_1}\,\eG{ - \mrs_2}\,\eG{1 + \mrs_1}\,\eG{1 + \mrs_2}\,
\frac{\eG{1 + \mrs_1 + \mrs_2}}{\eG{2 + \mrs_1 + \mrs_2}}
\nl
{}&\times&
\Bigl[ \uppsi(1 + \mrs_1) + \uppsi(1 + \mrs_2) \Bigr]\,( - \mrx_{-})^{- \mrs_1}\,( - \mrx_{+})^{ - \mrs_2} \spc
\eqa
where $\mrx_{\pm} = (1 \pm \beta)/2$, $\beta^2 = 1 - 4\,\mrm^2/s$.

The result can be generalized by using a formula for the higher derivatives of $\Gamma(\mrz)$ and its reciprocal,
due to Masayuki~\cite{Masa}:
\bqa
\frac{\mrd^\mrn}{\mrd \mrz^\mrn}\,\eG{\mrz} &=& \eG{\mrz}\,\sum_{\mrj=1}^{\mrn}\,\mrB_{\mrn\,,\,\mrj}
\lpar \uppsi\,,\,\uppsi^{(1)}\,,\,\dots\,,\,\uppsi^{(\mrn - \mrj)} \rpar \spc
\nl
\frac{\mrd^\mrn}{\mrd \mrz^\mrn}\,\frac{1}{\eG{\mrz}} &=& 
\frac{1}{\eG{\mrz}}\,\sum_{\mrj=1}^{\mrn}\,( - 1 )^\mrj\,\mrB_{\mrn\,,\,\mrj}
\lpar \uppsi\,,\,\uppsi^{(1)}\,,\,\dots\,,\,\uppsi^{(\mrn - \mrj)} \rpar \spc
\eqa
where $\uppsi^{(\mrn)}$ is the polygamma function and $\mrB_{\mrn\,,\,\mrj}$ are Bell polynomials of 
the second kind~\cite{Bell}. Given
\bq
\mrB_{\mrn\,,\,\mrj}\lpar \mrx_1\,,\,\dots\,,\,\mrx_{\mrn - \mrj + 1} \rpar \spc
\eq
the first polynomials are
\bq
\mrB_{1,1} = \mrx_1 \spc \quad
\mrB_{2,1} = \mrx_2 \spc \quad
\mrB_{2,2} = \mrx_1^2 \spc \quad
\mrB_{3,1} = \mrx_3 \spc \quad
\mrB_{3,2} = 3\,\mrx_1\,\mrx_2 \spc \quad
\mrB_{3,3} = \mrx_1^3 \spp
\eq
Using Bell polynomials we derive the following expansions, Cf. \Bref{Weinzierl:2004bn,Puhlfurst:2015vqx}:
\bqa
\eG{\mra + \mrb\,\ep} &=& \eG{\mra}\,\Bigl[
1 + \sum_{\mrn=1}^{\infty}\,\frac{(\mrb\,\ep)^\mrn}{\mrn\,!}\,\sum_{\mrj=1}^{\mrn}\,\mrB_{\mrn\,,\,\mrj}\lpar
\uppsi(\mra)\,,\,\dots\,,\,\uppsi^{(\mrn - \mrj)}(\mra) \rpar \Bigr] \spc
\nl
\eG{\mra\,\ep} &=& \frac{1}{\mra\,\ep}\,\Bigl[
1 + \sum_{\mrn=1}^{\infty}\,\frac{(\mra\,\ep)^\mrn}{\mrn\,!}\,\sum_{\mrj=1}^{\mrn}\,\mrB_{\mrn\,,\,\mrj}\lpar
\uppsi(1)\,,\,\dots\,,\,\uppsi^{(\mrn - \mrj)}(1) \rpar \Bigr] \spp
\eqa
Using the Riemann zeta function we derive
\bq
\psin{1} = ( - 1)^{\mrn + 1}\,\mrn\,!\;\zeta(\mrn + 1) \spc \qquad
\zeta(2\,\mrn) = ( - 1)^{\mrn + 1}\,(2\,\pi)^{2\,\mrn}\,\frac{\mrB_{2\,\mrn}}{2\,(2\,\mrn\,!)} \spc
\eq
where $\mrB_{2\,\mrn}$ are Bernoulli's numbers~\cite{HTF} and $\zeta(2\,\mrn + 1)$ is irrational.
The general expansion requires more general modified Fox functions, \eg
\bq
\mrH^{(\mrn)}_{\uppsi}\Bigl[ \mathbf{z}\,;\,(\mathbf{\alpha}\,\mathbf{a})\,;\,(\mathbf{\beta}\,,\,\mathbf{b})\,;\,
\gamma \Bigr] =
\Bigl[ \prod_{\mrj=1}^{\mrr}\,\int_{\mrL_\mri}\,\frac{\mrd \mrs_\mrj}{\tip} \Bigr]\,
\frac{\mrN}{\mrD}\,\prod_{\mrj=1}^{\mrr}\,( - \mrz_\mrj)^{\mrs_\mrj} \spc
\label{Hnpsi}
\eq
\bq
\mrN = \prod_{\mrj=1}^{\mrm}\,\eG{\alpha_\mrj + \sum_{\mrk=1}^{\mrr}\,\mra_{\mrj\,\mrk}\,\mrs_\mrk}\,
\sum_{\mrj=1}^{\mrn}\,\mrB_{\mrn\,,\,\mrj}
\lpar \uppsi(\gamma)\,,\,\dots\,,\,\uppsi^{(\mrn - \mrj)}(\gamma) \rpar \spc \quad
\mrD = \prod_{\mrj=1}^{\mrn}\,\eG{\beta_\mrj + \sum_{\mrk=1}^{\mrr}\,\mrb_{\mrj\,\mrk}\,\mrs_\mrk} \spc
\eq
with
\bq
\uppsi(\mrz) \sim \ln \mrz + \ord{\mrz^{-1}} \spc \qquad
\psin{\mrz} \sim \mrz^{ - \mrn} \spc
\eq
for $\mrz \to \infty$ in $\mid \marg(\mrz) \mid < \pi$.
Hypergeometric sums involving the digamma function have been discussed in \Bref{psis}.
MB integrals containing $\psin{\mrz}$ functions can also be computed by using the integral representations:
\bq
\uppsi(\mrz) - \uppsi(1) = \int_0^1\,\mrd \mrx\,\frac{1 - \mrx^{\mrz - 1}}{1 - \mrx} \spc
\qquad
\psin{\mrz} - \psin{1} = \int_0^1\,\mrd \mrx\,\frac{1 - \mrx^{\mrz - 1}}{1 - \mrx}\,\ln^{\mrn}\,\mrx \spc
\eq
with $\Re \mrz > 0$. For instance we obtain~\cite{Passarino:2001wy}:
\bq
\hyp{\frac{1 - \ep}{2}}{1}{\frac{1}{2}}{\mrz^2} = \frac{1}{1 - \mrz^2}\,\Bigl[
1 - \frac{\ep}{2}\,\mrz^2\,\ln\frac{1 + \mrz^2}{1 - \mrz^2} + \ord{\ep^2} \Bigr] \spc
\eq 
\bq
\int_{\mrL}\,\frac{\mrd \mrs}{\tip}\,\eG{ - \mrs}\,\eG{1 + \mrs}\,\uppsi(1 + \mrs)\,( - \mrz)^\mrs =
\frac{1}{1 - \mrz}\,\Bigl[ \uppsi(1) - \ln(1 - \mrz) \Bigr] \spp
\eq
\bq
\int_{\mrL}\,\frac{\mrd \mrs}{\tip}\,\eG{ - \mrs}\,\eG{1 + \mrs}\,\psin{1 + \mrs}\,( - \mrz)^\mrs =
\frac{\psin{1}}{1 - \mrz} + ( - 1)^\mrn\,\mrn\,!\,\frac{1}{\mrz}\,\mrS_{\mrn\,,\,1}(\mrz) \spc
\eq
where $\mrS_{\mrn\,,\,\mrp}$ are Nielsen's generalized polylogarithms~\cite{Kolbig:1983qt,Devoto:1983tc}.

A more general example of expansion is as follows:
\bq
\mrH_{\ep} = \Bigl[ \prod_{\mrj=1}^{2}\,\int_{\mrL_\mrj}\,\frac{\mrd \mrs_\mrj}{\tip} \Bigr]\,
\eG{ - \mrs_1}\,\eG{ - \mrs_2}\,\eG{\mra_1 + \ep + \mrs_1}\,\eG{\mra_2 + \ep + \mrs_2}\,
\frac{\eG{\mrb + \mrs_1 + \mrs_2}}{\eG{\mrc + \mrs_1 + \mrs_2}}\,( - \mrz_1)^{\mrs_1}\,( - \mrz_2)^{\mrs_2} \spp
\eq
After introducing
\bq
\mrR(\mrs_1\,,\,\mrs_2\,;\,\mrz_1\,,\,\mrz_2) =
\eG{ - \mrs_1}\,\eG{ - \mrs_2}\,\eG{\mra_1 + \mrs_1}\,\eG{\mra_2 + \mrs_2}\,
\frac{\eG{\mrb + \mrs_1 + \mrs_2}}{\eG{\mrc + \mrs_1 + \mrs_2}}\,( - \mrz_1)^{\mrs_1}\,( - \mrz_2)^{\mrs_2} \spc
\eq
\bq
\Uppsi^{(\mrn)} = \uppsi^{(\mrn)}(\mra_1 + \mrs_1) + \uppsi^{(\mrn)}(\mra_2 + \mrs_2) \spc
\eq
we obtain the following expansion:
\bqa
\mrH_{\ep} &=& \Bigl[ \prod_{\mrj=1}^{2}\,\int_{\mrL_\mrj}\,\frac{\mrd \mrs_\mrj}{\tip} \Bigr]\,
\mrR(\mrs_1\,,\,\mrs_2\,;\,\mrz_1\,,\,\mrz_2)\,\Bigl\{
1 + \ep\,\Uppsi^{(0)} +
\frac{\ep^2}{2\,!}\,\Bigl[ \Uppsi^{(0)}\,\Uppsi^{(0)} + \Uppsi^{(1)} \Bigr] 
\nl
{}&+& \frac{\ep^3}{3\,!}\,\Bigl[ \Uppsi^{(0)}\,\Uppsi^{(0)}\,\Uppsi^{(0)} +
3\,\Uppsi^{(0)}\,\Uppsi^{(1)} + \Uppsi^{(2)} \Bigr] + \ord{\ep^4} \Bigr\} \spp
\eqa
Numerical results will be discussed in \sect{exa}.
\section{Analytic continuation \label{AC}}
The connection between Feynman integrals and Fox functions
has been discussed in \Bref{Passarino:2024ugq}. Here we want to summarize the main steps.
First of all the Feynman integral is written in terms of the Feynman
parametrization; the first integral over the Feynman parameters is performed producing a generalyzed 
hypergeometric function, usually
a Lauricella $\mrF^{(\mrN)}_\mrD$ function~\cite{Ext,FDMB}. For the latter we use the
corresponding MB representation~\cite{FDMB} and then we compute the second integral and
repeat step{-}by{-}step the procedure. 
For instance 
\bq
\int_0^1 \mrd \mrx\,\mrx^{\mrb - 1}\,(1 - \mrx)^{\mrc - \mrb - 1}\,
\prod_{\mrj=1}^{2}\,(1 - \mrz_\mrj\,\mrx)^{- \mra_\mrj} =
\frac{\eG{\mrb}\,\eG{\mrc - \mrb}}{\eG{\mrc}}\,
\mrF^{(2)}_{\mrD}\lpar
\mra_1\,\,\mra_2\,;\,\mrb\;,\mrc\;\,\mrz_1\,\,\mrz_2 \rpar \spc
\eq
valid in the domain
\bq
\Lf^2 = \{\mathbf{z} \in \Cf^2\,:\,\mid \marg(1 - \mrz_\mrj) \mid < \pi\,,\,
\mrj= 1,2\} \spp
\eq
In the general case~\cite{FDMB} we will have
\bqa
\mrF^{(\mrN)}_{\sPD}\lpar \mra\,;\,\mrb_1\,,\dots\,,\mrb_{\mrN}\,;\,\mrc\,;\,
\mrz_1\,,\dots\,,\mrz_{\mrN}\rpar &=&
\frac{\eG{\mrc}}{\eG{\mra}\,\prod_{\mrj}\,\eG{\mrb_{\mrj}}}\,
\Bigl[ \prod_{\mrj=1}^{\mrN}\,\int_{\mrL_{\mrj}}\,\frac{\mrd \mrs_{\mrj}}{\tip} \Bigr]\,
\frac{\eG{\mra + \sum_{\mrj}\,\mrs_{\mrj}}}{\eG{\mrc + \sum_{\mrj}\,\mrs_{\mrj}}}
\nl
{}&\times&
\prod_{\mrj=1}^{\mrN}\,\eG{\mrb_{\mrj} + \mrs_{\mrj}}\,\eG{ - \mrs_{\mrj}}\,\lpar - \mrz_{\mrj} \rpar^{\mrs_{\mrj}} \spc
\label{FDMB}
\eqa
where $\mrL_{\mrj}$ is a deformed imaginary axis curved so that only the poles of $\eG{ - \mrs_{\mrj}}$ lie
to the right of $\mrL_{\mrj}$. 

There are subtle points and we will discuss them by means of a simple example: suppose to consider
\bq
\mrI = \int_0^1 \mrd \mrx\,
\mrx^{\mrb - 1}\,(1 - x)^{\mrc - \mrb - 1}\,
(1 - \mrz\,\mrx)^{- \mra} \spc
\label{exaf21}
\eq
which is the first integral to compute under the condition
$\Re \mrc > \Re \mrb > 0$ 
and $\mid \marg(1 - \mrz) \mid < \pi$
we obtain
\bq
\mrI = \eB{\mrb}{\mrc - \mrb}\,\hyp{\mra}{\mrb}{\mrc}{\mrz} \spp
\label{ItoF21}
\eq
Therefore, for $\mrz \in \Rf$ and $\mrz > 1$ we cannot use \eqn{ItoF21}.
Furthermore, assuming that \eqn{ItoF21} is valid, we write
\bq
\hyp{\mra}{\mrb}{\mrc}{\mrz} =
\frac{
      \eG{\mrc}
     }
     {
      \eG{\mra}\,\eG{\mrb}
     }\,
\int_{\mrL}\,\frac{\mrd \mrs}{\tip}\,\eG{ - \mrs}\,
\frac{
      \eG{\mra + \mrs}\,\eG{\mrb + \mrs}
      }
      {
       \eG{\mrc + \mrs}
      }\,
( - \mrz)^{\mrs} \spp
\label{F21MB}
\eq
where $\mid \marg{( - \mrz)} \mid < \pi$.
When $\mrz > 0$ \eqn{F21MB} cannot be used. However, in \eqn{exaf21} the Feynman prescription corresponds to
$\mrz \to \mrz - \mri\,\delta$, with $\delta \to 0_{+}$. Therefore we write
\bq
\mrI = ( - \mrz + \mri\,\delta)^{ - \mra}\,
\int_0^1 \mrd \mrx\,\mrx^{\mrb - 1}\,(1 - \mrx)^{\mrc - \mrb - 1}\,
\lpar \mrx - \frac{1}{\mrz} - \mri\,\delta \rpar^{ - \mra} \spp
\eq
This integral can be interpreted, for $\mrz > 1$, by using the 
Sokhotski{-}Plemelj{-}Fox theorem~\cite{Sthe,SPF} which is a statement~\cite{FPI} on the relationship between
the integral and the Cauchy Principal Value (CPV) or the Hadamard Finite-Part Integrals (FPI).
For instance, consider
\bq
\Upphi^{\pm}(\mrx_0) = {\mathcal H}\,\int_{\mra}^{\mrb} \mrd \mrx\,
\frac{\mrf(\mrx)}{( \mrx - \mrx_0 )^{\mrn + 1}} \pm
i\,\pi\,\frac{\mrf^{(\mrn)}(\mrx_0)}{\mrn\,!} \spp
\eq
where the integral takes on either the CPV ($\mrn = 0$) or the Hadamard FPI ($\mrn \ge 1$).
Let us consider the generalization where $\alpha$ is not an integer and define
\bq
\upphi^{\pm}_{\alpha}(\mrx_0) = \lim_{\delta \to 0}\,\int_{\mra}^{\mrb} \mrd \mrx\,
\lpar \mrx - \mrx_0 \pm \mri\,\delta \rpar^{-\alpha} \spc 
\label{mdiff}
\eq
with $\mrx_0 \in [\mra\,,\,\mrb]$. We obtain
\bq
\upphi^{\pm}_{\alpha}(\mrx_0) = \lim_{\delta \to 0}\,\lpar \pm \mri\,\delta \rpar^{- \alpha}\,\Bigl[
\mrX_\mrb\;{}_2\mrF_1\lpar \alpha\,,\,1\,;\,2\,;\, \pm i\,\frac{\mrX_\mrb}{\delta}\rpar -
\mrX_\mra\;{}_2\mrF_1\lpar \alpha\,,\,1\,;\,2\,;\, \mp i\,\frac{\mrX_\mra}{\delta}\rpar \Bigr] \spc
\eq
which is valid for $\mid \mathrm{arg} ( 1 \pm i\,\frac{\mrX_\mra}{\delta} ) \mid < \pi$ and
$\mid \mathrm{arg} ( 1 \mp i\,\frac{\mrX_\mrb}{\delta} ) \mid < \pi$, with $\mrX_{\mra(\mrb)}= \mra(\mrb) - \mrx_0$.
Using
\bq
{}_2\mrF_1\lpar \alpha\,,\,1\,;\,2\,;\,\mrz \rpar = \frac{1}{1 - \alpha}\,\Bigl[
( - \mrz )^{- \alpha}\,{}_2\mrF_1\lpar \alpha\,,\,\alpha - 1\,;\,2 - \alpha\,;\,\mrz^{-1}\rpar
 + \frac{1}{\mrz} \Bigr] \spc
\eq
which is valid for $\mid \mathrm{arg} ( - \mrz ) \mid < \pi$, we derive 
\bq
\upphi^{\pm}_{\alpha}(\mrx_0) = \frac{1}{1 - \alpha}\,\Bigl[ 
\lpar \mrb - \mrx_0 \pm i\,0 \rpar^{1 - \alpha} -
\lpar \mra - \mrx_0 \pm i\,0 \rpar^{1 - \alpha}\Bigr] \spc
\eq
for arbitrary values of $\alpha$. If $\mrf(\mrx)$ is analytic in $\mrx_0$ we obtain
\bqa
\Upphi^{\pm}(\mrx_0) &=&
\sum_{\mrn=0}^{\mrN - 1}\,\frac{\mrf^{(\mrn)}(\mrx_0)}{\mrn\,!}\,\upphi^{\pm}_{\alpha - \mrn}(\mrx_0) +
\int_{\mra}^{\mrb} \mrd \mrx\,{\overline{\mrf}}(\mrx)\,\lpar \mrx - \mrx_0 - i\,0\rpar^{- \alpha} \spc
\nl
{\overline{\mrf}}(\mrx) &=& \mrf(\mrx) - \sum_{\mrn=0}^{\mrN - 1}\,\mrf^{(\mrn)}(\mrx_0)\,(\mrx - \mrx_0)^\mrn \spc
\eqa
where $\mrN - \Re\,\alpha \ge 0$ and $\mrf^{(\mrn)}(\mrx_0)$ not singular for 
$\mrn \le \mrN - 1$.

We present a few examples, having introduced $\mrx_0 = 1/\mrz$.
\bq
\lim_{\delta \to 0}\,\int_0^1 \mrd \mrx\,(1 - \mrx)^2\,
(\mrx - \mrx_0 - \mri\,\delta)^{-1} =
\mrx_0 - \mrx_0\,(1 - \mrx_0)\,
\ln \frac{1}{\mrx_0 - 1 + i\,\pi}  \spp
\eq
The MB representation, originally defined for $\mrz < 0$ is convergent in
a larger domain if (as in this case) $\Re (\mrc - \mra - \mrb) > 0$,
giving the analytic continuation of $\mrI$ in that region. We have
verified the numerical agreement for $\mrz > 1$.

All the relations between Feynman integrals and Fox functions are based on this assumption which, in
general terms can be stated as follows: the convergence of a Fox function
is controlled by three parameters $\alpha, \beta$ and $\lambda$. For $\mrz < 0$ and
$\alpha = 2, \beta = 0$ we need $\lambda < - 1$. Returning to our simple example
let us consider the case $\mrz < 0$ and $\Re(\mrc - \mra - \mrb) \le 0$.
In this case we can (repetitively) use \eqn{F21cont} until all the hypergeometric
functions satisfy $\Re(\mrc - \mra - \mrb) > 0$. For these functions we
can use the MB representation with a result which is convergent also in the region
$\mrz > 0$, giving an analytic continuation. We illustrate the
procedure by considering the following example:
\bq
\mrI = \int_0^1 \mrd \mrx\,(1 - \mrx)^{-1/2}\,(1 - \mrz\,\mrx)^{-2} \spc
\eq
 where $\mrz$ must be understood as $\mrz - \mri\,\delta, \delta \to 0_{+}$. It follows
\bq
\mrI = x_0^2\,\int_0^1 \mrd \mrx\,
(1 - \mrx)^{-1/2}\,(\mrx - \mrx_0 - \mri\,\delta)^{-2} \spc \quad
\mrx_0 = 1/\mrz \spp
\eq
We derive
\bq
\mrI = \frac{1}{2}\,\mrx_0^2\,(1 - \mrx_0^2)^{-3/2}\,\Bigl[
\ln \frac{1}{\mrx_0} - 1) + i\,\pi \Bigr] -
\frac{1}{\mrx_0}\,(1 - \mrx_0)^{-1/2} - (1 - \mrx_0)^{-3/2} +
\mrI_{+} \spc
\eq
where we have defined the $\mrI_{+}$ distribution as follows:
\bq
\mrI_{+} = (1 - \mrx_0)^{-3/2}\,\mrJ \spc \qquad
\mrJ = \int_{\mrz_0}^{1}\,\frac{\mrd \mrz}{\mrz^2}\,\Bigl[ (1 - \mrz)^{-1/2} - 1 - \frac{1}{2}\,\mrz \Bigr] \spc
\eq
where $\mrz_0 = \mrx_0/(\mrx_0 - 1)$. To compute $\mrJ$ we introduce
$\mrzb$ where $0 < \mrzb < 1$ and obtain
\bqa
\mrJ &=& 1 + \frac{1}{\mrz_0} + \frac{1}{\mrz_0}\,(1 - \mrz_0)^{-1/2} -
\frac{1}{\mrzb}\,(1 - \mrzb)^{-1/2} + 2\,(1 - \mrzb)^{1/2}\,\hyp{2}{\frac{1}{2}}{\frac{3}{2}}{1 - \mrzb}
\nl
{}&+& \frac{1}{2}\,\ln \mrzb + \frac{3}{4}\,\Bigl[
\mrz_0\,\mathrm{Li}_{5/2}(\mrz_0) - \mrzb\,\mathrm{Li}_{5/2}(\mrzb) \Bigr]
\spc
\eqa
where we have introduced the function
\bq
\mathrm{Li}_{\mu}(\mrz) = \int_0^1\,\mrd \mrx\,
(1 - \mrz\,\mrx)^{- \mu}\,\ln \mrx =
- \,\upphi^{*}_{\mu}(\mrz\,,\,2\,,\,1) \spc 
\eq
where $\upphi^{*}_{\mu}$ is the generalized Lerch function~\cite{GLerch}.
When $\mrz < 0$ we can use the following relation:
\bqa
{}&{}& \hyp{2}{1}{\frac{3}{2}}{\mrz} =
\frac{\sqrt{\pi}}{8}\,\Bigl[
              \lpar 9 - \frac{7}{\mrz - 1} - \frac{1}{(\mrz - 1)^2} \rpar\,\mrH_1 -
\frac{15}{2}\,\lpar 2 + \frac{1}{\mrz - 1} - \frac{1}{(\mrz - 1)^2} \rpar\,\mrH_2 \Bigr] \spc
\nl
{}&{}& \mrH_{\mrn} = \mrH\lpar 2\,,\,1\,,\,\frac{5 + 2\,\mrn}{2}\,,\, - \mrz \rpar \spc \quad 
\mrH \lpar \mra\,,\,\mrb\,,\,\mrc\,,\, - \mrz \rpar =
\int_{\mrL} \frac{\mrd \mrs}{\tip}\,\eG{ - \mrs}\,
\frac{
      \eG{\mra + \mrs}\,\eG{\mrb + \mrs}
      }
      {
       \eG{\mrc + \mrs}
      }\,( - \mrz)^{\mrs} \spp
\label{MBac}
\eqa
The MB integrals in \eqn{MBac} converge also for $\mrz > 0$ giving an analytic
continuation in that region. Therefore, after using the contiguity relations,
we can still proceed with the algorithm, bu only above the corresponding normal threshold.
A complete description is given in \Bref{FFPR}.

To summarize: the original Euler{-}Mellin integral is defined in a domain $\mrD \in \Cf$. Outside this
domain it is defined as the Hadamard finite{-}part integral~\cite{Hreg}. The Euler{-}Mellin integral can
be rewritten in terms of a MB representation. By using contiguity relations (when needed)
the resulting sum of MB integrals gives the analytic continuation of the original integral above the normal
threshold. The two procedures agree and the agreement has been tested numerically.

To give an example of the connection between Feynman integrals and Fox functions we consider
a scalar one{-}loop three{-}point functions in arbitrary space{-}time dimensions~\cite{Bardin:1999ak}:
\bq
\mrJ_{\mrd} = \mrC_0\lpar \,0\,,\,0\,,\,\mrs\,;\,0\,,\,\mrM\,,\,0 \rpar
 = \pi^{\ep/2}\,\eG{1 - \frac{\ep}{2}}\,\int_0^1 \mrd \mrx_1\,\int_0^{\mrx_1} \mrd \mrx_2\,\Bigl[
\mrs\,\mrx_1\,\mrx_2 + \mrM^2\,\mrx_1 - ( \mrs + \mrM^2)\,\mrx_2 \Bigr]^{ - 1 + \ep/2} \spp
\eq
After performing the $\mrx_2$ integral we obtain
\bq
\mrJ_{\mrd} = \pi^{\ep/2}\,\eG{1 - \frac{\ep}{2}}\,\int_0^1 \mrd \mrx \; \mrx\,\mrb^{\ep/2 - 1}\,
\hyp{1 - \frac{\ep}{2}}{1}{2}{ - \frac{\mra}{\mrb}\,\mrx} \spc
\eq
where we have introduced
\bq
\mra = - \mrs\,( 1 - \mrx ) - \mrM^2 \spc \qquad
\mrb = \mrM^2\,\mrx \spp
\eq
Next we use the MB representation for the Gauss hypergeometric function and perform the 
$\mrx$ integration by using
\bq
\int_0^1 \mrd \mrx\,\mrx^{\ep/2}\,\Bigl(1 - \frac{\mrs}{\mrs + \mrM^2}\,\mrx \Bigr)^{\mrs_1} =
\frac{\eG{1 + \ep/2}}{\eG{2 + \ep/2}}\,
\hyp{ - \mrs_1}{1 + \frac{\ep}{2}}{2 + \frac{\ep}{2}}{\frac{\mrs}{\mrs + \mrM^2}} \spp
\eq
Using again a MB representation we obtain
\bqa
\mrJ_{\mrd} &=& \pi^{\ep/2}\,\mrM^{\ep - 2}\,\Bigl[ \prod_{\mrj=1}^{2}\,\int_{\mrL_\mrj}\,
\frac{\mrd \mrs_\mrj}{\tip} \Bigr]\,
\frac{\mrN}{\mrD}\,
\lpar - \frac{\mrs + \mrM^2}{\mrM^2} \rpar^{\mrs_1}\,
\lpar - \frac{\mrs}{\mrs + \mrM^2} \rpar^{\mrs_2} \spc
\nl
\mrN &=&
\eG{ - \mrs_2}\,\eG{1 + \mrs_1}\,\eG{\mrs_2 - \mrs_1}\,
\eG{1 - \frac{\ep}{2} + \mrs_1}\,\eG{1 + \frac{\ep}{2} + \mrs_2} \spc
\nl
\mrD &=& \eG{2 + \mrs_1}\,\eG{2 + \frac{\ep}{2} + \mrs_2} \spp
\eqa
Therefore, $\mrJ$ is a generalized Fox function. Note that the $\mrs_1$ integral has 
$\lambda_1 = - 1 + \sigma_2 - \ep/2$ while the $\mrs_2$ integral has $\lambda_2 = - 2 - \sigma_1$. For
$\ep = 0$ we have $-1 < \sigma_1 < 0$ and $- \sigma_1 < \sigma_2 < 0$.
If contiguity relations are needed we observe that the $\mrs_2$ integral can be rewritten as
\bq 
\frac{\eG{ - \mrs_1}}{1 + \frac{\ep}{2}}\,\hyp{ - \mrs_1}{1 + 
\frac{\ep}{2}}{2 + \frac{\ep}{2}}{\mrz_2} \spc
\eq
with $\mrz_2 = \mrs/(\mrs + \mrM^2)$. We can use \eqn{F21cont}.
The $\mrs_1$ integral can be rewritten as
\bq
\mrG^{1,2}_{2,2}\lpar \mrz_1\,;\,0\,,\,\frac{\ep}{2}\,;\,\mrs_2\,,\,- 1 \rpar =
\frac{\eG{1 - \mrs_2}}{\eG{1 - \frac{\ep}{2}}}\,
\hyp{1 - \mrs_2}{2}{1 - \frac{\ep}{2}}{ - \mrz_1} \spc
\eq
with $\mrz_1 = - 1 - \mrs/\mrM^2$. Again we can use \eqn{F21cont}.
\section{Singularities of the Fox function \label{Hls}}
When the $\mrH$ function is the univariate Meijer $\mrG$ function we know that~\cite{HTF}:
\begin{enumerate}

\item if $\mrp < \mrq$ the only singularities are at $\mrz= 0\,,\,\infty$; $\mrz = 0$ is a regular
singularity, $\mrz = \infty$ an irregular one;

\item if $\mrp = \mrq$, $\mrz = \infty$ is also a regular singularity; moreover, the point 
$\mrz = ( - 1)^{\mrp - \mrm - \mrn}$ is also a regular singularity.

\end{enumerate}
The (univariate) Fox function is, in general, multivalued but one{-}valued on the Riemann surface
of $\ln \mrz$~\cite{BRaa}; the analytic stucture follows from the conditions I{-}IV of \sect{UFF}. 
In particular, when $\alpha$ and $\beta$ defined in \eqn{cpar} are zero the nature of the Fox
function is determined by the parameters $\lambda$ and $\rho$, again defined in \eqn{cpar}

For multivariate Fox functions no fundamental system for the neighborhood of the singular points is
known in the literature. An exception is given by the Lauricella $\mrF^{(\mrN)}_{\mrD}$ functions; as
explained in Section~2 of \Bref{FDMB} they can be given in terms of one{-}dimensional MB integrals
obtaining a system of analytic continuation formulae, generalizing the well{-}known results for the
Gauss hypergeometric function. 
The Horn system of partial differential equations for a bivariate Fox function has been discussed in
Section~3 of \Bref{Passarino:2024ugq}.
This approach has been used in \Brefs{Kershaw:1973km,PhysRevD.9.370,PhysRevD.11.452}, where the system corresponding to 
the Feynman integral for a one{-}loop diagram has been constructed, see also \Bref{Ghs,Moriello:2019yhu,Armadillo:2022ugh}.

We will discuss one particular example:
\bqa
\mrH_{\AT} &=& \lpar - \zeta_0\,\zeta \rpar^{ - 1 + \ep/2}\,
\Bigl[ \prod_{\mrj=1}^{3}\,\int_{\mrL_\mrj}\,\frac{\mrd \mrs_\mrj}{2\,\mri\,\pi} \Bigr]\,\frac{\mrN}{\mrD}\, 
\mrz_0^{\mrs_1}\,( - \mrz_{-} )^{ - \mrs_2}\,( - \mrz_{+} )^{ - \mrs_3} \spc
\nl
\mrN &=&
\eG{ - \mrs_2}\,
\eG{ - \mrs_3}\,
\eG{\mrs_2 - \mrs_1}\,
\eG{\mrs_3 - \mrs_1}\,
\eG{1 + \mrs_3 + \mrs_2}\,
\eG{1 + \ep + \mrs_1}\,
\eG{1 - \frac{1}{2}\,\ep + \mrs_1} \spc
\nl
\mrD &=&
\eG{ - \mrs_1}\,
\eG{2 + \mrs_3 + \mrs_2}\,
\eG{2 + \ep + \mrs_1} \spc
\label{HAT}
\eqa
which is a generalized Fox function with parameters $\mrr = 3, \mrm = 7$ and $\mrn = 3$. 
In \eqn{HAT} we have 
\bq
\mrz_{+} + \mrz_{-} = 1 \spc \qquad
\zeta_0 = \frac{1}{\mrz_0 - 1} \spc \qquad
\zeta = \mrz_{-}\,(\mrz_{-} - 1) \spp
\eq
In order to find the leading singularity of $\mrH_{\AT}$ we introduce the matrix $\mrH$,
\bq
\mrH_{11} = \mrH_{22} = \mrh = - \zeta \spc \qquad
\mrH_{12} = \zeta + \frac{1}{2} \spc \qquad
\mrG = {\mathrm{det}}\,\mrH = - \zeta - \frac{1}{4} \spp
\eq
We define the following additional quantities:
\bq
{\overline{\mrX}}_1 = - \frac{1}{4} - \zeta\,(1 + \frac{1}{4}\,\zeta_0) \spc \quad
{\overline{\mrX}}_2 = \frac{1}{4}\,\zeta_0\,\zeta \spc \quad
\mrX_\mrj = \frac{{\overline{\mrX}}_\mrj}{\mrG} \spc \quad
\mrC = - \frac{1}{4}\,\zeta_0^2\,\zeta^2 \spp
\eq
The integral in \eqn{HAT} can be transformed into the following form:
\bqa
\mrH_{\AT} &=&
\frac{1}{\eG{1 - \ep/2}}\,\lpar \frac{\mrC}{\mrG} \rpar^{1 - \ep/2}\,\Bigl[
\mrX_1\,\mrX_2\,\mrH^{(1)}_{\AT} +
\mrX_1\,(\mrX_1 - \mrX_2)\,\mrH^{(2)}_{\AT} +
\mrX_2\,(1 - \mrX_1)\,\mrH^{(3)}_{\AT} 
\nl
{}&+&
(\mrX_1 - \mrX_2)\,(1 - \mrX_1)\,\mrH^{(4)}_{\AT} \Bigr] \spc
\eqa
\bqa
\mrH^{(1)}_{\AT} &=& \Bigl[ \prod_{\mrj=1}^{3}\,\int_{\mrL_\mrj}\,\frac{\mrd \mrs_\mrj}{\tip} \Bigr]\,
\frac{\mrN_\mra}{\mrD_\mra}\,
\mru_2^{\mrs_1}\,\mru_6^{\mrs_2}\,\mru_7^{\mrs_3} \spc
\qquad
\mrH^{(2)}_{\AT} = \Bigl[ \prod_{\mrj=1}^{4}\,\int_{\mrL_\mrj}\,\frac{\mrd \mrs_\mrj}{\tip} \Bigr]\,
\frac{\mrN_\mrb}{\mrD_\mrb}\,
\mru_3^{\mrs_1}\,\mru_4^{\mrs_2}\,\mru_5^{\mrs_3}\,\mru_1^{\mrs_4} \spc
\nl
\mrH^{(3)}_{\AT} &=& \Bigl[ \prod_{\mrj=1}^{3}\,\int_{\mrL_\mrj}\,\frac{\mrd \mrs_\mrj}{\tip} \Bigr]\,
\frac{\mrN_\mra}{\mrD_\mra}\,
\mrv_2^{\mrs_1}\,\mrv_6^{\mrs_2}\,\mrv_7^{\mrs_3} \spc
\qquad
\mrH^{(4)}_{\AT} = \Bigl[ \prod_{\mrj=1}^{4}\,\int_{\mrL_\mrj}\,\frac{\mrd \mrs_\mrj}{\tip} \Bigr]\,
\frac{\mrN_\mrb}{\mrD_\mrb}\,
\mrv_3^{\mrs_1}\,\mrv_4^{\mrs_2}\,\mrv_5^{\mrs_3}\,\mrv_1^{\mrs_4} \spp
\eqa
Furthermore we have
\bqa
\mrN_\mra &=&
\eG{ - \mrs_2}\,
\eG{ - \mrs_3}\,
\eG{\mrs_2 + \mrs_1}\,
\eG{\mrs_3 + \mrs_1}\,
\eG{1 + \mrs_3 + \mrs_2}\,
\eG{1 - \ep/2 - \mrs_1}\,
\eG{1 - \mrs_3 - \mrs_2 - 2\,\mrs_1}\,
\nl
\mrD_\mra &=&
\eG{\mrs_1}\,
\eG{2 + \mrs_3 + \mrs_2}\,
\eG{2 - \mrs_3 - \mrs_2 - 2\,\mrs_1}
\nl\nl
\mrN_\mrb &=&
\eG{ - \mrs_2}\,
\eG{ - \mrs_3}\,
\eG{ - \mrs_4}\,
\eG{\mrs_2 + \mrs_1}\,
\eG{\mrs_3 + \mrs_1}\,
\eG{1 + \mrs_3 + \mrs_2}\,
\eG{1 - \ep/2 - \mrs_1}\,
\eG{ - 1 + \mrs_4 - \mrs_3 - \mrs_2}
\nl
{}&\times&
\eG{1 + \mrs_4 - \mrs_3 - \mrs_2 - 2\,\mrs_1}
\nl
\mrD_\mrb &=&
\eG{\mrs_1}\,
\eG{ - 1 - \mrs_3 - \mrs_2}\,
\eG{2 + \mrs_3 + \mrs_2}\,
\eG{2 + \mrs_4 - \mrs_3 - \mrs_2 - 2\,\mrs_1} \spp
\eqa
Auxiliary quantities are
\bq
\mrR_{\pm} = - 1 - 2\,\zeta \pm \sqrt{ - \mrG} \spc
\eq
\bqa
\mru_1 &=& - \frac{\mrX_1}{(\mrX_1 - \mrX_2)}
\spc \quad
\mru_2 = \frac{\mrC}{\mrX_1^2\,\mrG\,\mrh}
\spc \quad
\mru_3 = \frac{\mrC\,\mrX_1^2}{\mrG\,\mrh}
\spc \quad
\mru_4 = - 2\,\frac{(\mrX_1 - \mrX_2)\,\mrh}{\mrX_1\,\mrR_{-}} \spc
\nl
\mru_5 &=& - 2\,\frac{(\mrX_1 - \mrX_2)\,\mrh}{\mrX_1\,\mrR_{+}}
\spc \quad
\mru_6 = 2\,\frac{\mrX_2\,\mrh}{\mrX_1\,\mrR_{-}}
\spc \quad
\mru_7 = 2\,\frac{\mrX_2\,\mrh}{\mrX_1\,\mrR_{+}} \spc
\nl
\nl
\mrv_1 &=& \frac{(1 - \mrX_1)}{(\mrX_1 - \mrX_2)}
\spc \quad
\mrv_2 = \frac{\mrC}{(1 - \mrX_1)^2\,\mrG\,\mrh}
\spc \quad
\mrv_3 = \frac{\mrC}{(1 - \mrX_1)\,\mrG\,\mrh}
\spc \quad
\mrv_4 = 2\,\frac{(\mrX_1 - \mrX_2)\,\mrh}{\mrR_{-}\,(1 - \mrX_1)} \spc
\nl
\mrv_5 &=& 2\,\frac{(\mrX_1 - \mrX_2)\,\mrh}{\mrR_{+}\,(1 - \mrX_1)}
\spc \quad
\mrv_6 = - 2\,\frac{\mrX_2\,\mrh}{\mrR_{-}\,(1 - \mrX_1)}
\spc \quad
\mrv_7 = - 2\,\frac{\mrX_2\,\mrh}{\mrR_{+}\,(1 - \mrX_1)} \spp
\eqa
From the work of \Brefs{Kershaw:1971rc,Ferroglia:2002mz,Passarino:2018wix} it follows that for
\bq
0 \,\le\, \mrX_2 \,\le\, \mrX_1 \,\le\, 1 \spc \qquad
\mrC \to 0\;(\mrG \not= 0) \spc
\eq
the function $\mrH_{\AT}$ is singular. The leading behavior can be found by using the results of 
\Bref{Passarino:2018wix}; for $\ep = 0$ we obtain
\bq
\mrH_{\AT} \,\sim\, - \frac{1}{2}\,\ln\frac{\mrC}{\mrG}\,\Bigl[ \mrI_1 + \mrI_2 \Bigr] \spp
\eq
After introducing 
\bq
\rho = - 1 - \frac{1}{2\,\zeta} \spc \qquad \tau = 2 + \frac{1}{2\,\zeta} \spc
\eq
\bq
\rho_1 = - \rho \spc \quad \rho_2 = 1 - \rho \spc \quad \tau_1 = - \tau \spc \quad \tau_2 = 1 - \tau \spc \quad
\lambda = \frac{\mrG}{\mrh} \spc
\eq
we obtain
\bq
\mrI_1 = \pi\,\mrG^{ - 1/2}\,\theta(\rho)\,\theta(1 - \rho) -
\sum_{\mrj=1}^{2}\,\frac{\mid \rho_\mrj \mid}{\mrh\,\rho_\mrj^2}\,
\hyp{1}{\frac{1}{2}}{\frac{3}{2}}{ - \frac{\lambda}{\mrh\,\rho_\mrj^2}} \spc
\eq
with $\mrI_2$ following after the substitution $\rho \to \tau$.

When $\mrC \not= 0$ and $\mrG \not= 0$ the $\mrH_{\AT}$ function develops an imaginary part for
$\zeta\,(1 + \zeta_0) > - 1/4$.
\section{Sinc expansion \label{SEXP}}
We give the basics elements of Sinc numerical methods following \Bref{Sinc}.
There is an extensive treatment of one{-}dimensional Sinc quadrature methods, see \Brefs{Sbook,Sinc}, but very little is
present in the literature for multi{-}dimensional Sinc quadrature methods~\cite{mSinc,GSinc}.
\paragraph{General aspects} \hspace{0pt} \\
Given a function $\mrf(\mrx)$ with $\mrx \in \Rf$ we intoduce the Cardinal function
\bq
\mrC\lpar \mrf\,,\,\mrh \rpar(\mrx) = \sum_{\mrk \in \Zf}\,\mrf(\mrk,\mrh)\,\sinc\lpar\frac{\mrx}{\mrh} - \mrk\rpar \spc
\qquad \sinc(\mrx) = \frac{\sin(\pi\,\mrx)}{\pi\,\mrx} \spp
\eq
The Sinc approximation over the interval $[\mra\,,\,\mrb]$ is defined by
\bq
\mrf(\mrx) \approx \sum_{\mrk}\,\mrf(\mrx_\mrk)\,\sinc\lpar \frac{\upphi(\mrx)}{\mrh} - \mrk\rpar \spc
\label{Sbas}
\eq
where $\upphi$ is a one{-}to{-}one mapping of $[\mra\,,\,\mrb]$ onto $\Rf$ and $\mrx_{\mrk} = \upphi^{-1}(\mrk \mrh)$.
Application of the Sinc approximation is based on the fact that (following the notations of \eqn{UFFtwo})
the absolute value of the integrand is comparable with
\bq
\exp\{ - \frac{1}{2}\,\alpha\,\pi\,\mid \mrt \mid - \phi\,\mrt\}\,
\mid \mrt \mid^{\beta\,\sigma + \lambda}\,\mrR^{ - \sigma}\,\rho^{\sigma} \spc
\eq
where $\mrz = \mrR\,\exp\{\mri\,\phi\}$ and $\mrs = \sigma + \mri\,\mrt$.
Therefore the explicit form of \eqn{Sbas} requires additional definitions~\cite{Sinc}. 
Let $\mrz \in \Cf$ and define the strip
\bq
\mrD_\mrd = \{\mrz \in \Cf\,:\,\mid \Im \mrz \mid < \mrd \} \spp
\eq
\bei

\item[\ovalbox{I}] The interval is $\Rf$. If $\mrz \in \mrD_\mrd$ and
\bqa
\Re \mrz &\le& 0 \spc \qquad \mid \mrf(\mrz) \mid \le \mrc_{-}\,\exp\{ - \alpha_{-}\,\mid \mrz \mid\} \spc
\nl
\Re \mrz &\ge& 0 \spc \qquad \mid \mrf(\mrz) \mid \le \mrc_{+}\,\exp\{ - \alpha_{+}\,\mid \mrz \mid\} \spc
\label{Slpo}
\eqa
then we use Sinc points $\upphi(\mrz) = \mrz$ and $\mrz_{\mrk} = \mrk \mrh$.
\item[\ovalbox{II}] The interval is $\Rf$. If $\mrz \in \mrD_\mrd$ with
\bq
\mrD_\mrd = \{\mrz \in \Cf\,:\,\mid \marg \Bigl\{ \sinh \Bigl[ \mrz + (1 + \mrz^2)^{1/2} \Bigr] \Bigr\} \mid < \mrd \spc
\eq
\bqa
\Re \mrz &\le& 0 \spc \qquad \mid \mrf(\mrz \mid \le \mrc_{-}\,\mid \mrz \mid^{ - \alpha_{-}} \spc
\nl
\Re \mrz &\ge& 0 \spc \qquad \mid \mrf(\mrz) \mid \le \mrc_{+}\,\exp\{ - \alpha_{+}\,\mid \mrz \mid\} \spc
\eqa
then we use
\bq
\upphi(\mrz) = \ln \Bigl\{\sinh \Bigl[ \mrz + (1 + \mrz^2)^{1/2} \Bigr] \Bigr\} \spc
\eq
\bq
\mrz_\mrk = \frac{1}{2}\,\lpar \mru_\mrk - \mru^{-1}_\mrk \rpar \spc \qquad
\mru_\mrk = \ln \Bigl[ \exp\{\mrk \mrh\} + (1 + \exp\{2\,\mrk \mrh\})^{1/2} \Bigr] \spp
\label{Slpt}
\eq
\item[\ovalbox{III}] The interval is $\Rf$. If $\mrD_\mrd$ is defined by
\bq
\mrD_\mrd = \{ \mrz \in \Cf\,:\,\mid {\mathrm arg} \Bigl[ \mrz + (1 + \mrz^2)^{1/2} \Bigr] \mid < \mrd \} \spc
\eq
take $\upphi(\mrz) = \ln \Bigl[ \mrz + ( 1 + \mrz^2)^{1/2} \Bigr]$; if $\mrz \in \mrD_\mrd$ and
\bqa
\Re \mrz &\le& 0 \spc \qquad \mid \mrf(\mrz) \mid \le \mrc_{-}\,( 1 + \mid \mrz \mid)^{- \alpha_{-}} \spc
\nl
\Re \mrz &\ge& 0 \spc \qquad \mid \mrf(\mrz) \mid \le \mrc_{+}\,( 1 + \mid \mrz \mid)^{- \alpha_{+}} \spc
\eqa
then the Sinc points are defined by
\bq
\mrz_\mrk = \sinh(\mrk\,\mrh) \spc \qquad
\frac{1}{\upphi^{\prime}(\mrz_\mrk)} = \cosh(\mrk\,\mrh) \spp 
\eq
\eei
In all cases we introduce a positive integer $\mrN$ and define
\bq
\mrM = \Bigl[ \frac{\alpha_{+}}{\alpha_{-}}\,\mrN \Bigr] \spc \qquad
\mrh = \Bigl( \frac{\mrd}{\alpha_{+}\,\mrN} \Bigr)^{1/2} \spc
\label{Spar}
\eq
where $[\,\mrx\,]$ is the integer part of $\mrx$.
Having defined all the auxiliary quantities we obtain
\bq
\mrf(\mrz) \approx \sum_{\mrk=-\mrM}^{\mrN}\,\mrf(\mrz_\mrk)\,\sinc\lpar \frac{\upphi(\mrz)}{\mrh} - \mrk \rpar \spc
\quad
\int_{\mra}^{\mrb}\,\mrf(\mrz) \approx \mrh\,\sum_{\mrk=-\mrM}^{\mrN}\,\mrf(\mrz_\mrk)\,\bigl[
\upphi^{\prime}(\mrz_\mrk) \bigr]^{-1} \spp
\eq
In the computation of a Fox function we integrate over the real variable $\mrt$ but the analytic continuation, 
$\mrt \in \Cf$, is needed in order to determine the parameter $\mrd$ which defines the step size $\mrh$.
The accuracy of the Sinc approximation on $\Rf$ is based on the fact that $\mrf$ is analytic and
uniformly bounded on the strip $\mrD_\mrd$.

Let us consider univariate Fox function, \eg
\bq
\mrH = \int_{\mrL}\,\frac{\mrd \mrs}{\tip}\,\mrf(\mrs) \spc \qquad
\mrf(\mrs) = \eG{ - \mrs}\,\frac{\eG{\mra + \mrs}\,\eG{\mrb + \mrs}}{\eG{\mrc + \mrs}}\,\mrz^{\mrs} \spp
\eq
In this example we assume that $\mra, \mrb$ are real and positive with $\mrb > \mra$. In general we write 
$\mrs = \sigma + \mri\,\mrt$ and $\mrz = \mrR\,\exp\{\mri\,\phi\}$ with
\bq
\mrH= \int_{- \infty}^{+ \infty}\,\frac{\mrd \mrt}{2\,\pi}\,\mrf(\mrt) \spp
\eq
The absolute value of the integrand is comparable with
\bq
\exp\{ - \frac{\alpha}{2}\,\mri\,\mid \mrt \mid - \phi\,\mrt\}\,\mid \mrt \mid^{\beta\,\sigma + \lambda}\,\mrR^{- \sigma}\,
\rho^{\sigma} \spc
\eq
when $\mid \mrt \mid$ is large. If $\alpha > 0$ and $\mid \phi \mid < \frac{1}{2}\,\alpha\,\pi$ we have
\bq
\mrt \to \pm \infty \quad \exp\{( - \frac{1}{2}\,\alpha\,\pi \mp \phi)\,\mid \mrt \mid\} \spc
\eq
Therefore we can write
\bqa
\mid \mrf(\mrt) \mid &\le& \mrk_{-}\,\exp\{ - \alpha_{-}\,\mid \mrt \mid\} \quad \mbox{for} \quad \Re \mrt \le 0 \spc
\nl
\mid \mrf(\mrt) \mid &\le& \mrk_{+}\,\exp\{ - \alpha_{+}\,\mid \mrt \mid\} \quad \mbox{for} \quad \Re \mrt \ge 0 \spc
\eqa
with
\bq
\alpha_{-} = \frac{1}{2}\,\alpha\,\pi  - \phi \spc \quad
\alpha_{+} = \frac{1}{2}\,\alpha\,\pi  + \phi \spp
\eq
In order to apply the results of \Bref{Sinc} we must consider $\mrt \in \Cf$, \ie $\mrt= \omega + i\,\tau$. The poles of the
integrand are at
\bq
\mra = \mrn \spc \quad \mrs = - \mra + \mrj \spc \quad \mrs = - \mrb + \mrk \spc
\eq
with $\mrn, \mrj, \mrk \in \Zf^{+}_{0}$.
Since now $\mrs = \sigma - \tau + i\,\omega$, and assuming $\mrb > \mra$ the integrand develops singularities for
$\tau = \mra + \sigma$ and $\tau = \sigma$. With $\sigma < 0$ we require $ \tau > - \mid \sigma \mid$ and
$\tau < \mra - \mid \sigma \mid$.
\paragraph{Optimization of Sinc approximation} \hspace{0pt} \\
Consider the following sum:
\bq
\sum_{\mrk_1=-\mrM_1}^{\mrN}\,\dots\,\sum_{\mrk_\mrn=-\mrM_\mrn}^{\mrN}\,
\mrR_\Gamma(\mrk_1\,,\,\dots\,,\,\mrk_\mrn)\,\Bigl[ \prod_{\mrj=1}^{\mrn}\,\mrz_\mrj^{\mrs_{\mrj}(\mrk_\mrj)} \Bigr] \spc
\eq
where $\mrs_\mrj(\mrk_\mrj) = \sigma_\mrj + i\,\mrk_\mrj\,\mrh_\mrj$ and $\mrR_\Gamma$ is the quotient of products of Euler
Gamma functions depending on $\mrs_\mrj(\mrk_\mrj)$. The sum requires 
\bq
(\mrN + \mrM_1 + 1)\,\dots\,(\mrN + \mrM_\mrn + 1)
\eq
evaluations of $\mrR_{\Gamma}$ functions. However,
\bq
\mrR_\Gamma( - \mrk_1\,,\,\mrk_2\,,\,\dots\,,\,\mrk_\mrn) =
{\overline{\mrR_\Gamma}}(\mrk_1\,,\, - \mrk_2\,,\,\dots\,,\, - \mrk_\mrn) \spc
\eq
\etc Therefore we do not need to compute from scratch the complex conjugate of a specific Gamma function.
It can be verified that, for three variables and $\mrN = 3, \mrM_\mrj = 2$, we only need $130$ evaluations instead of 
$343$. For $\mrN = 4$ and $\mrM_\mrj = 3$ we need $293$ evaluations instead of $729$ evaluations.
Of course this result is based on the assumption that $\mid \mrz_\mrj \mid < \pi$.
\paragraph{Accuracy of Sinc approximation} \hspace{0pt} \\
One{-}dimensional integrals: the error in the truncated Sinc expansion can
be determined~\cite{Sinc,eSinc} by using the parameter
\bq
\ep_{\mrN} = \mrN^{1/2}\,\exp\{ - ( \pi \mrd \alpha_{+}/\mrN )^{1/2} \} \spp
\eq
Given the norm
\bq
\mid \mid \mrf \mid \mid= \mathrm{sup}_{\mrx \in \Rf}\,\mid \mrf(\mrx) \mid \spc
\eq
the absolute norm of the difference between the exact value of the integral and
the Sinc approximation is bounded by $\mrC\,\ep_{\mrN}$, where $\mrC$ is a positive constant
independent of $\mrN$.
If the constants $\alpha_{\pm}$ and $\mrd$ cannot be accurately estimated then
$\mrh$ is defined by $\mrh = \gamma\,\mrN^{-1/2}$ where $\gamma$ is a
constant independent of $\mrN$ and $\ep_\mrN$ can be replaced by
$\exp\{- \delta\,\mrN^{1/2} \}$ where $\delta$ is a positive constant
independent of $\mrN$. Consider now the following integral:
\bq
\mrI = \Bigl[ \prod_{\mrj=1}^{2}\,\int_{\mrL_\mrj}\,\frac{\mrd \mrs_\mrj}{\tip} \Bigr]\,
\frac{
      \eG{ - \mrs_1}\,\eG{ - \mrs_2}\,\eG{\mra_1 + \mrs_1 + \mrs_2}\,\eG{\mra_2 + \mrs_1 - \mrs_2}
     }
     {
      \eG{\mrb_1 + \mrs_1}\,\eG{\mrb_2 + \mrs_2}
     }\,\mrz_1^{\mrs_1}\,\mrz_2^{\mrs_2} \spp
\eq
We see that for $\mid \mrt_1 \mid \to \infty$ and $\mrt_2$ finite the
parameters for the $\mrt_1$ integral can be determined exactly; the same
for the $\mrt_2$ integral when $\mid \mrt_2 \mid \to \infty$ and
$\mrt_1$ finite. Consider the ray $\mrt_2 = \lambda\,\mrt_1$ with
$\mrt_1 \to \infty$. It is easly seen that, when $\lambda \not= 1$ the
exponential behavior does not change. However, consider the integral in \eqn{sexa};
We see the following exponential behaviors:
\bqa
\mrt_1\,,\,\mrt_2 \to \infty \spc \quad \mrt_2 = \lambda\,\mrt_1
\quad &\mapsto& \quad
- \pi\,(1 + \lambda)\,\mrt_1 - (\phi_1 + \lambda\,\phi_2)\,\mrt_1 \spc
\nl
\mrt_1 \to \infty\,,\,\mrt_2 \to - \infty \spc \quad
\quad &\mapsto& \quad
- ( \pi - \phi_1 + \lambda\,\phi_2 ) \spc
\eqa
with $\lambda > 0$. It follows that the convergence is not at stake but the
error cannot be determined rigorously. Indeed, the values of $\mrM_1,
\mrh_1$ and $\mrM_2, \mrh_2$ depend on the region in the $\mrt_1{-}\mrt_2$ plane.
Therefore the choice of Sinc points based on ``sequential" integrals,
\ie
\bq
\mid \mrt_2 \mid < \rho \spc \quad \mid \mrt_1 \mid \to \infty
\quad \hbox{or} \quad
\mid \mrt_1 \mid < \rho \spc \quad \mid \mrt_2 \mid \to \infty \spc
\eq
is not optimal. This remains an open problem: 
for an accurate Sinc approximation of a function $\mrf$ on a contour $\Gamma$, we
require two conditions: (a) analyticity of $\mrf$ in a domain $\mrD$ with $\Gamma \in \mrD$, and
(b) a set of Lipschitz conditions~\cite{Lip} of $\mrf$ on $\Gamma$~\cite {Sinc}. The infinite{-}point Sinc
formula may be very accurate when the first condition is satisfied, even though
the second condition is not. In this case, the use of Sinc approximation
requires a large number of evaluation points in order to sum the series accurately. 
For a rigorous treatment of the problem see \Bref{Sth} where
near optimality of the Sinc approximation is established in a variety of spaces of functions analytic 
in a strip region about the real axis.
Let us consider an example
\bq
\mrI = \Bigl[ \prod_{\mrj=1}^{2}\,\int_{\mrL_\mrj}\,\frac{\mrd \mrs_\mrj}{\tip} \Bigr]\,
\eG{ - \mrs_1}\,\eG{ - \mrs_2}\,
\frac{
      \eG{\frac{2}{5} + \mrs_1 + \mrs_2}\,\eG{\frac{1}{2} + \mrs_1}\,\eG{\frac{1}{3} + \mrs_2}
     }
     {
      \eG{3 + \mrs_1 + \mrs_2}
     }\,\mrz_1^{\mrs_1}\,\mrz_2^{\mrs_2} \spp
\label{SKlatt}
\eq
Given $\mrs_\mrj = \sigma_\mrj + \mri\,\mrt_\mrj$ the integration domain is $\Gamma = \Rf^2$. If we put
$\mrt_\mrj = \omega_\mrj + i\,\tau_\mrj$ the domain of analycity $\mrD$ has a triangular shape in the
$\tau_1{-}\tau_2$ plane containing the origin; therefore condition (a) is satisfied but determining $\mrD_\mrd$
after imposing Lipschitz conditions is a problem outside the goal of this work. Addiional informations are based
on \Bref{multiSinc}; we stress once again that a fundamental ingredient in the Sinc methods is represented by the space of
complex{-}valued functions which are analytic in the strip
\bq
\mrD_\mrd := \{\mrz \in \Cf\, : \, \mid \Im \mrz \mid < \mrd \} \spp
\eq
The error bound in the multivariate case is discussed in \Bref{multiSinc} where, once again, the result depends on
determining the functions 
\bq
\mrf_\mrk(\mrz_\mrk) \,:\,\mrd_\mrk \to \Rf \spc
\eq
which are restrictions of $\mrz_1,\,\dots\,,\mrz_\mrn)$ onto $\mrd_\mrk$ with fixed remaining variables. 
\paragraph{Quasi{-}Monte Carlo method} \hspace{0pt} \\
Given a function $\mrf$ the QMC method samples $\mrf$ at $\mrN$ points of a regular lattice (Korobov lattice)
and introduces random variables $\mathbf{a}$, defining
\bq
\mrI_{\mrN}(\mathbf{a}) = \frac{1}{\mrN}\,\sum_{\mrj=1}^{\mrN}\,\mrf\lpar \frac{\mrj}{\mrN}\,\mathbf{v} +
\mathbf{a} \rpar \spc
\eq
where $\mathbf{v}$ is the Korobov lattice generator vector. For $\mrs$ random $\mathbf{a}$ vectors the
integral is approximated by
\bq
\mrI \approx \frac{1}{\mrs}\,\sum_{\mrj=1}^{\mrs}\,\mrI_{\mrN}(\mathbf{a}_{\mrj}) \spc
\eq
with a standard error estimation.

Our Sinc numerical results have been compared with an automatic multidimensional integration subroutine, which uses 
randomized Korobov rules~\cite{Koro,Kort,PK,Rand}. Korobov lattice rules are a suitable algorithm for integration 
in weighted Korobov spaces of analytic functions~\cite{pillichshammer2020notekorobovlatticerules}. 
Lattice rules are one important branch of quasi{-}Monte Carlo (QMC) methods traditionally applied
in the approximation of multivariate integrals over the unit cube $[0\,,\,1]^\mrd$
with periodic integrand functions, see the classical references~\cite{CRo,CRoo,CRt}.
Approaches exist in the literature~\cite{KRd} to apply quasi{-}Monte Carlo methods designed for
integration over the unit cube $[0\,,\,1]^\mrd$ to integration over $\Rf^\mrd$
analytic functions with exponential decay.

Quoting \Bref{Sinc},  ``Sinc functions live naturally in spaces of analytic functions
defined over bounded or unbounded domains''. More rigorously, for Sinc approximation, the space of functions
consists of functions that are analytic in a domain containing the (open) interval or contour of
approximation and are Lipschitz functions on the interval. Furthermore, Sinc numerical methods allow us
to treat Feynman integrals in the physical regions.

The agreement (see \sect{exa}) between non{-}optimal Sinc lattices and Korobov lattices is a supportive signal. 
Indeed, for the integral in \eqn{SKlatt} we obtain
\[
\begin{array}{lll}
 & \Re & \Im \\
\hline
&& \\
\mbox{Korobov}           &  189.7582(1) & - 0.140888(1) \\
&& \\
\mbox{Sinc}_{\mrN= 300}    &  189.758173  & - 0.140888297 \\
&& \\
\hline
\end{array}
\]
                                           
To give a rough estimate of the bounds we consider the following integral:
\bq
\mrI = \Bigl[ \prod_{\mrj=1}^{2}\,\int_{\mrL_\mrj}\,\frac{\mrd \mrs_\mrj}{\tip} \Bigr]\,
\eG{ - \mrs_1}\,\eG{ - \mrs_2}\,
\frac{
      \eG{1 + \mrs_1 + \mrs_2}\,\eG{1 + \mrs_1}\,\eG{1 + \mrs_2}
     }
     {
      \eG{2 + \mrs_1 + \mrs_2}
     }\,\mrz_1^{\mrs_1}\,\mrz_2^{\mrs_2} \spc
\eq
which is equivalent to the following exact result:
\bq
\mrE = \frac{1}{\mrz_1 - \mrz_2}\,\ln \frac{1 + \mrz_1}{1 + \mrz_2} \spp
\eq
We use 
\bq
\mrz_1 = \frac{1}{4} + 0.01\,\mri \spc \qquad \mrz_2 = \frac{1}{3} + 0.001\,\mri \spp
\eq
Let $\mrS_\mrk$ be the Sinc approximation of $\mrI$ with $\mrN= \mrk$. We define
\bq
\mrD_\mrk = - \ln \mid \mrE - \mrS_\mrk \mid \spc
\eq
and obtain
\[
\begin{array}{llll}
 & \Re & \quad \Im & \quad \mrD \\
\hline
&&& \\
\mrE      &  0.774444302 & \quad - 0.335796919\,\times\,10^{-2} & \\
&&& \\
\mrS_{10}  &  0.780760584 & \quad - 0.434946809\,\times\,10^{-2} & \quad - 5.053 \\
&&& \\
\mrS_{50}  &  0.774443278 & \quad - 0.335684314\,\times\,10^{-2} & \quad + 13.396 \\
&&& \\
\mrS_{100} &  0.774444300 & \quad - 0.335798174\,\times\,10^{-2} & \quad + 18.188 \\
&&& \\
\hline
\end{array}
\]
Consider a second integral,
\bq
\mrI_2 = \Bigl[ \prod_{\mrj=1}^{2}\,\int_{\mrL_\mrj}\,\frac{\mrd \mrs_\mrj}{\tip} \Bigr]\,
\eG{ - \mrs_1}\,\eG{ - \mrs_2}\,
\frac{
      \eG{1 + \mrs_1 + \mrs_2}\,\eG{1 + \mrs_1}\,\eG{2 + \mrs_2}
     }
     {
      \eG{3 + \mrs_1 + \mrs_2}
     }\,\mrz_1^{\mrs_1}\,\mrz_2^{\mrs_2} \spc
\eq
which is equivalent to the following exact result:
\bq
\mrE_2 = \frac{1}{\mrz_2 - \mrz_1}\,\Bigl[ \frac{1 + \mrz_1}{\mrz_2 - \mrz_1}\,
\ln\frac{1 + \mrz_1}{1 + \mrz_2} + 1 \bigr] \spp
\eq
We use 
\bq
\mrz_1 = - \frac{1}{4} \spc \qquad \mrz_2 = \frac{1}{3} + 0.01\,\mri \spp
\eq
We obtain
\[
\begin{array}{llll}
 & \Re & \quad \Im & \quad \mrD \\
\hline
&&& \\
\mrE_2      &  0.446121669  & \quad - 0.243886402\,\times\,10^{-2} & \\
&&& \\
\mrS_{10}   &  0.458589605  & \quad - 0.749132747\,\times\,10^{-2} & \quad + 0.835 \\
&&& \\
\mrS_{30}   &  0.450877031  & \quad - 0.241065289\,\times\,10^{-2} & \quad + 0.818 \\
&&& \\
\mrS_{50}   &  0.449243070  & \quad - 0.494585044\,\times\,10^{-2} & \quad + 0.814 \\
&&& \\
\mrS_{100}  &  0.443756775  & \quad - 0.137754729\,\times\,10^{-2} & \quad + 0.802 \\
&&& \\
\mrS_{300}  &  0.446528917  & \quad - 0.297180950\,\times\,10^{-2} & \quad + 0.808 \\
&&& \\
\mrS_{500}  &  0.445321290  & \quad - 0.280657519\,\times\,10^{-2} & \quad + 0.805 \\
&&& \\
\mrS_{1000} &  0.446391692  & \quad - 0.282246205\,\times\,10^{-2} & \quad + 0.808 \\
&&& \\
\hline
\end{array}
\]

The power behavior of the $\mrs_1$ integral is not satisfactory as seen especially in the imaginary part.
In this case it is better to use a contiguity relation. The $\mrI_2$ integral is proportial to an
$\mrF_1$ Appell function~\cite{Asur} and we can use
\bqa
\mrF_1\lpar 1\,;\,1\,,\,2\,;\,3\,;\, - \mrz_1\,,\,- \mrz_2 \rpar &=&
\frac{1}{12\,(1 + \mrz_1)}\,\Bigl[
(12 + 20\,\mrz_1 - 8\,\mrz_2)\,
\mrF_1\lpar 1\,;\,1\,,\,3\,;\,4\,;\, - \mrz_1\,,\,- \mrz_2 \rpar 
\nl
{}&+& 9\,(\mrz_2 - \mrz_1)\,
\mrF_1\lpar 1\,;\,1\,,\,4\,;\,5\,;\, - \mrz_1\,,\,- \mrz_2 \rpar \Bigr] \spc
\eqa
and use the MB representation for the Appell functions.
Indeed consider
\bq
\mrI_3 = \Bigl[ \prod_{\mrj=1}^{2}\,\int_{\mrL_\mrj}\,\frac{\mrd \mrs_\mrj}{\tip} \Bigr]\,
\eG{ - \mrs_1}\,\eG{ - \mrs_2}\,
\frac{
      \eG{1 + \mrs_1 + \mrs_2}\,\eG{1 + \mrs_1}\,\eG{4 + \mrs_2}
     }
     {
      \eG{5 + \mrs_1 + \mrs_2}
     }\,\mrz_1^{\mrs_1}\,\mrz_2^{\mrs_2} \spp
\eq
We obtain
\[
\begin{array}{lll}
 & \Re & \Im \\
\hline
&& \\
{\tt OKROBV} &   8.16687758(4) & - 0.0524562(2) \\
&& \\
\mrS_{10}     &   9.21579469    & - 0.152544264 \\
&& \\
\mrS_{30}     &   8.26254429    & - 0.0540061617 \\
&& \\
\mrS_{50}     &   8.19088582    & - 0.0525001759 \\
&& \\
\mrS_{100}    &   8.16884067    & - 0.0524777335 \\
&& \\
\mrS_{300}    &   8.16687874    & - 0.0524526209 \\
&& \\
\hline
\end{array}
\]
Where {\tt OKROBV} is an in{-}house automatic multidimensional integration routine
based on the program {\tt DKBRVC} from the package {\tt MVNDST}, written by
Alan Genz, which uses randomized Korobov rules~\cite{PK,Rand}.

It is worth noting that $\mrI_3$ should not be confused with $\mrI_2$, it is one of the
integrals to be computed after using the contiguity relations for $\mrI_2$.

A difficult case is represented by the following integral:
\bq
\mrJ = \Bigl[ \prod_{\mrj=1}^{3}\,\int_{\mrL_\mrj}\,\frac{\mrd \mrs_\mrj}{\tip} \Bigr]\,
\frac{
      \eG{\frac{1}{2} + \mathbf{s}}
     }
     {
      \eG{\frac{5}{2} + \mathbf{s}}
     }\,
     \Bigl[ \prod_{\mrj=1}^{3}\,\eG{ - \mrs_j}\,\eG{1 + \mrs_\mrj} \Bigr]\,
     \Bigl[ \prod_{\mrj=1}^{3}\,\mrz_\mrj^{\mrs_\mrj} \Bigr] \spc
\eq
where $\mathbf{s} = \mrs_1 + \mrs_2 + \mrs_3$ and where $\mrz_\mrj < 0$. We use
$\mrz_1 = -1.11$, $\mrz_2 = - 2.12$ and $\mrz_3 = - 1.95$ and derive
\[
\begin{array}{lll}
 & \Re & \Im \\
\hline
&& \\
{\tt OKROBV} & - 1.4144(2) & - 6.7965(3) \\
&& \\
\mrS_{5}     & - 1.5936    & - 6.4457 \\
&& \\
\mrS_{100}   & - 1.6303    & - 6.4368 \\
&& \\
\mrS_{500}   & - 1.4296    & - 6.7174 \\
&& \\
\hline
\end{array}
\]
showing a slow convergence rate due to the power behavior of the integrand. Also in this case it is better to use 
contiguity relations; for instance, given 
\bq
\mrJ^{\prime} = \Bigl[ \prod_{\mrj=1}^{3}\,\int_{\mrL_\mrj}\,\frac{\mrd \mrs_\mrj}{\tip} \Bigr]\,
\frac{
      \eG{\frac{1}{2} + \mathbf{s}}
     }
     {
      \eG{\frac{9}{2} + \mathbf{s}}
     }\,
     \Bigl[ \prod_{\mrj=1}^{3}\,\eG{ - \mrs_j}\,\eG{1 + \mrs_\mrj} \Bigr]\,
     \Bigl[ \prod_{\mrj=1}^{3}\,\mrz_\mrj^{\mrs_\mrj} \Bigr] \spc
\eq
we obtain
\[
\begin{array}{lll}
 & \Re & \Im \\
\hline
&& \\
{\tt OKROBV}   & 18.8774462(5) & - 4.163375(3) \\
&& \\
\mrS_{500}     & 18.8778074    & - 4.163513 \\
&& \\
\hline
\end{array}
\]
\paragraph{Comparison of lattice rules with exact results} \hspace{0pt} \\
Given a  Lauricella function
\bq
\mrF^{(\mrN)}_{\mrD}\lpar \mrb\,;\,\mra_1\,\dots\,\mra_{\mrN}\,;\,\mrc\,;\,\mathbf{z} \rpar \spc
\eq
where all the $\mra_\mrj$ parameters are positive integers we can use partial fraction decomposition in order
to reduce the Lauricella function to a combination of Gauss hypergeometric functions. 
In the following two examples we will need
\bq
\hyp{2}{\frac{1}{2}}{\frac{3}{2}}{\mrz} = \frac{1}{2}\,\hyp{1}{\frac{1}{2}}{\frac{3}{2}}{\mrz} +
\frac{1}{2}\,\frac{1}{1 - \mrz} \spc \quad
\hyp{1}{\frac{1}{2}}{\frac{3}{2}}{\mrz} = \frac{1}{2\,\mru}\,\ln \frac{1 + \mru}{1 - \mru} \spc
\eq
with $\mru = \sqrt{\mrz}$.

Consider the following integral:
\bq
\mrH = \Bigl[ \prod_{\mrj=1}^{2}\,\int_{\mrL_\mrj}\,\frac{\mrd \mrs_\mrj}{\tip} \Bigr] \,
\frac{
      \eG{\frac{1}{2} + \mrs_1 + \mrs_2}
     }
     {\eG{\frac{3}{2} + \mrs_1 + \mrs_2}
     }\,
     \eG{ - \mrs_1}\,\eG{2 + \mrs_1}\,\eG{ - \mrs_2}\,\eG{1 + \mrs_2}\,\mrz_1^{\mrs_1}\,\mrz_2^{\mrs_2} \spp
\eq
Setting $\mrz_1 = 0.15 + 0.01\,\mri$, $\mrz_2 = 0.55 - 0.01\,\mri$ and $\sigma_1 = \sigma_2 = - 0.1$ we can compare the
exact result with the Korobov lattice and the Sinc lattice.
\[
\begin{array}{lrl}
&& \\
\hline
&& \\
\mbox{Exact}    &                                    & 62.6024046    - 0.183034716\,\mri  \\
&& \\
\mbox{Sinc}     &  \quad 355152\;\mbox{calls} \quad  & 62.6024046    - 0.183034712\,\mri \\
&& \\
\mbox{Korobov}  & \quad 3071856\;\mbox{calls} \quad  & 62.6024045(2) - 0.18303475(5)\,\mri \\                                
&& \\
\hline
\end{array}
\]
A second example is given by the following integral:
\bq
\mrH = \Bigl[ \prod_{\mrj=1}^{2}\,\int_{\mrL_\mrj}\,\frac{\mrd \mrs_\mrj}{\tip} \Bigr] \,
\frac{
      \eG{\frac{1}{2} + \mathbf{s}}
     }
     {\eG{\frac{3}{2} + \mathbf{s}}
     }\,
     \prod_{\mrj=1}^{4}\,\eG{ - \mrs_\mrj}\,\eG{1 + \mrs_\mrj}\,\mrz_\mrj^{\mrs_\mrj} \spc
\eq
where $\mathbf{s}= \sum_\mrj\,\mrs_\mrj$. Setting $\sigma_\mrj = - 0.1$ and
\bq
\mrz_1 = 0.11 + 0.01\,\mri \spc \quad
\mrz_2 = 0.22 + 0.01\,\mri \spc \quad
\mrz_3 = 0.33 + 0.01\,\mri \spc \quad
\mrz_4 = 0.44 + 0.01\,\mri \spc
\eq
we obtain  $2317.89109 - 20.7965668\,\mri$ as the exact result. Korobov lattice cannot reach a satisfactory accuracy,
\eg with $69970770$ calls to the integrand we obtain $2334.3 \pm 7.5$ for the real part and
$- 20.97 \pm 0.17$ for the imaginary part. Sinc lattice produces the following results:
\[
\begin{array}{lr}
& \\
\hline
\mbox{result} & \mbox{N. of calls} \\
& \\
2427.00901 - 28.0863730\,\mri    & \quad  160000 \\
& \\
2320.67471 - 20.7568558\,\mri    & \quad 12744000 \\
& \\
2318.11021 - 20.7970070\,\mri    & \quad 97020000 \\
& \\
\hline
\end{array}
\]
The conclusion is that  (non{-}optimal) Sinc lattice is faster than Korobov lattice. With increasing number of variables 
Korobov lattice cannot reach an arbitrary accuracy even if we have not investigated ``periodizing transformations'' of
the integrand~\cite{Per}, \ie methods involving a variable substitution to the integral, dimension{-}by{-}dimension, 
to end up with a periodic integrand function while preserving the value of the integral. A second approach~\cite{BP} is
known as method of ``Bernoulli polynomials''. However, the main goal of this work was to discuss Sinc numerical methods.
\section{Hypergeometric methods \label{hmb}}
When computing a Fox function the integrand shows oscillations; indeed we can write
\bq
\eG{\mrx + \mri\,\mry} = \mid \eG{\mrx + \mri\,\mry} \mid\,\,\exp\{\mri\,\phi\} \spc
\eq
with oscillations given by
\bq
\phi = - \mry\,\int_0^{\infty} \mrd \mrt\,\frac{\exp\{ - \mrx\,\mrt\}}{1 - \exp\{ - \mrt\}}\,
\Bigl[ \sinc( \mry\,\mrt) - 1 \Bigr] =
\uppsi(\mrx)\,\mry - \frac{1}{3\,!}\,\uppsi^{(2)}(\mrx)\,\mry^3 + \ord{\mry^5} \spc
\eq
where $\uppsi^{(\mrn)}$ is the polygamma function.
The computational problem with a Fox function of $\mrr$ variables is represented by the oscillations of
the integrand, requiring a proibitive cost in number of points in the Korobov lattice when $\mrr$ is large.
When using the Sinc lattice for large $\mrr$ we need a fine tuning of the parameters in order to
obtain a quasi{-}optimal Sinc lattice.

A Fox function of $\mrr$ variables receives the largest contributions from the hypercube 
$[ - \mrd\,,\,\mrd ]^{\mrr}$. Therefore our strategy will be the following: given
\bq
\mrH = \Bigl[ \prod_{\mrj=1}^{\mrr}\,\int_{\mrL_\mrj}\,\frac{\mrd \mrs_\mrj}{\tip} \Bigr]\,
\mrR(\mathbf{s}\,,\,\mathbf{z}) \spc
\eq
we define $\mrR_\mra$ as a function which can be integrated analytically for $\mathbf{t} \in [ - \mrd\,,\,\mrd ]^{\mrr}$
and where the error in the approximation can be made arbitrarily small. The integral outside
$[ - \mrd\,,\,\mrd ]^{\mrr}$ is then computed with the usual methods, taking into account the decay rate of
the integrand, following from
\bq
\mid \eG{\mrx + \mri\,\mry} \mid \sim \exp\{ - \pi/2\,\mid \mry \mid\}\,
\mid \mry \mid^{1/2 - \mrx}\, \qquad \mid \mry \mid \to \infty \spp
\eq
The hypergeometric methods~\cite{RPBo,RPBt} have been discussed in details in \Bref{HAL}; we present a short summary:
we write the Gamma function as the sum of the two incomplete Gamma functions,
\bq
\eG{\mrz} = \sG{\mrz}{\mrN} + \bG{\mrz}{\mrN} \spc
\eq
where we use
\bq
\sG{\mrz}{\mrN} = \frac{\mrN^{\mrz}\,\exp\{-\mrN\}}{\mrz}\,\sum_{\mrk=0}^{\infty}\,
\frac{\mrN^{\mrk}}{\poch{\mrz + 1}{\mrk}} \spc \quad
\bG{\mrz}{\mrN} \sim \exp\{-\mrN\}\,\mrN^{\mrz - 1}\,\sum_{\mrl=0}^{\infty}\,
\frac{\poch{1 - \mrz}{\mrk}}{( - \mrN )^{\mrk}} \spc
\eq
where for $\Gamma$ we have used the asymptotic expansion. If we truncate the first series after $\mrK$ terms and the
second series after $\mrL$ terms then the errors are~\cite{HAL}
\bq
\ep_{\upgamma} \approx \exp\{-\mrN\}\,\frac{\mrN^{\mrK}}{\mrK\,!} \spc \qquad
\ep_{\Gamma} \approx \exp\{-\mrN\}\,\frac{\mrL\,!}{\mrN^{\mrL}} \spc
\eq
and the optimal values for $\mrK$ and $\mrL$ can be determined by imposing $\ep_{\upgamma} = \ep_{\Gamma} = 2^{-\mrp}$.

Let us consider a simple example:
\bq
\mrH = \int_{\mrL}\,\frac{\mrd \mrs}{\tip}\,\eG{ - \mrs}\,\eG{\mra + \mrs}\,\mrz^{\mrs} \spc \qquad
\mrs = \sigma + \mri\,\mrt \spc \quad - \mra < \sigma < 0 \spp
\eq
First we write
\bq
\sum_{\mrk=0}^{\mrK}\,\frac{\mrN^{\mrk}}{\poch{\mrz + 1}{\mrk}} = 1 +
\sum_{\mrk=1}^{\mrK}\,\mrN^{\mrk}\,\sum_{\mrj=1}^{\mrk - 1}\,\frac{\mrC(\mrK\,,\,\mrj)}{\mrz + \mrj} \spc
\eq
where the coefficients are given by
\bq
\mrC(\mrK\,,\,\mrj) = \Bigl[ \prod_{\mrl=1\,,\,\mrl\not=\mrj}^{\mrK}\,(\mrl - \mrj) \Bigr]^{-1} \spp
\eq
Similarly, we write
\bq
\sum_{\mrl=0}^{\mrL}\,( - \mrN)^{- \mrl}\,\poch{1 - \mrz}{\mrl} = 1 +
\sum_{\mrl=1}^{\mrL}\,( - \mrN)^{ - \mrl}\,\sum_{\mrj=0}^{\mrl}\,
\Bigl[ \begin{array}{c} \mrl \\ \mrj \end{array} \Bigr]\,
(1 - \mrz)^{\mrj} \spc 
\eq
where we have introduced Stirling numbers. Next we define
\bq
\zeta = - \marg (\mrz) + \mri\,\ln \mid \mrz \mid \spp
\eq
The first few terms in the computation of $\mrH_{\mrd}$,
\bq
\mrH_{\mrd} = \int_{- \mrd}^{+ \mrd}\,\frac{\mrd \mrt}{2\,\pi}\,\eG{ - \mrs}\,\eG{\mra + \mrs}\,\mrz^{\mrs} \spc
\eq
require the following functions:
\bqa
\Upphi_{1}(1\,,\,+) &=&
\Upphi_1(1,1,2\,;\, - \mri\,\frac{\mrd}{\sigma},\zeta\,\mrd) +  
\Upphi_1(1,1,2\,;\,\mri\,\frac{\mrd}{\sigma}, - \zeta\,\mrd) \spc 
\nl
\Upphi_{1}(1\,,\,-) &=& 
\Upphi_1(2,1,3\,;\, - \mri\,\frac{\mrd}{\sigma},\zeta\,\mrd) - 
\Upphi_1(2,1,3\,;\,\mri\,\frac{\mrd}{\sigma}, - \zeta\,\mrd) \spc
\nl
\Upphi_{1}(2\,,\,+) &=&
\Upphi_1(1,1,2\,;\, - \mri\,\frac{\mrd}{\sigma_\mra},\zeta\,\mrd) + 
\Upphi_1(1,1,2\,;\,\mri\,\frac{\mrd}{\sigma_\mra}, - \zeta\,\mrd) \spc
\nl
\Upphi_{1}(2\,,\,-) &=&
\Upphi_1(2,1,3\,;\, - \mri\,\frac{\mrd}{\sigma_\mra},\zeta\,\mrd) - 
\Upphi_1(2,1,3\,;\,\mri\,\frac{\mrd}{\sigma_\mra}, - \zeta\,\mrd) \spc
\nl
\Upphi_{1}(3\,,\,+) &=&
\Upphi_1(1,1,2\,;\, - \mri\,\frac{\mrd}{\sigma_{-}}, - \zeta\,\mrd) + 
\Upphi_1(1,1,2\,;\,\mri\,\frac{\mrd}{\sigma_{-}},\zeta\,\mrd) \spc
\nl
\Upphi_{1}(3\,,\,-) &=&
\Upphi_1(2,1,3\,;\, - \mri\,\frac{\mrd}{\sigma_{-}}, - \zeta\,\mrd) - 
\Upphi_1(2,1,3\,;\,\mri\,\frac{\mrd}{\sigma_{-}},\zeta\,\mrd) \spc
\nl
\Upphi_{1}(4\,,\,+) &=&
\Upphi_1(1,1,2\,;\, - \mri\,\frac{\mrd}{\sigma_{\mra\,+}},\zeta\,\mrd) + 
\Upphi_1(1,1,2\,;\,\mri\,\frac{\mrd}{\sigma_{\mra\,+}}, - \zeta\,\mrd) \spc
\nl
\Upphi_{1}(4\,,\,-) &=&
\Upphi_1(2,1,3\,;\, - \mri\,\frac{\mrd}{\sigma_{\mra\,+}},\zeta\,\mrd) - 
\Upphi_1(2,1,3\,;\,\mri\,\frac{\mrd}{\sigma_{\mra\,+}}, - \zeta\,\mrd) \spc
\eqa
where $\Upphi_1$ is one of the Horn{-}Humbert confluent, hypergeometric, functions of two 
variables~\cite{HTF,Hornc} and where we have defined the following quantities:
\bq
\sigma_{\mra} = \sigma + \mra \spc \quad
\sigma_{\pm} = 1 \pm \sigma \spc, \quad
\sigma_{\mra\,\pm} = 1 \pm \sigma \pm \mra \spp
\eq
Furthermore we need another function,
\bq
\upgamma(\mrn\,,\,\pm) = \upgamma(\mrn\,,\, - \zeta\,\mrd) \pm \upgamma(\mrn\,,\,\zeta\,\mrd) \spc
\eq
which is a combination of two incomplete Gamma functions. We obtain
\bq
\mrH_{\mrd} = \exp\{ - 2\,\mrN\}\,\frac{\mrN^{\mra}\,\mrz^{\sigma}}{2\,\pi}\,\mrh_{\mrd} \spc
\eq
\bqa
\mrh_{\mrd} &=&
       - \frac{1}{2}\,(1 - \mrN)\,\mrd^2\,\mrN^{-2}\,\sigma^{-1}\,\Upphi_1(1,-)
       + \frac{1}{2}\,(1 + \mrN)\,\mrd^2\,\mrN^{-2}\,\sigma_{\mra}^{-1}\,\Upphi_1(2,-)
\nl          
{}&-&    \frac{1}{2}\,\mrd^2\,\mrN^{-1}\,\sigma_{-}^{-1}\,\Upphi_1(3,-)
       - \frac{1}{2}\,\mrd^2\,\mrN^{-1}\,\sigma_{\mra\,+}^{-1}\,\Upphi_1(4,-)
\nl         
{}&-& \Bigl[ - \sigma_{\mra\,-} + \mrN + \frac{1}{\poch{\mra}{1}}\,\mrN^2 + 
                \frac{1}{\poch{\mra}{2}}\,\mrN^3 \Bigr]\,\mri\,\mrd\,\mrN^{-2}\,\sigma^{-1}\,(\mrN - 1)\,\Upphi_1(1,+)
\nl          
{}&-& \Bigl[\sigma_{+}\ - \mrN + \frac{1}{\poch{\mra}{1}}\,\mrN^2 - \frac{1}{\poch{\mra}{2}}\,\mrN^3 \Bigr]\,
              \mri\,\mrd\,\mrN^{-2}\,\sigma_{\mra}^{-1}\,(\mrN + 1)\,\Upphi_1(2,+)
\nl          
{}&+& \Bigl[\sigma_{\mra\,-} - \mrN - \frac{1}{\poch{\mra}{1}}\,\mrN^2 - \frac{1}{\poch{\mra}{2}}\,\mrN^2\,(\mrN - 1) + 
                2\,\frac{1}{\poch{\mra}{3}}\,\mrN^3 \Bigr]\,\mri\,\mrd\,\mrN^{-1}\,\sigma_{-}^{-1}\,\Upphi_1(3,+)
\nl          
{}&+& \Bigl[ \sigma_{+}\ - \mrN + \frac{1}{\poch{\mra}{1}}\,\mrN^2 - \frac{1}{\poch{\mra}{2}}\,\mrN^2\,(\mrN + 1) + 
                2\,\frac{1}{\poch{\mra}{3}}\,\mrN^3 \Bigr]\,\mri\,\mrd\,\mrN^{-1}\,\sigma_{\mra\,+}^{-1}\,\Upphi_1(4,+)
\nl          
{}&-& \Bigl[ \sigma_{\mra\,-}\,\sigma_{+}\ - \mrN\,\sigma_{+}\ - \mrN\,\sigma_{\mra\,-} + \mrN^2 \Bigr]\,
                \mri\,\zeta^{-1}\,\mrN^{-4}\,\upgamma(1,-)
\nl          
{}&-& (\sigma_{+}\ - \sigma_{\mra\,-}\,\zeta^{-2}\,\mrN^{-4}\,\upgamma(2,-)
      - \mri\,\zeta^{-3}\,\mrN^{-4}\,\upgamma(3,-) + \mathrm(h. o.) \spp
\eqa
The procedure can be improved and we will give a simple example: given
\bq
\mrR(\mathbf{s}\,,\,\mathbf{z}) = \frac{\eG{\mrb + \mrs_1 + \mrs_2}}{\eG{\mrc + \mrs_1 + \mrs_2}}\,
\prod_{\mrj=1}^2\,\eG{ - \mrs_\mrj}\,\eG{\mra_\mrj + \mrs_\mrj}\,\mrz_1^{\mrs_1}\,\mrz_2^{\mrs_2} \spc
\eq
consider the $\mrt_1{-}\mrt_2$ plane and the region $\mid \mrt_1 \mid < \mrd$ and $\mid \mrt_2 \mid > \mrd$.
First we use
\bq
\frac{1}{\eG{\mrc + \mrs_1 + \mrs_2}} = - \pi\,(\mrc + \mrs_1 + \mrs_2)\,\sin(\mrc + \mrs_1 + \mrs_2)\,
\eG{ - \mrc - \mrs_1 - \mrs_2} \spc
\eq
and write
\bq
\eG{ - \mrc - \mrs_1 - \mrs_2} = \frac{1}{\poch{-\mrc - \mrs_1 - \mrs_2}{\mrn}}\,\eG{\mrc^{\prime} - \mrs_1 - \mrs_2} \spc
\eq
with $\mrc^{\prime} = \mrn - \mrc$ and $\Re (\mrc^{\prime} - \mrs_1 - \mrs_2) > 0$. At this point we expand
\bqa
\frac{\eG{\mrb + \mrs_1 + \mrs_2}}{\eG{\mrb + \mrs_2}} &=&
(\mrb + \mrs_2)^{\mrs_1}\,\Bigl[ 1 + \frac{1}{2}\,\frac{\mrs_1\,(\mrs_1 - 1)}{\mrb + \mrs_2} + \;\dots \Bigr]
\nl
\frac{\eG{\mrc^{\prime} - \mrs_1 - \mrs_2}}{\eG{\mrc^{\prime} - \mrs_2}} &=&
(\mrc^{\prime} - \mrs_2)^{ - \mrs_1}\,\Bigl[ 1 + \frac{1}{2}\,
\frac{\mrs_1\,(\mrs_1 + 1)}{\mrc^{\prime} - \mrs_2} + \;\dots \Bigr] \spp
\eqa
Finally we use the hypergeometric expansion for the Gamma functions depending on $\mrs_1$ and perform the integration
over $\mrt_1$ in the interval $[ - \mrd\,,\,+\mrd]$.

In order to test the hypergeometric methods we have computed $\Gamma = \mid \eG{1 + \mri\,\mry} \mid^2$ and compared with
$\mrE = \mry\,\pi/\sinh(\mry\,\pi)$. In the following table we present $1 - \Gamma/\mrE$ in percent.

{\footnotesize{
\[
\begin{array}{lccc}
&&& \\
\hline
&&& \\
\mry = 0.1 &&& \\
&&&\\
\mrK \quad & \quad 15 \quad & \quad 20 \quad & \quad 25 \\
\mrN = 10 \quad & \quad 5.33 \quad & \quad 0.14 \quad & \quad 0.001 \\
\mrN = 15 \quad & \quad 55.9 \quad & \quad 10.3 \quad & \quad 0.66 \\
&&& \\
\hline
&&& \\
\mry = 1.1 &&& \\
&&& \\
\mrN = 10 \quad & \quad 4.27 \quad & \quad 0.09 \quad & \quad - 0.001 \\
\mrN = 15 \quad & \quad 53.9 \quad & \quad 9.2  \quad & \quad 0.51 \\
&&& \\
\hline
\mry = 2.1 &&& \\
&&& \\
\mrN = 10 \quad & \quad 1.57 \quad & \quad -0.03 \quad & \quad - 0.04 \\
\mrN = 15 \quad & \quad 48.3 \quad & \quad 6.1   \quad & \quad 0.16 \\
&&& \\
\hline
\mry = 3,1 &&& \\
&&& \\
\mrN = 10 \quad & \quad - 2.42 \quad & \quad - 0.13 \quad & \quad - 0.28 \\
\mrN = 15 \quad & \quad 37.8   \quad & \quad 1.18   \quad & \quad -0.32 \\
&&& \\
\hline
\end{array}
\]
}}
 
\section{Factorization/decomposition \label{FD}}
Computing a Lauricella function or a more general Fox function
corresponding to few variables is never a problem. However, the CPU time
needed in the case of a large number of variables requires additional
work. 

In this Section we introduce an algorithm for factorization/decomposition of
multivariate Fox functions
Our algorithm is a special example of a more general problem~\cite{DEC,ADEC} having to do
with the characterization of ``decomposable" multivariate functions. The
problem can be formualed  using the following example:
\bq
 \mrE\, \subset\, \prod_{\mrj=1}^{4}\,\mrS_\mrj \spc \quad
 \mrh : \mrE \to \Rf \spc \quad
 \mrf_\mrj^{\mrn} : \mrS_{\mrj}\,\times\,\mrS_{\mrj + 1} \to \Rf
\eq
The question is as follows: if
\bq
 \lim_{\mrn \to
\infty}\,\sum_{\mrj=1}^{3}\,\mrf_{\mrj}^{\mrn}(\mrs_\mrj\,,\,\mrs_{\mrj +
1}) =
 \mrh(\mrs_1,\,\dots\,,\mrs_4)
\eq
for all $\mrs_\mrj \in \mrE$, do there exists $\mrf_\mrj :
\mrS_\mrj\,\times\,\mrS_{\mrj + 1} \to \Rf$ such that
\bq
  \sum_{\mrj=1}^{3}\,\mrf_\mrj(\mrs_\mrj\,,\,\mrs_{\mrj+1}) =
\mrh(\mrs_1,\,\dots\,,\mrs_4)
\eq
for all $\mrs_\mrj \in \mrE$? For an answer to this question see \Bref{DEC}.

Another example~\cite{ADEC} concerns the construction of general decomposition
formulas  of functions of $\mrn$ variables into sums of $2^{\mrn}$ terms
where each term depends on a group of variables indexed by a particular
subset of $\mrN = \{1,\,\dots\,,\mrn\}$,
\bq
 \mrf = \sum_{\mathbf{u} \in \mrN} \,\mrf_{\mathbf{u}} \spc
 \eq
where $\mrf_{\mathbf{u}}$ depends only on the subset of variables
$\{\mrx_\mrj\,:\,\mrj \in \mathbf{u}\}$, see \Bref{ADEC}.
Stated differently the general problem consists in the approximation of a
multi{-}variate function $\mrf(\mrx_1,\,\dots\,,\mrx_\mrn)$ in the set of
separable functions
\bq
  \mrM = \{ \mru\,:\, \mru(\mathbf{x}) =
  \phi_1(\mrx_1),\,\dots\,,\phi(\mrx_{\mrn}) \} \spc
\eq
We have not analyzed the computational cost of various
decomposition algorithms. Alternative stratefies for reducing the number of contours in the multi-fold Mellin-Barnes 
integrals  can be found in \Bref{diaz2025simplewayreducenumber}. 

We will illustrate our strategy for the case of a Lauricella function of $\mrn + \mrn$ variables. Consider the
following integral
\bq
\mrI_{\mrn + \mrm} =
\Bigl[\prod_{\mrj=1}^{\mrn + \mrn}\,\int_{\mrL_\mrj}\,
\frac{\mrd \mrs_\mrj}{2\,\mri\,\pi} \Bigr]\,
\frac{\Gamma(\mra + \mathbf{s})}{\Gamma(\mrc + \mathbf{s})}\,\Bigl[
\prod_{\mrj=1}^{\mrn + \mrm}\,\Gamma( - \mrs_\mrj)\,
\Gamma(\mrb_\mrj + \mrs_\mrj)\,\mrz^{\mrs_\mrj} \Bigr] \spc
\label{Inm}
\eq
where we have introduced
\bq
 \mathbf{s} = \mathbf{s_L} + \mathbf{s_H} \spc \quad
 \mathbf{s_L} = \sum_{\mrj=1}^{\mrn}\,\mrs_{\mrj} \spc \quad
 \mathbf{s_H} = \sum_{\mrj=\mrn+1}^{\mrm}\,\mrs_{\mrj} \spp
\eq
We can use the following relation,
\bq
\frac{1}{\Gamma(\mrx + \mry)} =
\frac{\mrB(\mrx\,,\,\mry)}{\Gamma(\mrx)\,\Gamma(\mry)} \spc
\quad
\mrB(\mrx\,,\,\mry) = \int_0^1\,\mrd \mrv\,\mrv^{\mrx - 1}\,(1 -
\mrv)^{\mry - 1} \spc
\eq
and also the one for the reciprocal of the Euler Beta function~\cite{rBeta},
\bq
 \Gamma(\mrx + \mry) =
\frac{\Gamma(\mrx)\,\Gamma(\mry)}{\mrB(\mrx\,,\,\mry)} \spc   \quad
 \frac{1}{\mrB(\mrx\,,\,\mry)} =
 (\mrx + \mry - 1)\,\mrJ(\mrx\,,\,\mry)
\label{invb}
\eq
where, assuming $\Re \mrx > 1$ and $\Re \mry > 1$, we have
\bq
 \mrJ = \sum_{\mrk=0}^{\infty}\,\frac{1}{(\mrk\,!)^2}\,(1 - \mrx)_{\mrk}\,
 (1 - \mry)_{\mrk}
\eq
Alternatively, we can use
\bq
  \mrB(\mrx\,,\,\mry) = \sum_{\mrk=0}^{\infty}\,
  \frac{(1 - \mrx)_{\mrk}}{(\mry + \mrk)\,\mrk\,!} \spc
\eq
where $(\mrz)_\mrn$ is the raising factorial.

In this way we obtain a factorization, \ie the $\mrn + \mrm$ integral is
transformed into products of integrals. 

In our case we introduce 
\bq
\mrc_\mrn = \frac{\mrn}{\mrn + \mrm}\,\mrc \spc \qquad
\mrc_\mrm = \frac{\mrm}{\mrn + \mrm}\,\mrc \spc
\eq
and derive
\bq
\frac{1}{\Gamma(\mrc + \mathbf{s})} =
\frac{1}{\Gamma(\mrc/2 + \mathbf{s_L})\,\Gamma(\mrc/2 + \mathbf{s_H})}\,
\int_0^1\,\mrd \mrv \,
\mrv^{\mrc_\mrn + \mathbf{s_L} - 1}\,
(1 - \mrv)^{\mrc_\mrm + \mathbf{s_H} - 1  } \spc
\eq
with the following conditions:
\bq
\Re\,\mrc_\mrn + \sum_{\mrj=1}^{\mrn}\,\sigma_\mrj > 0 \spc \qquad
\Re\,\mrc_\mrm + \sum_{\mrj=\mrn+1}^{\mrm}\,\sigma_\mrj > 0 \spc \spp
\eq
Using \eqn{Inm} we derive
\bqa
\mrI_{\mrn + \mrm} &=& \int_0^1 \mrd \mrv\,\mrv^{\mrc_\mrn - 1}\,(1 - \mrv)^{\mrc_\mrm - 1}\,
\mrJ_{\mrn + \mrm}(\mrv) \spc
\nl
\mrJ_{\mrn + \mrm}(\mrv) &=& \mrd\,\mrS_{\mrn + \mrm}\,\eG{\mra + \mathbf{s}}\,
\Bigl\{ \prod_{\mrj=1}^{\mrn}\,\eG{ - \mrs_\mrj}\,\eG{\mrb_\mrj + \mrs_\mrj}\,(\mrv\,\mrz_\mrj)^{\mrs_\mrj} \Bigr\}\,
\Bigl\{ \prod_{\mrj=\mrn+1}^{\mrm}\,\eG{ - \mrs_\mrj}\,\eG{\mrb_\mrj + \mrs_\mrj}\,
\Bigl[(1 - \mrv)\,\mrz_\mrj \Bigr]^{\mrs_\mrj} \Bigr\} \spp
\eqa
We also introduce
\bqa
\mra_\mrn &=& \frac{\mrn}{\mrn + \mrm}\,\mra \spc \qquad
\Re\,\mra_\mrn + \sum_{\mrj=1}^{\mrn}\,\sigma_\mrj > 1 \spc
\nl
\mra_\mrm &=& \frac{\mrm}{\mrn + \mrm}\,\mra \spc \qquad
\Re\,\mra_\mrm + \sum_{\mrj=\mrn+1}^{\mrm}\,\sigma_\mrj > 1 \spp
\label{betac}
\eqa
If the conditions in \eqn{betac} are not satisfied we can shift the integration contours, 
$\{\sigma\} \to \{\sigma^{\prime}\}$ and subtract the poles of the integrand in the corresponding strips.
Using \eqn{invb} we derive
\bq
\eG{\mra + \mathbf{s}} = (\mra - 1 + \mathbf{s}_\mrL + \mathbf{s}_\mrH )\,
\eG{\mra_\mrn + \mathbf{s}_\mrL}\,\eG{\mra_\mrm + \mathbf{s}_\mrH}\,
\sum_{\mrk=0}^{\infty}\,\lpar \frac{1}{\mrk\,!} \rpar^2\,
\lpar 1 - \mra_\mrn - \mathbf{s}_\mrL \rpar_\mrk\,
\lpar 1 - \mra_\mrm - \mathbf{s}_\mrH \rpar_\mrk \spp
\eq
The result is
\bq
\mrJ_{\mrn + \mrm}(\mrv) = \sum_{\mrk=0}^{\infty}\,\lpar \frac{1}{\mrk\,!} \rpar^2\,
\mrJ^{\mrk}_{\mrn + \mrm}(\mrv) \spc \quad
\mrJ^{\mrk}_{\mrn + \mrm}(\mrv) = \mrJ^{\mrL}_{\mrn\,,\,\mrk}(\mrv)\,
\mrJ^{\mrH}_{\mrm\,,\,\mrk}(\mrv)\,(\mra - 1 + \mathbf{s}_\mrL + \mathbf{s}_\mrH ) \spp
\eq
The four blocks needed to compute $\mrI_{\mrn + \mrm}$ are,
\bqa
\mrF^{\mrL}_{\mrn\,,\,\mrk}(\mrv) &=& \mrd\,\mrS_\mrL\,
\frac{\eG{\mra_\mrn + \mathbf{s}_\mrL}}
     {\eG{\mrc_\mrn + \mathbf{s}_\mrL}}\,
\lpar 1 - \mra_\mrn - \mathbf{s}_\mrL \rpar_\mrk\,
\Bigl\{ \prod_{\mrj=1}^{\mrn}\,\eG{ - \mrs_\mrj}\,\eG{\mrb_\mrj + \mrs_\mrj}\,
( \mrv\,\mrz_\mrj)^{\mrs_\mrj} \Bigr\} \spc
\nl
\mrF^{\mrH}_{\mrm\,,\,\mrk}(\mrv) &=& \mrd\,\mrS_\mrH\,
\frac{\eG{\mra_\mrm + \mathbf{s}_\mrH}}
     {\eG{\mrc_\mrm + \mathbf{s}_\mrH}}\,
\lpar 1 - \mra_\mrm - \mathbf{s}_\mrH \rpar_\mrk\,
\Bigl\{ \prod_{\mrj=\mrn+1}^{\mrm}\,\eG{ - \mrs_\mrj}\,\eG{\mrb_\mrj + \mrs_\mrj}\,
\Bigl[ (1 - \mrv)\,\mrz_\mrj \Bigr]^{\mrs_\mrj} \Bigr\} \spc
\nl
\mrG^{\mrL}_{\mrn\,,\,\mrk}(\mrv) &=& \mrd\,\mrS_\mrL\,
\mathbf{s}_\mrL\,\frac{\eG{\mra_\mrn + \mathbf{s}_\mrL}}
     {\eG{\mrc_\mrn + \mathbf{s}_\mrL}}\,
\lpar 1 - \mra_\mrn - \mathbf{s}_\mrL \rpar_\mrk\,
\Bigl\{ \prod_{\mrj=1}^{\mrn}\,\eG{ - \mrs_\mrj}\,\eG{\mrb_\mrj + \mrs_\mrj}\,
( \mrv\,\mrz_\mrj)^{\mrs_\mrj} \Bigr\} \spc
\nl
\mrG^{\mrH}_{\mrm\,,\,\mrk}(\mrv) &=& \mrd\,\mrS_\mrH\,
\mathbf{s}_\mrH\,\frac{\eG{\mra_\mrm + \mathbf{s}_\mrH}}
     {\eG{\mrc_\mrm + \mathbf{s}_\mrH}}\,
\lpar 1 - \mra_\mrm - \mathbf{s}_\mrH \rpar_\mrk\,
\Bigl\{ \prod_{\mrj=\mrn+1}^{\mrm}\,\eG{ - \mrs_\mrj}\,\eG{\mrb_\mrj + \mrs_\mrj}\,
\Bigl[ (1 - \mrv)\,\mrz_\mrj \Bigr]^{\mrs_\mrj} \Bigr\} \spc
\label{fourb}
\eqa
where we have introduced
\bq
\mrd \mrS_{\mrL} = \prod_{\mrj=1}^{\mrn}\,\frac{\mrd \mrs_{\mrj}}{2\,\mri\,\pi} , \qquad
 \mrd \mrS_{\mrH} = \sum_{\mrj=\mrn+1}^{\mrm}\,
\frac{\mrd \mrs_{\mrj}}{2\,\mri\,\pi} \spp
\eq
It is interesting to observe that, by using
\bq
(1 - \mrx)_\mrk = \sum_{\mrj=0}^{\mrk}\,\mrd_{\mrk\,,\,\mrj}\,\mrx^\mrj \spc \qquad
\mrd_{\mrk\,,\,\mrj} = \sum_{\mri=\mrj}^{\mrk}\,( - 1)^{\mrj}\,\mrs(\mrk\,,\,\mri)\,\Pch{\mri}{\mrj} \spc
\eq
(where $\mrs(\mrk\,,\,\mri)$ are Stirling numbers of the first kind) we obtain
\bq
\eG{\mrx}\,(1 - \mrx)_\mrk = \sum_{\mrj=0}^{\mrk}\,\mrd_{\mrk\,,\,\mrj}\,\eG{\mrx + \mrj} \spp
\label{Gameq}
\eq
Therefore the decomposition can be written directly in terms of Lauricella functions, \eg
\bq
\mrF^{\mrL}_{\mrn\,,\,\mrk}(\mrv) =
\sum_{\mrj=0}^{\mrk}\,\mrd_{\mrk\,,\,\mrj}\,
\frac{\eG{\mra_\mrn + \mrj}\,\prod_{\mri=1}^{\mrn}\,\eG{\mrb_\mri}}
     {\eG{\mrc_\mrn}}\,\mrF^{(\mrn)}_{\mrD}\lpar
\mra_\mrn + \mrj\,;\,\mathbf{b}_\mrL\,;\,\mrc_\mrn\,;\,
\mrv\,\mrz_1,\,\dots\,,\mrv\,\mrz_\mrn \rpar \spp
\eq
Using \eqn{Gameq} inside \eqn{fourb} and splitting $\mathbf{s}_\mrL$ and/or $\mathbf{s}_\mrH$ (\eg
splitting $\mathbf{s}_\mrL$ into two blocks) we can iterate the procedure.

We have tested the partial decomposition by considering the following
example:
\bq
\mrI_{2\,\mrn}(\mra\,;\,\mathbf{b}\,;\,\mrc\,;\,\mathbf{z}) =
\int_0^1\,\mrd \mrx\,\mrx^{\mra -1}\,(1 - \mrx)^{\mrc - \mra -
1}\,\prod_{\mrj=1}^{2\,\mrn}\,(1 + \mrz_\mrj\,\mrx)^{- \mrb_\mrj} \spp
\eq
We can use the corresponding relation to a Lauricella function,
\bq
\mrI_{2\,\mrn} = \frac{\Gamma(\mra)\,\Gamma(\mrc -
\mra)}{\Gamma(\mrc)}\,\mrF^{(2\,\mrn)}_{\mrD}(\mra\,;\,\mathbf{b}\,;\,\mrc\,;\,-
\mathbf{z}) \spc
\eq
and
\bq
 \mrF^{(2\,\mrn)}_{\mrD}(\mra\,;\,\mathbf{b}\,;\,\mrc\,;\, - \mathbf{z}) =
 \frac{\Gamma(\mrc)}{\Gamma(\mra)\,\prod_{\mrj=1}^{2\,\mrn}\,\Gamma(\mrb_\mrj)}\,\mrH^{(2\,\mrn)}(\mra\,;\,\mathbf{b}\,;\,\mrc\,;\,\mathbf{z}) \spc
\eq
with $\mrH$ defined by
\bq
\mrH^{(2\,\mrn)}  = \Bigl[
\prod_{\mrj=1}^{2\,\mrn}\,\int_{\mrL_\mrj}\,\frac{\mrd
\mrs_\mrj}{2\,\mri\,\pi} \Bigr]\,\frac{\Gamma(\mra +
\mathbf{s})}{\Gamma(\mrc + \mathbf{s})}\,\prod_{\mrj=1}^{2\,\mrn}
\,\Gamma( - \mrs_\mrj)\,\Gamma(\mrb_ \mrj +
\mrs_\mrj)\,\mrz_\mrj^{\mrs_\mrj} \spp
\eq
We introduce
\bq
  \mrz_\mrj =\mrz_{\mrL} , \quad \mrj= 1,\dots,\mrn , \qquad
  \mrz_{\mrj} = \mrz_{\mrH} , \quad \mrj= \mrn,\dots,2\,\mrn \spc
\eq
and $\mrb_\mrj = 1/\mrp$. Therefore we obtain
\bq
 \mrH^{(2\,\mrn)}(\mra\,;\,\mrp^{-1}\,\dots\,\mrp^{-1}\,;\,\mrc\,;\,\mrz_{\mrL}\,\dots\,\mrz_{\mrL}
\,;\,\mrz_{\mrH}\,\dots\,\mrz_{\mrH}) =   \frac{
 \prod_{\mrj=1}^{2\,\mrn}\,\Gamma(\mrp^{-1})}{\Gamma^2(\mrn/\mrp)}\,
 \mrH^{(2)}(\mra\,;\,\mrn/\mrp\,\,,\mrn/\mrp\,\;\,\mrc\,;\,\mrz_\mrL\,,\,\mrz_\mrH) \spc
\eq
and we can test he partial decomposition of $\mrH^{(2\,\mrn)}$ by compting
the much simpler function $\mrH^{(2)}$. 

There are three ingredients in our procedure. The MB integrals converge
absolutely if the conditions requested by the theorem of \Bref{HS} are
satisfied, The $\mrv$ integral can be evaluated by means of Gauss{-}Jacobi
quadrature rules, an ideal approach for high accuracy computation. The
remaining ingredient is based on the series
\bq
 \mrS = \sum_{\mrk=0}^{\infty}\,
 \frac{1}{(\mrk\,!)^2}\,(1 - \mrx)_\mrk\,(1 - \mry)_\mrk
\eq
The absolute convergence of the series for $\Re \mrx > 1$, $\Re \mry > 1$
is based on the following result:
\bq
  \mrS = {}_2\,\mrF_1(1 - \mrx\,,\,1 - \mry\,;\,1\,;\,1)
  \eq
where
\bq
 {}_2\,\mrF_1(\alpha\,,\,\beta\,;\,\gamma\,;\,1) =
 \frac{
 \Gamma(\gamma)\,\Gamma(\gamma - \alpha - \beta)}
 {\Gamma(\gamma - \alpha)\,\gamma(\gamma - \beta)}
 \eq
valid for $\Re(\gamma - \alpha - \beta) > 0$.
We can add that
\bq
 \mru_\mrk = \frac{1}{(\mrk\,!)^2}\,(1 - \mrx)_\mrk\,(1 - \mry)_\mrk) \qquad
 \lim_{\mrk \to \infty}\,\frac{\mru_{\mrk+1}}{\mru_\mrk} = 1 \spc
 \eq
so that the convergence is sublinear (logarithmic).
When the series is truncated we find that depending on $\Im \mrx$ and/or
$\Im \mry$, the series has a slow rate of convergence. In particular, when
$\Im \mrx$ and/or $\Im \mry$ are large, almost $40$ terms are needed to
obtain a reliable result.

Going back to our special example we have $\mrH^{(10)}$ and $\mrH^{(2)}$; 
we select $\mra = 2.8, \mrc = 5$ and $\mrp = 1/3$.
There are three parameters to be used: $\mrN_\mrs$, the number of points to be used in the Sinc
quadrature of the MB integrals; $\mrN_{\mrq}$, the number of points to be
used in the Gauss{-}Jacobi quadrature; $\mrK$, the truncation in
evaluating the series. We first define
\bq
  \mrHb^{(2)} = \frac{\mrH^{(2)}}{\Gamma^2(5/3)} \qquad
  \mrHb^{(10)} = \frac{\mrH^{(10)}}{\Gamma^{10}(1/3)} =
  \sum_{\mrk=0}^{\mrK}\,\mrHb^{(10)}_{\mrk} \spc
\eq
and compare the two. The result for $\mrHb^{(2)}$, corresponding to 
$\mrz_\mrL = 1.15 + 0.1\,\mri$, $\mrz_\mrH = 1.61 + 0.1\,\mri$ and $\mrN_{\mrs} = 1000$ is
\bq
\mrHb^{(2)} = 0.0127207 \, - \, 0.0011116\,\mri \spp
\eq
For $\mrHb^{(10)}$ we use $\mrN_{\mrq} = 10$ and $\mrK = 11$. Missing the optimal choice for the Sinc lattice we
study the Sinc quadrature in the $\mrN_\mrs {-} \mrd$ plane. With $\mrM_\mrs = 15$ and $\mrd = 0.2$ we obtain
\bq
\mrHb^{(10)} = 0.0128886 \, - \, 0.0011715\,\mri \spc
\eq
as a consequence of the slow rate of convergence of the series. This fact can be seen by considering the first terms in 
the series 
\[
\begin{array}{lrr}
&& \\
\hline
&& \\
\mrk  & \Re \mrHb^{(10)} & \Im \mrHb^{(10)} \\
&& \\
  0 \quad & \quad      0.128938619\,\times\,10^{-1} \quad & \quad    - 0.120647974\,\times\,10^{-2} \\
  1 \quad & \quad     - 0.153018706\,\times\,10^{-3} \quad & \quad     0.939846380\,\times\,10^{-4} \\
  2 \quad & \quad     - 0.201374573\,\times\,10^{-3} \quad & \quad    0.192418920\,\times\,10^{-5} \\
  3 \quad & \quad      0.179600257\,\times\,10^{-4} \quad & \quad   - 0.181837526\,\times\,10^{-4} \\
  4 \quad & \quad      0.718373026\,\times\,10^{-4} \quad & \quad    - 0.154486869\,\times\,10^{-4} \\
  5 \quad & \quad      0.701196651\,\times\,10^{-4} \quad & \quad    - 0.104264498\,\times\,10^{-4} \\
  6 \quad & \quad      0.565057226\,\times\,10^{-4} \quad & \quad    - 0.665003667\,\times\,10^{-5} \\
  7 \quad & \quad      0.430984602\,\times\,10^{-4} \quad & \quad    - 0.418760004\,\times\,10^{-5} \\
  8 \quad & \quad      0.324168397\,\times\,10^{-4} \quad & \quad    - 0.263832214\,\times\,10^{-5} \\
  9 \quad & \quad      0.244236318\,\times\,10^{-4} \quad & \quad    - 0.166785257\,\times\,10^{-5} \\
 10 \quad & \quad      0.185520025\,\times\,10^{-4} \quad & \quad    - 0.105606714\,\times\,10^{-5} \\
 11 \quad & \quad      0.142446037\,\times\,10^{-4} \quad & \quad    - 0.666639882\,\times\,10^{-6} \\
&& \\
\hline
\end{array}
\]   
In order to improve the result we will use well{-}known methods of
acceleration~\cite{weniger}. We found the use of Ames transform~\cite{Ames} particularly simple and
efficient. Given
\bq
 \mrS = \sum_{\mrk=1}^{\infty}\,\mrx_\mrk \qquad\mrS_\mrK =
\sum_{\mrk=1}^{\mrK}\,\mrx_\mrk \spc
\eq
the Ames transform is based on
\bq
 \mrS^{\mrA} = \frac{1}{2}\,\sum_{\mrk=1}^{\mrK}\,
 \frac{\mrX_{\mrk}}{2^{\mrk}} + \mrR_\mrK \spc
\eq
\bq
 \mrX_\mrk = \sum_{\mrj=1}^{\mrk}\,\frac{\mrj}{\mrk}\,
 \Pch{\mrk}{\mrj}\,\mry_\mrj \spc \qquad
 \mry_{1,2} = 2\,\mrx_{1,2}, \quad \mry_\mrj = \mrx_ \mrj \;\; \mrj > 2 \spc
\eq
\bq
\mrR_\mrk = \frac{1}{2^\mrk}\,\sum_{\mrj=1}^{\mrk + 1}\,
\Pch{\mrk}{\mrj-1}\,\Bigl( \mry_\mrj + \mry_{\mrj + 1}
+ \,\dots\,\Bigr) \spp
\eq
The remainder $\mrR_\mrK$ is such that $\lim_{\mrK \to \infty}\,\mrR_\mrK
= 0$ and the stopping error at any term is less than the last term added.
We have also compared the Ames transform and the Shanks transform~\cite{Shanks} given by
\bq
\mrS_{\mrK} = \sum_{\mrk=0}^{\mrK}\,\mrx_\mrk \spc \qquad
\mrS^{\prime}_{\mrK} = \mrS_{\mrK + 1} - \frac{
(\mrS_{\mrK + 1} - \mrS_{\mrK})^2}{\mrS_{\mrK + 1} - 2\,\mrS_{\mrK} + \mrS_{\mrk - 1}} =
\mrS^{\prime}_{\mrK}(\mrS) \spc
\eq
with $\mrS^{\prime}_0 = \mrS_0$. The transform can be iterated, \eg
$\mrS^{\prime\prime}_{\mrK} = \mrS^{\prime}_{\mrK}(\mrS^{\prime})$. 
We also introduce
\bq
\delta_\mrr = 100\,\Bigl( 1 - \frac{\Re\,\mrHb^{(10)}}{\Re\,\mrHb^{(2)}} \Bigr) \spc \quad
\delta_\mri = 100\,\Bigl( 1 - \frac{\Im\,\mrHb^{(10)}}{\Im\,\mrHb^{(2)}} \Bigr) \spp
\eq
We derive the following result:
\begin{enumerate}

\item With $\mrz_\mrL = 1.15 + 0.1\,\mri$, $\mrz_\mrH = 1.61 + 0.1\,\mri$, selecting $\mrN_\mrs = 15$ and $\mrd = 0.2$ 
we obtain
\bqa
\Re\,\mrHb^{(10)} &=&          0.0128886 \spc \qquad \delta_\mrr = - 1.32 \spc
\nl
\Re\,\mrHb^{(10)}_{\mrA} &=&   0.0127148 \spc \qquad \delta_\mrr = - 0.46 \spp
\eqa
\item Selecting $\mrN_\mrs = 18$ and $\mrd = 0.2$ we obtain
\bqa
\Im \,\mrHb^{(10)} &=&        - 0.0011570 \spc \qquad \delta_\mri = - 4.09 \spc
\nl
\Im \,\mrHb^{(10)}_{\mrA} &=& - 0.0011134 \spc \qquad \delta_\mri = - 0.17 \spp
\eqa

\end{enumerate}

As a final check we have taken $\mrd$ as a random variable and derived an estimate on the 
$\mrd\,${-}error corresponding to $\mrN_\mrs = 15$:
\[
\begin{array}{lll}
&& \\
 \quad & \quad \Re & \Im \\
&& \\
\mrHb^{(2)}_{\mrA}  \quad & \quad  0.01272    & -\, 0.001112 \\
\mrHb^{(10)}_{\mrA} \quad & \quad  0.01274(8) & -\, 0.001133(6) \\
&& \\
\end{array}
\]

showing that the choice $\mrK = 11$ already provides a good approximation when using the Ames acceleration.
We have verified that the result is stable when increasing $\mrN_\mrq$.
\section{Auxiliary functions  \label{AFUN}}
In evaluating MB integrals the object of interest is
\bq
\mrR = \frac
{\prod_{\mrj=1}^{\mrm}\,\eG{\mrx_{\mrj}}}
{\prod_{\mrj=1}^{\mrn}\,\eG{\mry_{\mrj}}} \spc
\eq
where $\mathbf{x} \in \Cf$ and $\mathbf{y} \in \Cf$. We always use the relation
\bq
\mrR_{\mre} = \exp\{\ln \mrR\} = 
\exp\{\sum_{\mrj=1}^{\mrm}\,\ln\eG{\mrx_\mrj} - \sum_{\mrj=1}^{\mrn}\,\ln\eG{\mry_\mrj}\} \spp
\eq
This equation is correct, although in general
\bq
\ln \mrR \not= \sum_{\mrj=1}^{\mrm}\,\ln\eG{\mrx_\mrj} - \sum_{\mrj=1}^{\mrn}\,\ln\eG{\mry_\mrj} \spp
\eq
This is based on the fact that $\ln(\mra\,\mrb) = \ln(\mra) + \ln(\mrb) + \eta(\mra\,,\,\mrb)$,
but $\eta = 0$ or $\eta = \pm 2\,\mri\,\pi$.

There are several algorithms to compute $\ln \eG{\mrz}$, The Stirling series~\cite{HTF}, the formulas of
Lanczos and Spouge~\cite{Spouge}, Binet rising factorial series~\cite{Binet}, Burnside's formula~\cite{Burn} 
and many others; for a complete discussion we refer to \Bref{HAL}.

Our implementation of the Stirling series is as follows: given $\mrz = \mrx + \mri\,\mry$ and $\mrx < 0$ we use the relation
\bq
\eG{ - \mrz}\,\eG{\mrz} = - \frac{\pi}{\mrz}\,\mathrm{csc}(\pi\,\mrz) \spp
\eq
Next we use
\bq
\eG{\mrz} = \frac{\eG{\mrz + \mrn}}{(\mrz)_{\mrn - 1}} \spc
\eq
until we have $\mid \mrz + \mrn \mid > \ep^{-1}$, where $\ep \muchless 1$.
Here $( \mrz )_{\mrn}$ is the raising factorial. Finally we use
\bq
\ln\eG{\mrz} = (\mrz - \frac{1}{2})\,\ln \mrz - \mrz + \frac{1}{2}\,\ln(2\,\pi) +
\sum_{\mrn=1}^{\mrm}\,\frac{\mrB_{2\,\mrn}}{2\,\mrn\,(2\,\mrn - 1)}\,\mrz^{1-2\,\mrn} +
\ord{\mrz^{-1 - 2\,\mrm}} \spc
\eq
with $\mid \marg \,\mrz\, \mid < \pi$ and where $\mrB_{2\,\mrn}$ are Bernoulli's numbers. 
For fixed $\mra$ we can also use~\cite{Barnes}
\bq
\ln\eG{\mra + \mrz} = ( \mrz + \mra - \frac{1}{2} )\,\ln\mrz - \mrz + \frac{1}{2}\,\ln(2\,\pi) +
\sum_{\mrn=1}^{\mrm}\,( - 1)^{\mrn + 1}\,\frac{\mrB_{\mrn + 1}(\mra)}{\mrn\,(\mrn + 1)}\,\mrz^{ - \mrn} +
\ord{\mrz^{ - \mrm - 1}} \spc
\eq
with $\mid \marg \,\mrz\, \mid < \pi$ and where $\mrB_{\mrn}(\mra)$ are Bernoulli's polynomials~\cite{HTF}.
With $\ep^{-1} = 10$ and $\mrm = 4$ we have performed the following comparisons:
\[
\begin{array}{lll}
&& \\
\hline
&& \\
\mry  \quad & \quad 5 \quad & \quad 0.005 \\
\mid \eG{\frac{1}{2} + \mri\,\mry} \mid ^2 
\quad & \quad 9.4688688023924418\,\times\,10^{-7} \quad & \quad 3.1032293122516417 \\
&& \\
\frac{\pi}{\cosh(\mry\,\pi)}
\quad & \quad 9.4688688023964567\,\times\,10^{-7} \quad & \quad 3.1032293122545678 \\
&& \\
\hline
\end{array}
\]

\[
\begin{array}{llll}
&&& \\
\hline
&&& \\
\mrn \quad & \quad 1 \quad & \quad 2 \quad & \quad 3 \\
\eG{\frac{1}{2} - \mrn}
\quad & \quad -3.5449077018127881
\quad & \quad  2.3632718012086649
\quad & \quad -0.94530872048343373 \\
&&& \\
(-4)^\mrn\frac{\mrn\,!}{(2\,\mrn)\,!}\,\sqrt{\pi}
\quad & \quad -3.5449077018110318     
\quad & \quad  2.3632718012073544     
\quad & \quad -0.94530872048294179 \\
&&& \\
\hline
\end{array}
\]     

For small values of $\mid \mrz \mid$ we can use the Taylor expansion~\cite{HAL},
\bq
\ln\,\eG{\mrz} = - \ln \mrz +
\sum_{\mrn=1}^{\infty}\,\mrc_\mrn\,(\mrz - 1)^{\mrn} \spc
\eq
valid for $\mid \mrz - 1\mid < 2$ and with
\bq
\mrc_1 = 1 - \gamma \spc \qquad
\mrc_\mrn = ( - 1)^{\mrn}\frac{\zeta(\mrn) - 1}{\mrn} \spc \;\;
\mrn \ge 2 \spc
\eq
where $\gamma$ is the Euler{-}Mascheroni constant and $\zeta(\mrz)$ is the
Riemann zeta function. For larger values of $\mid \mrz \mid$ we will use
\bq
\ln\,\eG{\mrz} = \ln(\mrz - 1) + \ln\,\eG{\mrz - 1} \spc
\eq
recursively, covering the domains $\mid \mrz - 2 \mid < 2$ \etc

In addition, the computation of $\exp\{\mrx\}$ is performed as follows:
\bq
\mry= \frac{\mrx}{2^\mrk} \spc \quad \mrz = \exp\{\mry\} \spc \quad \exp\{\mrx\} = \mrz^{2^\mrk} \spp
\eq
The computation of $\sin \mrx$ and $\cos \mrx$ proceeds by using $\mry = \mrx/32$, $\mrs = \sin \mry$ and
$\mrc= \cos \mry$, then
\bqa
\mrs_2 = 2\,\mrs\,\mrc \spc \quad &\dots& \quad \sin(\mrx) = 2\,\mrs_{16}\,\mrc_{16} \spc
\nl
\mrc_2 = 2\,\mrc^2 - 1 \spc \quad &\dots& \quad \cos(\mrx) = 2\,\mrc_{16}^2 - 1 \spp
\eqa  
The stopping criterion is commonly defined as follows: given
\bq
\mrS = \sum_{\mrn=0}^{\infty}\,\mra_\mrn\,\mrz^{\mrn} \spc \qquad
\mrS_{\mrN} = \sum_{\mrn=0}^{\mrN}\,\mra_\mrn\,\mrz^{\mrn} \spc 
\eq
we use $\mrS_{\mrN}$ and stop computing termes when $\mid \mra_{\mrN + 1} \mid < \mathrm{tol}\,\mid \mrS_{\mrN} \mid$
for some $\mathrm{tol}$. Alternatively, we can terminate the computation of the series using the more
stringent condition that two successive terms are small. A more general discussion on the truncation error
term can be found in \Bref{Tmem}; for instance, if $\mrT_{\mrN}(\mrz)$ is the truncation error term in
$\ln \eG{\mrz}$ when using Stirling's approximation, one possible estimate~\cite{Spira} is given by
\bqa
\mid \mrT_{\mrN} &\le& 2\,\mid \frac{\mrB_{2\,\mrN}}{2\,\mrN - 1} \mid\,\mid \Im \mrz \mid^{1-2\,\mrN} \spc \qquad
\Re \mrz < 0, \spc \quad \Im \mrz \not=0 \spc
\nl
\mid \mrT_{\mrN} &\le& \mid \frac{\mrB_{2\,\mrN}}{2\,\mrN - 1} \mid\,\mid \mrz \mid^{1-2\,\mrN} \spc \qquad
\Re \mrz \ge 0 \spp 
\eqa
\paragraph{Choice of the algorithm} \hspace{0pt} \\
Selection of the algorithms for computing $\ln \eG{\mrz}$. Given
$\mrz= \mrx + \mri\,\mry$ and $\mrx < 0$ we use
\bq
\eG{\mrz} = - \frac{\pi}{\mrz\,\sin(\pi\,\mrz)}\,\frac{1}{\eG{- \mrz}} \spc
\eq
so that, in computing $\eG{\mrz}$ we have $\Re \mrz > 0$.
\bei
\item[\ovalbox{S}] Stirling series. We introduce
\bq
\mrn_\mrs = \Bigl[ - \mrx + (\ep^{-2} - \mry^2)^{1/2} \Bigr] \spc
\eq
obtaining
\bq
\eG{\mrz} = \bigl[ \prod_{\mrj=0}^{\mrn_\mrs - 1} (\mrz + \mrj) \bigr]^{-1}\,
\eG{\mrx + \mrn_\mrs + \mri\,\mry} \spp
\eq
\item[\ovalbox{T}] Taylor series. We introduce
\bq
\mrn_\mrt = \Bigl[ \mrx - 1 + ( 1 - \mry^2)^{1/2} \Bigr] \spp
\eq
The Taylor series requires $\mid \mrz - 1 \mid < 2$ but we prefer to use
$\mid \mrz - 1 \mid < 1$ and use
\bq
\eG{\mrz} = \Bigl[ \prod_{\mrj=1}^{\mrn_\mrt} (\mrz - \mrj) \Bigr]\,
\eG{\mrx - \mrn_\mrt + \mri\,\mry} \spp
\eq
When the choice is not obvious we adopt the following strategy: for
$\mrn_\mrs \le \mrn_\mrt$ we use the Stirling series, otherwise we use
the Taylor series.

We can also use the Taylor series in the region $\mid \mrz - 1 \mid < 2$. If
$2 < \mid \mrz - 1 \mid < 3$ we can use
\bq
\eG{\mrz} = \pi^{-1/2}\,2^{\mrz - 1}\,\eG{\frac{\mrz}{2}}\,\eG{\frac{\mrz + 1}{2}} \spc
\eq
and repeat the algorithm if needed.
\eei
We also need to compute rising and falling sequential products,
\bq
(\mrz)_\mrn = \prod_{\mrj=0}^{\mrn-1}\,(\mrz + \mrj) \spc \qquad
(\mrz)^\mrn = \prod_{\mrj=0}^{\mrn-1}\,(\mrz - \mrj) \spp
\eq
We will write
\bq
(\mrz)_\mrn = (\mrn - 1)\,!\;\mrz\,\prod_{\mrj=1}^{\mrn-1}(1 + \frac{\mrz}{\mrj})
\spc \qquad
(\mrz)^\mrn = ( - 1)^{\mrn-1}\,(\mrn - 1)\,!\;
\mrz\,\prod_{\mrj=1}^{\mrn-1}\,(1 - \frac{\mrz}{\mrj}) \spc
\eq
and take the corresponding logarithms where, for large $\mrn$ we use~\cite{tBurn}
\bq
\ln \mrn\,! = \mrn\,\ln \mrn - \mrn + \frac{1}{2}\,\ln(2\,\pi\,\mrn) +
\frac{1}{12\,\mrn} + \ord{\mrn^{-3}}
\quad \mbox{or} \quad
\mrn\,! \sim \sqrt{2\,\pi}\,\lpar \frac{\mrn + 1/2}{\mre} \rpar^{\mrn + 1/2} \spp
\eq
In this work, the symbol $(\mrz)^{\mrn}$ is used to
represent the falling factorial, and the symbol $(\mrz)_{\mrn}$
is used for the rising factorial. In fact, in the theory of special functions
(in particular the hypergeometric function) and in the standard reference
work Abramowitz and Stegun~\cite{abramowitz+stegun}, the Pochhammer symbol is used to represent the rising factorial.
The opposite conventions are used in combinatorics~\cite{comb}. 

Furthermore, in computing $\ln(1 + \mrz)$ with $\mid \mrz \mid \to 0$ we always use
\bq
\ln(1 + \mrz) = \mrz\,\mrR_\mrc \Bigr( (1 + \frac{\mrz}{2})^2\,,\,1 + \mrz \Bigr) \spp
\eq
where $\mrR_\mrc$ is one of Carlson's elliptic integrals~\cite{Carlson}.

In computing the $\uppsi$ function we adopt the following strategy: when $\Re \mrz < 0$ we use
\bq
\uppsi(\mrz) = \uppsi( - \mrz) - \pi\,\cot(\pi\,\mrz) - \frac{1}{\mrz} \spp
\eq
Next we use
\bq
\uppsi(\mrz)= \uppsi(\mrz + \mrn) + \sum_{\mrj=0}^{\mrn - 1}\,\frac{1}{\mrz + \mrj} \spc
\eq
until $\mid \mrz + \mrn \mid > \ep^{-1}$ for some $\ep \muchless 1$. Finally we use
\bq
\uppsi(\mrz) = \ln \mrz - \frac{1}{2\,\mrz} - \sum_{\mrj=1}^{\mrN}\,\frac{\mrB_{2\,\mrj}}{2\,\mrj}\,\mrz^{ - 2\,\mrj} \spp
\eq
A similar strategy can be adopted in the computation of polygamma functions by using the results of
Section~$6.4$ of \Bref{abramowitz+stegun}:
\bq
\psin{\mrz} = ( - 1)^\mrn\,\psin{1 - \mrz} - \pi\,\frac{\mrd^\mrn}{\mrd \mrz^\mrn}\,\cot( \pi \mrz) \spc \quad
\psin{\mrz} = \psin{\mrz + \mrk} + ( - 1)^{\mrn + 1}\,\mrn\,!\;\sum_{\mrj=0}^{\mrk - 1}\,
\frac{1}{(\mrz + \mrj)^{\mrn + 1}} \spc
\eq
\bq
\psin{\mrz} \sim ( - 1)^{\mrn - 1}\,\Bigl[
\frac{(\mrn - 1)\,!}{\mrz^\mrn} + \frac{\mrn\,!}{2\,\mrz^{\mrn + 1}} +
\sum_{\mrj=1}^{\infty}\,\mrB_{2\,\mrj}\,\frac{(2\,\mrj + \mrn - 1)\,!}{(2\,\mrj)\,!}\,\mrz^{- \mrn - 2\,\mrj} \spc
\eq
for $\mid \marg(\mrz) \mid < \pi$.
\paragraph{Ratio of $Gamma$ functions} \hspace{0pt} \\
In order to reduce the number of Gamma function we consider the following expansion~\cite{HTgam}:
\bq
\frac{\eG{\mrz + \mra}}{\eG{\mrz + \mrb}} \sim
\sum_{\mrn=0}^{\infty}\,\mrC_{\mrn}(\mra\,,\,\mrb)\,\mrz^{\mra - \mrb - \mrn} \spc
\eq
where the coefficients satisfy
\bq
\mrC_\mrn(\mrc\,,\,\mrb) = \frac{1}{\mrn}\,\sum_{\mrk=0}^{\mrn - 1}\,\Bigl[
\frac{
      \eG{\mrc - \mrk + 1}
     }
     {
      (\mrn - \mrk + 1)\,!\;\eG{\mrc - \mrn + 1}
     } -
( - 1)^{\mrn + \mrk}\,\mrc\,\mrb^{\mrn - \mrk} \Bigr]\,\mrC_\mrk(\mrc\,,\,\mrb) \spc
\eq
where $\mrc= \mra - \mrb$ and where the first coefficients are:
\bqa
\mrC_1 &=& \Bigl[ \mrb + \frac{1}{2}\,(\mrc - 1) \Bigr]\,\mrc \spc
\nl
\mrC_2 &=& - \frac{1}{24}\,(1 - 3\,\mrc)\,(\mrc - 2) + \frac{1}{2}\,(\mrc - 1)\,\mrb + \frac{1}{2}\,\mrb^2 \spc
\nl
\mrC_3 &=&
       \frac{1}{2160}\,(10 - 26\,\mrc - 9\,\mrc^2 + 15\,\mrc^3)\,(\mrc - 2)
          - \frac{1}{36}\,(1 - 2\,\mrc^2)\,(\mrc - 2)\,\mrb
          - \frac{1}{36}\,(2 + 5\,\mrc - 12\,\mrc^2 + 6\,\mrc^3)\,\mrb^2
\nl
{}&+&       \frac{1}{6}\,(3 - 2\,\mrc)\,\mrc\,\mrb^3 \spc
\eqa
with $\mrc= \mra - \mrb$. We implement the algorithm as follows:
\bq
\frac{\eG{\mrz + \mra}}{\eG{\mrz + \mrb}} \approx
\frac{(\mrz + \mrb)^{\mrn-1}}{(\mrz + \mra)^{\mrn-1}}\,
\frac{\eG{\mrz + \mra + \mrn}}{\eG{\mrz + \mrb + \mrn}} \spc
\eq
for large enough values of $\mrn$ the agreement with the Stirling series is excellent but the algorithm is slow,
by approximately a factor two, w.r.t. the Stirling series.

Another result is given by the following equation~\cite{Gquot}:
\bq
\frac{
      \prod_{\mrj=1}^{\mrn}\,\eG{\alpha_\mrj}
     }
     {
      \prod_{\mrj=1}^{\mrn}\,\eG{\beta_\mrj}
     } \sim
\prod_{\mrk=0}^{\mrk_{\max}}\,
\frac{
      (\beta_1 + \mrk)\,\dots\,(\beta_{\mrn} + \mrk)
     }
     { 
      (\alpha_1 + \mrk)\,\dots\,(\alpha_{\mrn} + \mrk)
     } \spc
\eq
where $\alpha_\mrj\,,\,\beta_\mrj$ are nonzero complex numbers, none of which are negative integers and
$\sum\,\alpha_\mrj = \sum\,\beta_\mrj$.
Alternatively we can use Section~(1.3) of \Bref{HTF},
\bq
\prod_{\mrk=1}^{\mrn}\,\frac{\eG{1 - \beta_\mrk}}{\eG{1 - \alpha_\mrk}} =
\prod_{\mrk=1}^{\infty}\,
\frac{
      \prod_{\mrj=1}^{\mrn}\,(\mrk - \alpha_\mrj)
     }
     {
      \prod_{\mrj=1}^{\mrn}\,(\mrk - \beta_\mrj)} \spp
\eq
The product is convergent, however the rate of convergence is very slow, requiring high values 
for $\mrk_\max$.

Incomplete Gamma functions are defined by
\bqa
\sG{\mrz}{\mrx} &=& \int_0^{\mrx} \mrd \mrt \exp\{ - \mrt \}\,\mrt^{\mrz - 1} =
\exp\{ - \mrx\}\,\frac{\mrx^{\mrz}}{\mrz}\,
{}_{\scriptstyle{1}}\,\mrF_{\scriptstyle{1}}\lpar 1\,,\,1 + \mrz\,,\,\mrx \rpar \spc
\nl
\bG{\mrz}{\mrx} &=& \int_{\mrx}^{\infty} \mrd \mrt \exp\{ - \mrt \}\,\mrt^{\mrz - 1} =
\exp\{ - \mrx\}\,\mrx^{\mrz - 1}\,
{}_{\scriptstyle{2}}\,\mrF_{\scriptstyle{0}}\lpar 1\,,\,1 - \mrz\,,\, - \frac{1}{\mrx} \rpar \spp
\eqa
\paragraph{Horn{-}Humbert functions} \hspace{0pt} \\
The Horn{-}Humbert series $\Upphi_1$ is defined by the double series
\bq
\Upphi_1(\alpha\,,\,\beta\,,\,\gamma\,;\,\mrx\,,\,\mry) =
\sum_{\mrm,\mrn=0}^{\infty}\,
\frac{\poch{\alpha}{\mrm + \mrn}\,\poch{\beta}{\mrm}}{\poch{\gamma}{\mrm + \mrn}}\,
\frac{\mrx^{\mrm}\,\mry^{\mrn}}{\mrm\,!\;\mrn\,!} \spc \quad \mid \mrx \mid < 1 \spp
\eq
The analytic continuation is given by the following integral,
\bq
\Upphi_1(\alpha\,,\,\beta\,,\,\gamma\,;\,\mrx\,,\,\mry) =
\frac{1}{\eB{\alpha}{\beta}}\,\int_0^1 \mrd \mrt
\exp\{\mry\,\mrt\}\,\mrt^{\alpha - 1}\,(1 - \mrt)^{\beta - 1}\,(1 - \mrx\,\mrt)^{ - \gamma} \spc
\eq
valid for $ \Re \alpha > 0$, $\Re \beta > 0$ and $\mid \marg(1 - \mrx) \mid < \pi$.
The function can also be written as
\bq
\Upphi_1(\alpha\,,\,\beta\,,\,\gamma\,;\,\mrx\,,\,\mry) = 
\sum_{\mrn=0}^{\infty}\,\frac{\mry^{\mrn}}{\mrn\,!}\,\eB{\alpha + \mrn}{\beta}\,
\hyp{\gamma}{\alpha + \mrn}{\alpha + \beta + \mrn}{\mrx} \spp
\label{PHser}
\eq
Numerical values for $\Upphi_1$ can be obtained by using the Gauss{-}Jacobi quadrature,
\ie we compute the Gauss{-}Jacobi knots ($\mrk$) and weights ($\mrw$) for the interval $[0\,,\,1]$ and
exponents $\alpha - 1$ and $\beta - 1$ and obtain
\bq
\int_0^1 \mrd \mrt
\exp\{\mry\,\mrt\}\,\mrt^{\alpha - 1}\,(1 - \mrt)^{\beta - 1}\,(1 - \mrx\,\mrt)^{ - \gamma} =
\sum_{\mrn=1}^{\mrN}\,\mrw_\mrn\,\mrF(\mrk_\mrn) \spc \quad
\mrF(\mrt) = \exp\{ \mry\,\mrt - \gamma\,\ln(1 - \mrx\,\mrt)\} \spp
\eq
If we use \eqn{PHser} and $\alpha, \beta$ and $\gamma$ are positive integers we are in the
logatithmic case. For instance, when $\alpha = \beta = \gamma = 1$ we have
\bq
\hyp{1}{\mrn + 1}{\mrn + 2}{\mrx} = ( \mrn + 1)\,(\mrF_{\infty} + \mrF_{\mrn-1} + \mrL) \spc
\label{Hln}
\eq
\bqa
\mrF_{\infty} &=&
\mrx^{-\mrn-1}\,\sum_{\mrk=1}^{\infty}\,
\frac{\poch{1}{\mrn+\mrk}\,(\mrk - 1)\,!}{(\mrn + \mrk)\,\;\mrk\,!}\,\mrx^{-\mrk} \spc
\nl
\mrF_{\mrn-1} &=&
- \frac{1}{\mrx}\,\sum_{\mrk=0}^{\mrn-1}\,
\frac{(\mrn-\mrk-1)\,!\;\poch{1}{\mrk}}{(\mrn-\mrk)\,!\;\mrk\,!}\,\mrx^{-\mrk} \spc
\nl
\mrL &=&
- \mrx^{-\mrn-1}\,\frac{\poch{1}{\mrn}}{\mrn\,!}\,\Bigl[
\ln( - \mrx) + \uppsi(\mrn+1) - \uppsi(1) \Bigr] \spp
\eqa
As a next step we define multivariate $\Upphi_1$ functions. The first one is
\bq
\Upphi^{(2)}_1(\mathbf{a}\,,\,\mathbf{b}\,,\,\mathbf{c}\,;\,\mathbf{x}\,,\,\mathbf{y}) =
\int_0^1 \mrd \mrt_1\,\mrd \mrt_2\,
\Bigl[ \prod_{\mrj=1}^{2}\,\exp\{\mry_{\mrj}\,\mrt_{\mrj}\}\,\mrt_{\mrj}^{\mra_{\mrj} - 1}\,
(1 - \mrt_{\mrj})^{\mrb_{\mrj} - 1}\Bigr]\,
(1 - \mrx_1\,\mrt_1)^{-\mrc_1}\,\Bigl[ 1 - \mrx_2\,(\mrt_1 + \mrt_2)\Bigr]^{-\mrc_2} \spp
\label{Phi2int}
\eq
If $\mid \mrx_1 \mid < 1$ and $\mid \mrx_2 \mid < 1/2$ we can write
\bqa
\Upphi^{(2)}_1(\mathbf{a}\,,\,\mathbf{b}\,,\,\mathbf{c}\,;\,\mathbf{x}\,,\,\mathbf{y}) &=&
\sum_{\mrm_1,\mrm_2=0}^{\infty}\,\frac{\mry_1^{\mrm_1}\,\mry_2^{\mrm_2}}{\mrm_1\,!\;\mrm_2\,!}\,
\sum_{\mrj_1,\mrj_2=0}^{\infty}\,\sum_{\mrj_3=0}^{\mrj_2}\,
\eB{\mra_2 + \mrm_2 + \mrj_3}{\mrb_2}
\nl
{}&\times&
\eB{\mra_1 + \mrm_1 + \mrj_1 + \mrj_2 - \mrj_3}{\mrb_1}\,
\frac{\ff{-\mrc_1}{\mrj_1}\,\ff{-\mrc_2}{\mrj_2}\,\ff{\mrj_2}{\mrj_3}}{\mrj_1\,!\;\mrj_2\,!\;\mrj_3\,!}\,
( - \mrx_1 )^{\mrj_1}\,( - \mrx_2 )^{\mrj_2} \spc
\eqa
where $\eB{\mra}{\mrb}$ is the Euler Beta function and $\ff{\alpha}{\mrk}$ is the falling factorial.

We will not present general results for the analytic continuation and only the relevant case, where 
$\mrb_\mrj = \mrc_\mrj = 1$ and $\mra_\mrj = \mrn_\mrj +1$ integers, will be discussed. 
After expanding the exponentials in \eqn{Phi2int} we obtain
\bq
\Upphi^{(2)}_1 = \sum_{\mrm,\mrn=0}^{\infty}\,\frac{\mry_1^{\mrm}\,\mry_2^{\mrn}}{\mrm\,!\;\mrn\,!}\,
\int_0^1 \mrd \mrt_1\,\frac{\mrt_1^{\mrm + \mrn_1}}{1 - \mrx_1\,\mrt_1}\,
\int_0^1 \mrd \mrt_2\,\frac{\mrt_2^{\mrn + \mrn_2}}{1 - \mrx_2\,(\mrt_1 + \mrt_2)} \spp
\eq
Performing first the $\mrt_1$ integral we arrive at
\bq
\Upphi^{(2)}_1 = \sum_{\mrm,\mrn=0}^{\infty}\,\frac{\mry_1^{\mrm}\,\mry_2^{\mrn}}{\mrm\,!\;\mrn\,!}\,
\mrI(\mrN_1\,,\,\mrN_2\,;\,\mrx_1\,,\,\mrx_2) \spc
\eq
where $\mrm + \mrn_1 = \mrN_1$, $\mrn + \mrn_2 = \mrN_2$. Furthermore,
\bqa
\mrI(\mrN_1\,,\,\mrN_2\,;\,\mrx_1\,,\,\mrx_2) &=& \frac{1}{(\mrN_1 + 1)\,(\mrN_2 + 1)}\,
\sum_{\mrk_1,\mrk_2=0}^{\infty}\,
\frac{\poch{\mrN_1 + 1}{\mrk_1 + \mrk_2}\,\poch{1}{\mrk_1}\,\poch{1}{\mrk_2}}
     {\poch{\mrN_1 + 2}{\mrk_1 + \mrk_2}}\,\frac{\mrx_1^{\mrk_1}\,\mrx_2^{\mrk_2}}{\mrk_1\,!\;\mrk_2\,!}
\nl
{}&\times&
\hyp{1 - \mrk_2}{\mrN_2 + 1}{\mrN_2 + 2}{\mrx_2} \spp
\eqa
If we exchange the order of integration and perform the $\mrt_2$ integral we obtain
\bq
\Upphi^{(2)}_1 =
\sum_{\mrm,\mrn=0}^{\infty}\,\frac{\mry_1^{\mrm}\,\mry_2^{\mrn}}{\mrm\,!\;\mrn\,!}\,
\frac{1}{\mrn + \mrn_2 + 1}\,
\int_0^1 \mrd \mrt\,\frac{\mrt^{\mrm + \mrn_1}}{(1 - \mrx_1\,\mrt)\,(1 - \mrx_2\,\mrt)}\,
\hyp{1}{\mrn+\mrn_2+1}{\mrn + \mrn_2 + 2}{\frac{\mrx_2}{1 - \mrx_2\,\mrt}} \spp
\eq
Performing a partial fraction decomposition
\bq
\frac{1}{(1 - \mrx_1\,\mrt)\,(1 - \mrx_2\,\mrt)} =
\frac{1}{\mrx_1 - \mrx_2}\,\,\Bigl( \frac{\mrx_1}{1 - \mrx_1\,\mrt} - \frac{\mrx_2}{1 - \mrx_2\,\mrt} \Bigr) \spc
\eq
we arrive at the following integral:
\bq
\mrI^{\mrj}_{\Upphi} = \int_0^1 \mrd \mrt\,
\frac{\mrt^{\mrN_1}}{1 - \mrx_{\mrj}\,\mrt}\,\hyp{1}{\mrN_2 + 1}{\mrN_2 + 2}{\frac{\mrx_2}{1 - \mrx_2\,\mrt}} \spp
\eq
After using \eqn{Hln} we are left with a set of elementary integrals; after performing them we finally arrive at the 
third representation by series of the function.

If we have a power series in $\mrr$ variables,
\bq
\mrf(\mathbf{z}) = \sum_{\{\mrn_j\}=0}^{\infty}\,\mrA(\mathbf{n})\,\prod_{\mrj=1}^{\mrr}\,\mrz_\mrj^{\mrn_\mrj} \spc
\eq
and if
\bq
\frac{\mrA(\mathbf{n} + \mathbf{e}_\mrj)}{\mrA(\mathbf{n})} = \frac{\mrP(\mathbf{n})}{\mrQ(\mathbf{n})} \spc
\label{Hornc}
\eq
with $\mathbf{e}_\mrj = (0,\dots,1_\mrj,\dots,0)$ and $\mrP, \mrQ$ are finite polynomials then $\mrf$ 
satisfies~\cite{Horn35,HTF} 
\bq
\Bigl[ \mrQ_\mrj\lpar \mathbf{z}\,\frac{\partial}{\partial \mathbf{z}} \rpar\,\mrz_\mrj^{-1} -
\mrQ_\mrj\lpar \mathbf{z}\,\frac{\partial}{\partial \mathbf{z}} \rpar \Bigr]\,\mrf = 0 \spc
\qquad \mrj = 1,\dots,\mrr \spp
\eq
For $\Upphi_1$ we obtain
\bq
\mrx\,(1 - \mrx)\,\Upphi_{1,\mrx \mrx} +
\mry\,(1 - \mrx)\,\Upphi_{1,\mrx \mry} +
\Bigl[ \gamma - (\alpha + \beta + 1)\,\mrx \Bigr]\,\Upphi_{1,\mrx} -
\beta\,\gamma\,\Upphi_{1,\mry} -
\alpha\beta\,\Upphi_1 = 0 \spc
\eq
\bq
\mry\,\Upphi_{1,\mry \mry} +
\mrx\,\Upphi_{1,\mrx \mry} +
(\gamma - \mry)\,\Upphi_{1,\mry} -
\mrx\,\Upphi_{1,\mrx} -
\gamma\,\Upphi_1 = 0 \spp
\eq
For $\Upphi^{2}_1$ the situation is different; indeed it can be written as a multiple series involving
hypergeometric functions. The condition in \eqn{Hornc} is violated and it is not possible to write
a system of linear partial differential equations in a closed form, \ie $\Upphi^{(2)}_1$ is not a hypergeometric Horn
function of four variables.

Another function is 
\bq
\Upphi^{(3)}_{1\,\mrA}(\mathbf{a}\,,\,\mathbf{b}\,,\,\mathbf{c}\,;\,\mathbf{x}\,,\,\mathbf{y}) =
\Bigl[ \prod_{\mrj=1}^{2}\,\int_0^1 \mrd \mrt_\mrj\,\mrt_\mrj^{\mrn_\mrj} 
\exp\{\mry_\mrj\,\mrt_\mrj\} \Bigr]\,
(1 - \mrx_1\,\mrt_1)^{-1}\,
(1 - \mrx_2\,\mrt_1 - \mrx_3\,\mrt_2)^{-1} \spp
\eq
After expanding the exponental we are left with the following integral
\bq
\mrJ =
\Bigl[ \prod_{\mrj=1}^{2}\,\int_0^1 \mrd \mrt_\mrj\,\mrt_\mrj^{\mrN_\mrj} \Bigr]\,
(1 - \mrx_1\,\mrt_1)^{-1}\,
(1 - \mrx_2\,\mrt_1 - \mrx_3\,\mrt_2)^{-1} \spc
\eq
which can be written as follows:
\bqa
\mrJ &=& \sum_{\mrn=0}^{\infty}\,
\Bigl( \begin{array}{c} -1 \\ \mrn \end{array} \Bigr)\,
( - \mrx_1 )^{\mrn}\,
\mrI(\mrn\,,\,\mrN_1\,,\,\mrN_2) \spc
\qquad \mid \mrx_1 \mid < 1 \spc
\nl
\mrI &=& \frac{\eG{\mrn + \mrN_1 + 1}\,\eG{\mrN_2 + 1}}
              {\eG{\mrn + \mrN_1 + 2}\,\eG{\mrN_2 + 2}}\,
\mrF_2\lpar1\,,\,\mrn + \mrN_1 + 1\,,\,\mrN_2 + 1\,,\,
\mrn + \mrN_1 + 2\,,\,\mrN_2 + 2\,;\,\mrx_2\,,\,\mrx_3 \rpar \spc
\eqa
where $ \mrF_2$ is one of the Appell functions~\cite{HTF,Asur}.

The generalization to more variables introduces new functions, \eg
\bqa
\Upphi^{(3,1)}_{1\,\mrB}(\mathbf{a}\,,\,\mathbf{b}\,,\,\mathbf{c}\,;\,\mathbf{x}\,,\,\mathbf{y}) &=&
\int_0^1 \mrd \mrt_1\,\mrd \mrt_2\,\mrd \mrt_3
\Bigl[ \prod_{\mrj=1}^{3}\,\exp\{\mry_{\mrj}\,\mrt_{\mrj}\}\,\mrt_{\mrj}^{\mra_{\mrj} - 1}\,
(1 - \mrt_{\mrj})^{\mrb_{\mrj} - 1}\Bigr]\,
(1 - \mrx_1\,\mrt_1)^{-\mrc_1}\,
(1 - \mrx_2\,\mrt_2)^{-\mrc_2}
\nl
{}&\times&
\Bigl[ 1 - \mrx_3\,(\mrt_1 + \mrt_2 + \mrt_3)\Bigr]^{-\mrc_3} \spc
\nl
\Upphi^{(3,2)}_{1\,\mrB}(\mathbf{a}\,,\,\mathbf{b}\,,\,\mathbf{c}\,;\,\mathbf{x}\,,\,\mathbf{y}) &=&
\int_0^1 \mrd \mrt_1\,\mrd \mrt_2\,\mrd \mrt_3
\Bigl[ \prod_{\mrj=1}^{3}\,\exp\{\mry_{\mrj}\,\mrt_{\mrj}\}\,\mrt_{\mrj}^{\mra_{\mrj} - 1}\,
(1 - \mrt_{\mrj})^{\mrb_{\mrj} - 1}\Bigr]\,
(1 - \mrx_1\,\mrt_1)^{-\mrc_1}\,
\Bigl[ 1 - \mrx_2\,(\mrt_1 + \mrt_2)\Bigr]^{-\mrc_2}
\nl
{}&\times&
\Bigl[ 1 - \mrx_3\,(\mrt_1 + \mrt_3)\Bigr]^{-\mrc_3} \spp
\eqa
We can see that there will be $3$ $\Upphi^{(4,\mrj)}_1$ functions,
$4$ $\Upphi^{(5,\mrj)}_1$ functions \etc
\section{Examples \label{exa}}
In this Section we consider several examples; the Sinc lattice is defined in terms of an integer $\mrN$ in
\eqn{Spar}. The Sinc lattice points are defined in \eqn{Slpo} or in \eqn{Slpt}, depending on the
aymptotic behavior of the integrand. The bottleneck of the Sinc lattice is that it requires
$\ord{(2\,\mrN + 1)^{\mrr})}$ calls to the integrand, where $\mrr$ is the number of variables.

An high speed evaluation~\cite{deDoncker:2018nqe,deDoncker:2020gub,Borowka:2018goh}
of Fox functions requires that the Sinc lattice rules are implemented in
{\tt{CUDA}} for many{-}core computations on GPU's. Focusing a future line of work, it is expected to apply the
methodology, allowing for high speed evaluation of Feynman integrals in the physical region.
Meanwhile we have based our numerical results on the shared memory programming 
OpenMP~\cite{dagum1998openmp}. 

There are several cases where the number of MB integrals can be reduced. Consider the following example where
we have $6$ MB integrals:
\bqa
\mrH_6 &=& \Bigl[ \prod_{\mrj=1}^{6}\,\int_{\mrL_\mrj}\,\frac{\mrd \mrs_\mrj}{\tip}\,
                \eG{ - \mrs_j}\,\mrz_\mrj^{\mrs_\mrj} \Bigr]\,\frac{\mrN}{\mrD} \spc
\nl
\mrN &=&
\eG{2 + \mrs_5 + \mrs_3}\,
\eG{3 + \mrs_6 + \mrs_3}\,
\eGs{2 + \mrs_1 + \mrs_2}\,
\eG{1 + \mrs_1 + \mrs_2}\,
\eG{1 + \mrs_1 + \mrs_2 + \mrs_3}
\nl
{}&\times& 
\eG{1 + \mrs_3 + \mrs_4}\,
\eG{\frac{1}{2} + \mrs_4 + \mrs_5 + \mrs_6} \spc
\nl
\mrD &=&
\eG{1 + \mrs_3}\,
\eG{1 + \mrs_2 + \mrs_3}\,
\eGs{1 + \mrs_1 + \mrs_2}\,
\eG{2 + \mrs_1 + \mrs_2 + \mrs_3}\,
\eG{\frac{3}{2} + \mrs_4 + \mrs_5 + \mrs_6} \spp
\eqa
We can reduce $\mrH_6$ to MB integrals with $3$ variables at the price of introducing a series. 
Assuming that $\mid \mrz_3 \mid < 1$, the result is
\bq
\mrH_6 = \sum_{\mrn=0}^{\infty}\,(\mrn + 1)^2\,(\mrn + 2)\,\mrz_3^{\mrn}\,\mrJ_{\mrn}\,\spp
\eq
\bqa
\mrJ_{\mrn} = &-&
\sum_{\mrj=0}^{\mrn}\,\sum_{\mrk=0}^{\mrn - \mrj}\,\mrC_1(\mrn\,,\,\mrj\,,\,\mrk\,,\,\mrz_4\,,\,\mrz_5\,,\,\mrz_6)\,
\mrH_{31}(\mrn\,,\,\mrj\,,\,\mrk\,,\,\mrz_4\,,\,\mrz_5\,,\,\mrz_6) 
\nl {}&+&
\sum_{\mrj=0}^{\mrn+1}\,\sum_{\mrk=0}^{\mrn + 1 - \mrj}\,\mrC_2(\mrn\,,\,\mrj\,,\,\mrk\,,\,\mrz_4\,,\,\mrz_5\,,\,\mrz_6)\,
\mrH_{32}(\mrn\,,\,\mrj\,,\,\mrk\,,\,\mrz_4\,,\,\mrz_5\,,\,\mrz_6) 
\nl {}&+&
\sum_{\mrj=0}^{\mrn+1}\,\sum_{\mrk=0}^{\mrn + 2}\,\mrC_3(\mrn\,,\,\mrj\,,\,\mrk\,,\,\mrz_4\,,\,\mrz_5\,,\,\mrz_6)\,
\mrH_{33}(\mrn\,,\,\mrj\,,\,\mrk\,,\,\mrz_4\,,\,\mrz_5\,,\,\mrz_6) 
\nl {}&+&
\sum_{\mrj=0}^{\mrn}\,\sum_{\mrk=0}^{\mrn + 2}\,\mrC_4(\mrn\,,\,\mrj\,,\,\mrk\,,\,\mrz_4\,,\,\mrz_5\,,\,\mrz_6)\,
\mrH_{33}(\mrn\,,\,\mrj\,,\,\mrk\,,\,\mrz_4\,,\,\mrz_5\,,\,\mrz_6) \spp
\eqa
We introduce the following notations:
\bq
\int\,\mrd\,\mrS_{1 2 \mrj} = 
\int_{\mrL_1}\,\frac{\mrd \mrs_1}{\tip}\, 
\int_{\mrL_2}\,\frac{\mrd \mrs_2}{\tip}\, 
\int_{\mrL_\mrj}\,\frac{\mrd \mrs_\mrj}{\tip}\,
\Gamma(-\mrs_1\,,\,-\mrs_2\,,\,-\mrs_\mrj)\,
\mrz_1^{\mrs_1}\,\mrz_2^{\mrs_2}\,\mrz_\mrj^{\mrs_\mrj} \spc
\eq
\bq
\Gamma(-\mrs_1\,,\,-\mrs_2\,,\,-\mrs_\mrj)=
\eG{-\mrs_1}\,\eG{-\mrs_2}\,\eG{-\mrs_\mrj} \spp
\eq
The $3{-}$fold MB integrals are:
\bqa
\mrH_{31} &=& \int\,\mrd\,\mrS_{1 2 4}\,
\frac{
\eG{\frac{1}{2} + \mrs_4}\,
\eGs{2 + \mrs_2 + \mrs_1}\,
\eG{1 + \mrs_2 + \mrs_1 + \mrn}\,
\eG{1 - \mrk - \mrj + \mrs_4 + \mrn}\,
}{
\eG{\frac{3}{2} + \mrs_4}\,
\eG{1 + \mrs_2 + \mrn}\,
\eG{1 + \mrs_2 + \mrs_1}\,
\eG{2 + \mrs_2 + \mrs_1 + \mrn}\,
}
\nl
\mrH_{32} &= & \int\,\mrd\,\mrS_{1 2 5}\,
\frac{
\eG{\frac{1}{2} + \mrs_5}\,
\eGs{2 + \mrs_2 + \mrs_1}\,
\eG{1 + \mrs_2 + \mrs_1 + \mrn}\,
\eG{2 - \mrk - \mrj + \mrs_5 + \mrn}\,
}{
\eG{\frac{3}{2} + \mrs_5}\,
\eG{1 + \mrs_2 + \mrn}\,
\eG{1 + \mrs_2 + \mrs_1}\,
\eG{2 + \mrs_2 + \mrs_1 + \mrn}\,
}
\nl
\mrH_{33} &=& \int\,\mrd\,\mrS_{1 2 6}\,
\frac{
\eG{\frac{1}{2} + \mrs_6}\,
\eGs{2 + \mrs_2 + \mrs_1}\,
\eG{1 + \mrs_2 + \mrs_1 + \mrn}\,
\eG{3 - \mrk + \mrs_6 + \mrn}\,
}{
\eG{\frac{3}{2} + \mrs_6}\,
\eG{1 + \mrs_2 + \mrn}\,
\eG{1 + \mrs_2 + \mrs_1}\,
\eG{2 + \mrs_2 + \mrs_1 + \mrn} \spc
}
\eqa
with coefficients defined by
\bqa
\mrC_1 &=&
\mrp(\mrk + \mrj + \mrn\,,\,-1)\,\frac{\eG{\mrn + 1}}{\eG{1 - \mrk - \mrj + \mrn}}\,\mrB(1 + \mrj + \mrn\,,\,1 + \mrn)\,
\mrB(2 + \mrk + \mrn\,,\,2 + \mrn)\,
\mrp(5 + 2\,\mrn\,,\,\mrz_4)\,\mrp(\mrj\,,\,\mrz_5)\,\mrp(\mrk\,,\,\mrz_6)
\nl {}&\times&
\mrp( - 2 - \mrj - \mrn\,,\, \mrz_4 - \mrz_5)\,\mrp( - 3 - \mrk - \mrn\,,\,\mrz_4 - \mrz_6)\,
\nl
\mrC_2 &=&
\mrp(\mrk\,,\,-1)\,\frac{\eG{\mrn + 1}}{\eG{2 - \mrk - \mrj + \mrn}}\,\mrB(\mrj + \mrn\,,\,\mrn)\,
\mrB(2 + \mrk + \mrn\,,\,2 + \mrn)\,
\mrp(\mrj\,,\,\mrz_4)\,\mrp(4 + 2\,\mrn\,,\,\mrz_5)\,\mrp(\mrk\,,\,\mrz_6)
\nl &\times&
\mrp( - 1 - \mrj - \mrn\,,\, \mrz_4 - \mrz_5)\,\mrp( - 3 - \mrk - \mrn\,,\, \mrz_5 - \mrz_6)\,
\nl
\mrC_3 &=&
\mrp( - \mrj + 3\,\mrn\,,\,-1)\,\frac{\eG{\mrn + 1}}{\eG{3 - \mrk + \mrn}}\,\mrB(\mrj + \mrn\,,\,\mrn)\,
\mrB(1 + \mrk - \mrj + \mrn\,,\,1 - \mrj + \mrn)
\mrp(\mrj\,,\,\mrz_4)\,\mrp(1 + \mrk + \mrn\,,\,\mrz_5)\,\mrp(2 - \mrj + \mrn\,,\,\mrz_6)
\nl {}&\times&
\mrp( - 1 - \mrj - \mrn\,,\, \mrz_4 - \mrz_5)\,\mrp( - 2 - \mrk + \mrj - \mrn\,,\, \mrz_5 - \mrz_6)\,
\nl
\mrC_4 &=&
\frac{\eG{\mrn + 1}}{\eG{3 - \mrk + \mrn}}\,\mrB(1 + \mrj + \mrn\,,\,1 + \mrn)\,\mrB(\mrk - \mrj + \mrn\,,\, - \mrj + \mrn)
\mrp(2 + \mrk + \mrn\,,\,\mrz_4)\,\mrp(\mrj\,,\,\mrz_5)\,\mrp(1 - \mrj + \mrn\,,\,\mrz_6)
\nl {}&\times&
\mrp( - 2 - \mrj - \mrn\,,\, \mrz_4 - \mrz_5)\,\mrp( - 1 - \mrk + \mrj - \mrn\,,\,\mrz_4 - \mrz_6) \spc
\eqa
where we have defined
\bq
\mrp(\mrn\,,\,\mrz)= \mrz^\mrn\, \spc \qquad
\mrB(\mrn\,,\,\mrk) = \frac{\mrn\,!}{\mrk\,!\;(\mrn - \mrk\,!)} \spp
\eq
The strategy consists in identifying a Lauricella function $\mrF^{(3)}_{\mrD}$, use the Euler{-}Mellin representation,
perform a set of partial fraction decompositions and use again the Mellin{-}Barnes representation.

The results obtained using a Sinc approximation are always compared
with those obtained using {\tt OKROBV}. 
In the following we give results for $\overline{\mrH} = (2\,\pi)^{\mrr}\,\mrH$, where $\mrr$ is the number of
variables and ${\overline{\mrH}}_{\mrV}$ denotes numerical integration via {\tt OKROBV}. 
\bei

\item[\ovalbox{1}]
\bq
\mrH = \int_{\mrL}\,\frac{\mrd \mrs}{\tip}\,
\frac{
\eG{ - \mrs}\,\eG{0.6 - \mrs}\,\eG{0.41 + \mrs}\,\eG{0.59 + \mrs}\,\eG{0.64 + \mrs}
     }
     {
\eG{0.8 - \mrs}\,\eG{1.8 + \mrs}\,\eG{1.25 + \mrs}
     }\,\mrz^{\mrs} \spp
\eq
The corresponding parameters are $\alpha = 2, \beta = 0$ and $\lambda= - 2.61$. The countour $\mrL$ has
$\sigma = - 0.25$. With $\mrz = 0.51 + 0.1\,\mri$ we obtain
\bq
{\overline{\mrH}}_{\mrV} = 40.81335(3) \;- 0.35114129(2)\,\mri \spc
\eq
with a Sinc result
\sline
\bqa
\mrN &=& 100 \qquad  40.8138092 \;- 0.350783157\,\mri
\nl
\mrN &=& 200 \qquad  40.8133491 \;- 0.351120669\,\mri
\nl
\mrN &=& 300 \qquad  40.8133450 \;- 0.351139515\,\mri
\nl
\mrN &=& 400 \qquad  40.8133454 \;- 0.351141087\,\mri
\nl
\mrN &=& 500 \qquad  40.8133456 \;- 0.351141268\,\mri
\eqa
\sline
With $\mrz = - 0.51$ we obtain
\bq
{\overline{\mrH}}_{\mrV} = 45.442731(5) \;- 2.517246(3) \,\mri
\eq                           
with a Sinc result
\sline
\bqa
\mrN &=& 100 \qquad  45.4431548 \;- 2.51734571\,\mri  
\nl
\mrN &=& 200 \qquad  45.4426150 \;- 2.51710345\,\mri
\nl
\mrN &=& 300 \qquad  45.4428094 \;- 2.51730573\,\mri
\nl
\mrN &=& 400 \qquad  45.4427649 \;- 2.51734779\,\mri
\nl
\mrN &=& 500 \qquad  45.4427144 \;- 2.51717746\,\mri
\eqa
\sline
The rate of convergence can be improved by using contiguity relations; for instance, if we replace
$\eG{0.41 + \mrs}$ with $\eG{- 0.59 + \mrs}$ we obtain
\sline
\bqa
\mrN &=& 100 \qquad  - 69.4080201 \;+ 17.4551589\,\mri
\nl
\mrN &=& 200 \qquad  - 69.4080209 \;+ 17.4551563\,\mri
\nl
\mrN &=& 300 \qquad  - 69.4080202 \;+ 17.4551570\,\mri
\nl
\mrN &=& 400 \qquad  - 69.4080202 \;+ 17.4551569\,\mri
\nl
\mrN &=& 500 \qquad  - 69.4080206 \;+ 17.4551567\,\mri
\eqa
\sline
\item[\ovalbox{2}]
\bq
\mrH = \Bigl[ \prod_{\mrj=1}^{2}\,\int_{\mrL_\mrj}\,\frac{\mrd \mrs_\mrj}{\tip} \Bigr]\,
\frac{
\eG{ - \mrs_1}\,\eG{ - \mrs_2}\,\eG{1.41 + \mrs_1 + \mrs_2}\,\eG{1.52 + \mrs_1}\,\eG{1.63 + \mrs_2}
     }
     {
\eG{5.95 + \mrs_1 + \mrs_2}
     }\,\mrz_1^{\mrs_1}\,\mrz_2^{\mrs_2} \spp
\eq
The integration countours correspond to $\sigma_1 = - 0.251$ and $\sigma_2 = - 0.152$.
The corresponding parameters are:
\bqa
\alpha_1 &=& 2 \spc \quad \beta_1 = 0 \spc \quad \lambda_1 = - 4.91 \spc
\nl
\alpha_2 &=& 2 \spc \quad \beta_2 = 0 \spc \quad \lambda_2 = - 0.502 \spp
\eqa
With $\mrz_1 = 0.11 + 0.01\,\mri$ and $\mrz_2 = 0.12 + 0.01\,\mri$ we obtain
\sline
\bqa
\Re\,{\overline{\mrH}}_{\mrV} &=& 0.232722851 \,\pm\, (0.325\,\times\,10^{-9}) \spc 
\nl            
\Im\,{\overline{\mrH}}_{\mrV} &=& (- 0.160991856\,\times\,10^{-2})\,\pm\, (0.402\,\times\,10^{-10}) \spp
\eqa
\sline
The Sinc result is
\sline
\bqa
\mrN &=& 100 \qquad  0.232723113 \;- 0.160991939\,\times\,10^{-2}\,\mri
\nl
\mrN &=& 500 \qquad  0.232722850 \;- 0.160991854\,\times\,10^{-2}\,\mri
\nl
\mrN &=& 900 \qquad  0.232722850 \;- 0.160991854\,\times\,10^{-2}\,\mri
\eqa
\sline
With $\mrz_1 = - 0.11$ and $\mrz_2 = 0.12$ we obtain
\sline
\bq
\Re\,{\overline{\mrH}}_{\mrV} = 0.251543018 \,\pm\, 0.395\,\times\,10^{-7} \spc
\eq
\sline
while the result for the imaginary part is not converging. The Sinc results are
\sline
\bqa
\mrN\;  500 &\qquad&   0.251544852 \;- 0.304024324\,\times\,10^{-5}\,\mri
\nl
\mrN\;  3000 &\qquad&  0.251542719 \;+ 0.653893363\,\times\,10^{-7}\,\mri
\eqa
\sline
Although we have a reliable prediction for the real part, the imaginary part is out of control. We can use contiguity 
relations; after repeated applications we compute integrals where, for instance, 
\bq
\eG{\mra_{1,2} + \mrs_{1,2}} \to \eG{\mra_{1,2} - 5 + \mrs_{1,2}} \spc
\eq
giving the following results: 
\sline
\bq
\Re\,{\overline{\mrH}}_{\mrV} =  0.0164853556 \,\pm\, 0.350\,\times\,10^{-9} \spc \quad            
\Im\,{\overline{\mrH}}_{\mrV} =  0.0177740112 \,\pm\, 0.106\,\times\,10^{-8} \spc
\eq
\sline
where the Sinc results give
\sline
\bq
\mrN = 1000 \qquad  0.0164853552 - 0.0177740110\,\mri
\eq            
\sline
\item[\ovalbox{3}]
\bq
\mrH = \Bigl[ \prod_{\mrj=1}^{3}\,\int_{\mrL_\mrj}\,\frac{\mrd \mrs_\mrj}{\tip} \Bigr]\,
\eG{1.4 + \mathbf{s}}\,
\frac{
      \eG{ - \mrs_1}\,\eG{1.5 + \mrs_1}\,\eG{ - \mrs_2}\,\eG{1.6 + \mrs_2}\,\eG{ - \mrs_3}\,\eG{1.7 + \mrs_3}
     }
     {
      \eG{1.8 + \mrs_1}\,\eG{1.9 + \mrs_2}\,\eG{1.5 + \mrs_3}
     }\,
     \prod_{\mrj=1}^{3}\,\mrz_\mrj^{\mrs_\mrj} \spp
\eq
With $\mrz_1 = 0.11 + 0.01\,\mri$, $\mrz_2 = 0.92 + 0.01\,\mri$ and $\mrz_3 = 0.33 + 0.01\,,i$ we obtain
\sline
\bq
\Re\,{\overline{\mrH}}_{\mrV} =   65.3308646  \,\pm\, 0.136\,\times\,10^{-3} \spc \quad            
\Im\,{\overline{\mrH}}_{\mrV} = -  1.15503936 \,\pm\, 0.130\,\times\,10^{-4} \spc
\eq
\sline
where the Sinc results give
\sline
\bqa
\mrN = 30 &\qquad&   65.3838763 \;- 1.15557321\,\mri
\nl
\mrN = 50 &\qquad&   65.3372624 \;- 1.15512931\,\mri
\nl
\mrN= 100 &\qquad&   65.3309512 \;- 1.15503547\,\mri
\eqa     
\sline
\item[\ovalbox{4}]
\bq
\mrH = \Bigl[ \prod_{\mrj=1}^{3}\,\int_{\mrL_\mrj}\,\frac{\mrd \mrs_\mrj}{\tip} \Bigr]\,
\eG{1.4 + \mathbf{s}}\,
\frac{
      \eG{ - \mrs_1}\,\eG{1.5 + \mrs_1}\,\eG{ - \mrs_2}\,\eG{1.6 + \mrs_2}\,\eG{ - \mrs_3}\,\eG{1.7 + \mrs_3}
     }
     {
      \eG{6.8 + \mrs_1}\,\eG{1.9 + \mrs_2}\,\eG{1.5 + \mrs_3}
     }\,
     \prod_{\mrj=1}^{3}\,\mrz_\mrj^{\mrs_\mrj} \spp
\eq
With $\mrz_1 = - 2.11$, $\mrz_2 = 0.92 + 0.1\,\mri$ and $\mrz_3 = 0.33 + 0.1\,\mri$ we obtain
\sline
\bq
\Re\,{\overline{\mrH}}_{\mrV} =    0.195177105  \,\pm\, 0.530\,\times\,10^{-3} \spc \quad            
\Im\,{\overline{\mrH}}_{\mrV} =  - 0.0404843040 \,\pm\, 0.523\,\times\,10^{-3} \spc
\eq
\sline
where the Sinc results give
\sline
\bqa
\mrN = 30  &\qquad&    0.195458204 \;- 0.0404677606\,\mri
\eqa
\sline
\item[\ovalbox{5}]
\bq
\mrH = \Bigl[ \prod_{\mrj=1}^{4}\,\int_{\mrL_\mrj}\,\frac{\mrd \mrs_\mrj}{\tip} \Bigr]\,
\frac{
      \eG{1.1 + \mathbf{s}}
     }
     {
      \eG{5 + \mathbf{s}}
     }\,
\prod_{\mrj=1}^{4}\,\eG{ - \mrs_\mrj}\,\eG{\mra_\mrj + \mrs_\mrj}\,\mrz_\mrj^{\mrs_\mrj} \spc
\eq
with
\bq
\mra_1 = 0.11 \spc \quad
\mra_2 = 0.22 \spc \quad
\mra_3 = 0.33 \spc \quad
\mra_4 = 0.44 \spc \qquad 
\mrz_\mrj = \mra_\mrj + 0.01\,\mri \spp
\eq
\bq
{\overline{\mrH}}_{\mrV} =  - 0.254000345\,\times\,10^5 + 0.193449783\,\times\,10^4\,\mri \spc            
\eq
where the Sinc results give
\sline
\bqa
\mrN = 30  &\qquad&   - 0.257302543\,\times\,10^5 \;+ 0.195892166\,\times\,10^4\,\mri
\nl
\mrN = 50 &\qquad&    - 0.254483013\,\times\,10^5 \;+ 0.193845936\,\times\,10^4\,\mri
\nl
\mrN = 70  &\qquad&   - 0.254033569\,\times\,10^5 \;+ 0.193516252\,\times\,10^4\,\mri
\eqa
\sline
Changing $\mrz_1$ into $\mrz_1 = - 0.11$ gives
\sline
\bq
{\overline{\mrH}}_{\mrV} =   - 0.207139065\,\times\,10^5 \;+  0.243313533\,\times\,10^5\,\mri \spc            
\eq
\sline
where the Sinc results give
\sline
\bqa
\mrN = 30  &\qquad&    - 0.209012094\,\times\,10^5 \;+  0.245774532\,\times\,10^5
\nl
\mrN = 50  &\qquad&    - 0.207455758\,\times\,10^5 \;+ 0.243642375\,\times\,10^5
\eqa
\sline
\item[\ovalbox{6}] Next we condider
\bq
\mrH = \Bigl[ \prod_{\mrj=1}^{4}\,\int_{\mrL_\mrj}\,\frac{\mrd \mrs_\mrj}{\tip} \Bigr]\,
\frac{
      \eG{\frac{1}{2} + \mathbf{s}}
     }
     {
      \eG{\frac{11}{2} + \mathbf{s}}
     }\,
\prod_{\mrj=1}^{4}\,\eG{ - \mrs_\mrj}\,\eG{\mra_\mrj + \mrs_\mrj}\,\mrz_\mrj^{\mrs_\mrj} \spc
\eq
with $\mra_\mrj = 1$, $\sigma_\mrj = - 0.1$ and
\bq
\mrz_1 = - 2.11 \spc \quad
\mrz_2 = 0.22 + 0.1\,\mri \spc \quad
\mrz_3 = 0.33 + 0.1\,\mri \spc \quad
\mrz_4 = 0.44 + 0.1\,\mri \spc 
\eq
obtaining
\[
\begin{array}{lrrr}
&&& \\
\hline
&&& \\
\mbox{Lattice} \quad & \quad \mbox{Calls} \quad & \quad \Re \quad & \quad \Im \\
&&& \\
\hline
&&& \\
\mbox{Korobov} \quad & \quad  281437986 \quad & \quad  62.98(9) \quad & \quad - 8.91(5) \\
&&& \\
\mbox{Sinc}    \quad & \quad  11298540  \quad & \quad  62.9799396 \quad & \quad - 8.91142516 \\  
&&& \\
               \quad & \quad  86972936  \quad & \quad  62.9370547 \quad & \quad - 8.91706193 \\
&&& \\
               \quad & \quad 331085208  \quad & \quad  62.9329160 \quad & \quad - 8.91725259 \\
&&& \\
\hline
\end{array}
\]
\item[\ovalbox{7}] Finally we condider one example with $6$ MB integrals:
\bqa
\mrH &=& \Bigl[ \prod_{\mrj=1}^{6}\,\int_{\mrL_\mrj}\,\frac{\mrd \mrs_\mrj}{\tip} 
\mrz_\mrj^{\mrs_\mrj} \Bigr]\,\frac{\mrN}{\mrD} \spc
\nl
\mrN &=&
\Bigl[ \prod_{\mrj=1}^{2}\,\eG{\mrs_\mrj} \Bigr]\,
\Bigl[ \prod_{\mrj=4}^{6}\,\eG{ - \mrs_\mrj} \Bigr]\,
\eG{\mrs_5 - \mrs_3}\,\eG{\mrs_6 - \mrs_3}\,\eG{2 + \mrs_1 + \mrs_2}
\nl
{}&\times&
\eG{1 + \mrs_1 + \mrs_2 + \mrs_3}\,\eG{1 + \mrs_4 + \mrs_5 + \mrs_6}\,
\eG{1 - \mrs_1 - \mrs_2 + \mrs_3 + \mrs_4} \spc
\nl
\mrD &=&
\eG{ - \mrs_3}\,\eG{1 + \mrs_1 + \mrs_2}\,\eG{2 + \mrs_1 + \mrs_2 + \mrs_3}\,
\eG{2 + \mrs_4 + \mrs_5 + \mrs_6} \spp
\eqa
The integration contours are defined by $\sigma_\mrj = - 0.1\,\mrj$. We select
$\mrx_1 = 1.11, \mrx_2 = 1.22\;\dots$ $\;\mrx_6 = 1.66$, with $\mry_\mrj = 0.01 + 0.001\,\mrj$.
The QMC routine returns $(0.459 \pm 0.073)\,\times\,10^{-3}$ for the real part and
and $( - 0.186 \pm 0.017)\,\times\,10^{-4}$ for the imaginary part. The Sinc results are
\[
\begin{array}{lll}
&& \\
\hline
&& \\
\mrN  \quad & \quad 10^3\,\times\,\Re \mrH \quad & \quad 10^4\,\times\,\Im \mrH \\
&& \\
\hline
&& \\
20 \quad & \quad  0.454582945 \quad & \quad  - 0.181266753 \\
25 \quad & \quad  0.459768680 \quad & \quad  - 0.184452728 \\
30 \quad & \quad  0.461671474 \quad & \quad  - 0.184955317 \\
35 \quad & \quad  0.462484578 \quad & \quad  - 0.184829107 \\ 
&& \\
\hline
\end{array}
\]
We observe the large errors in the Korobov lattice; the results from the Sinc lattice follow only
after a fine tuning of the corresponding parameters.
If we consider the difference between two consecutive approximants for $10^3\,\times\,\Re \mrH$ we obtain
$0.0052$, $0.0019$ and $0.008$ which can be used to impose variable precision control~\cite{Tmem}.
\eei
\paragraph{Behavior on the unit disk} \hspace{0pt} \\
This behavior is described in details in \Bref{Passarino:2024ugq}; here we consider the following example:
\bq
\mrH = \Bigl[ \prod_{\mrj=1}^{3}\,\int_{\mrL_\mrj}\,\frac{\mrd \mrs_\mrj}{\tip}\,\mrz_\mrj^{\mrs_\mrj}\,
\frac{\mrN}{\mrD} \spc
\eq
\bqa
\mrN &=&
\eG{1 + \mrs_1}\,\eG{ - \mrs_2}\,\eG{ - \mrs_3}\,\eG{\ep + \mrs_2}
\nl
{}&\times&
\eG{ - 1 + \ep - \mrs_1}\,\eG{2 - \ep + \mrs_1}\,\eG{ - \mrs_1 + \mrs_2 + \mrs-3}\,\eG{1 + \ep + \mrs_2 + \mrs_3} \spc
\nl
\mrD &=& \eG{2 + \mrs_1}\,\eG{1 + \ep + \mrs_2}\,\eG{3 + 2\,\ep - \mrs_1 + \mrs_2 + \mrs-3} \spp
\eqa
Introducing $\mrs_\mrj= \sigma_\mrj + \mri\,\mrt_\mrj$ the integration contours are defined by
\bq
- 1 + \ep < \sigma_1 < 0 \spc \quad
\max(\sigma_1\,,\,- \ep) < \sigma_2 < 0 \spc \quad
\max(\sigma_1 - \sigma_2\,,\,- 1 - \ep - \sigma_2) < \sigma_3 < 0 \spc
\eq
requiring $\ep < 1$. The $\mrs_1$ integral is the Meijer $\mrG^{2,2}_{3,3}$ function, with parameters~\cite{HTF}
\bqa
\mra_1 &=& 0 \spc \qquad 
\mra_2 = - 1 + \ep \spc \qquad
\mra_3 = 3 + 2\,\ep + \mrs_2 + \mrs_3 \spc
\nl
\mrb_1 &=& - 1 + \ep \spc \qquad
\mrb_2 = \mrs_2 + \mrs_3 \spc \qquad
\mrb_3 = - 1 \spp
\eqa
The Meijer $\mrG^{2,2}_{3,3}$ has a regular singularity at $\mrz_1 = - 1$. We give results for $\ep = 1/4$ and
\bq
\mrz_2 = 0.15 + 0.01\,\mri \spc \qquad \mrz_3 = 1.33 + 0.01\,\mri \spc
\eq
\bq
\sigma_1 = - \frac{3}{8} \spc \qquad
\sigma_2 = - \frac{1}{8} \spc \qquad
\sigma_3 = - \frac{1}{6} \spp
\eq
In the following a negative value for $\mrz_1$ must be understood as $\mrz_1 + \mri\,\delta$ with
$\delta \to 0_{+}$.
\[
\begin{array}{lrr}
&& \\
\hline
&& \\
\mrz_1\quad  & \quad \Re \quad & \quad \Im \\
&& \\
\hline
&& \\
- 1.100  \quad & \quad  30.468 \quad & \quad  264.424 \\
- 1.050  \quad & \quad  27.861 \quad & \quad  272.974 \\
- 1.010  \quad & \quad  25.436 \quad & \quad  280.269 \\
- 1.005  \quad & \quad  25.109 \quad & \quad  281.211 \\
- 1.000  \quad & \quad  24.776 \quad & \quad  282.161 \\
- 0.995  \quad & \quad  24.437 \quad & \quad  283.117 \\
- 0.990  \quad & \quad  24.092 \quad & \quad  284.080 \\
- 0.950  \quad & \quad  21.098 \quad & \quad  292.045 \\
- 0.900  \quad & \quad  16.690 \quad & \quad  302.672 \\
&& \\
\hline
\end{array}
\] 
For completeness we also give the results around $\mrz_1 = 0$
\[
\begin{array}{lrr}
&& \\
\hline
&& \\
\mrz_1 \quad & \quad \Re \quad & \quad \Im \\
&& \\
\hline
&& \\
- 1.0 \quad & \quad     24.776 \quad & \quad  282.161 \\
- 0.5 \quad & \quad  -  69.645 \quad & \quad  416.998 \\
- 0.1 \quad & \quad  - 471.429 \quad & \quad   510.325 \\
+ 0.1 \quad & \quad  - 452.051 \quad & \quad     0.342 \\  
+ 0.5 \quad & \quad  - 252.980 \quad & \quad     0.266 \\
+ 1.0 \quad & \quad  - 183.740 \quad & \quad     0.213 \\
&& \\
\hline
\end{array}
\] 
and also results for $\mid \mrz_1 \mid \to \infty$
\[
\begin{array}{lrr}
&& \\
\hline
&& \\
\mrz_1 \quad & \quad \Re \quad & \quad \Im \\
&& \\
\hline
&& \\
- 10^3 \quad & \quad     1.427 \quad & \quad    1.680 \\
- 10^2 \quad & \quad     6.894 \quad & \quad    9.446 \\
- 10 \quad & \quad      27.251 \quad & \quad    52.906 \\
-  1 \quad & \quad      24.776 \quad & \quad   282.161 \\
+  1 \quad & \quad   - 183.740 \quad & \quad     0.213 \\
+ 10 \quad & \quad   -  49.823 \quad & \quad     0.070 \\
+ 10^2 \quad & \quad -  10.808 \quad & \quad     0.016 \\
+ 10^3 \quad & \quad -   2.122 \quad & \quad     0.003 \\
&& \\
\hline
\end{array}
\] 
\paragraph{$\ep$ expansion} \hspace{0pt} \\
Following the results of \sect{ModFF} we consider
\bq
\mrI = \frac{\eG{\frac{5}{2}}}{\eG{\frac{1}{2} + \ep}\,\eG{\frac{1}{3} + \ep}}\,
\mrF^{(2)}_{\mrD}\lpar 1\,;\,\frac{1}{2} + \ep\,,\,\frac{1}{3} + \ep\,;\,\mrc\,;\, - \mrz_1\,,\, - \mrz_2 \rpar \spc
\label{eexp}
\eq
with $\mrc = 5/2$ and $\mrz_1 = 1/4 + 0.01\,\mri$, $\mrz_2 = 3/2 + 0.01\,\mri$ we obtain
\bq
\mrI = \sum_{\mrn=0}^{3}\,\mrh_\mrn\,\ep^{\mrn} + \ord{\ep^4} \spc \qquad
\mrh_\mrn = \mrx_\mrn + \mri\,\mry_\mrn \spp
\eq
The coefficients are
\[
\begin{array}{rrrr}
&&& \\
\hline
&&& \\
\mrx_0 =     2.962142 \quad&\quad 
\mrx_1 =  - 16.573431 \quad&\quad
\mrx_2 =    68.762056 \quad&\quad 
\mrx_3 = - 247.309805 \\
&&& \\
\hline
&&& \\
\mry_0 = - 0.006941151 \quad&\quad
\mry_1 =   0.024133104 \quad&\quad
\mry_2 = - 0.076026624 \quad&\quad
\mry_3 =   0.222240110 \\
&&& \\
\hline
\end{array}
\]
As an example we have computed $\mrI$ with no expansion and $\ep = 0.01, 0.1$:
\[
\begin{array}{lrr}
&& \\
\hline
&& \\
\ep \quad & \quad \Re \quad & \Im \\
&& \\
\hline
&& \\
0.01 \quad & \quad 2.80304490 \quad & \quad \; - 0.00670719357 \\
     \quad & \quad 2.80303617 \quad & \quad \; - 0.00670720040 \\
0.1  \quad & \quad 1.80763442 \quad & \quad \; - 0.00511579304 \\
     \quad & \quad 1.74510923 \quad & \quad \; - 0.00506586674 \\
&& \\
\hline
\end{array}
\]      
where the first entry is the ``exact'' result and the second is the result truncated at $\ord{\ep^3}$.
Higher orders in the $\ep$ expansion can also be ``predicted'' by using well{-}known 
techinques~\cite{David:2013gaa}.
For instance, using the Weniger $\delta{}${-}transform~\cite{Weniger:1997zz} (see also the
Levin $\tau{}${-}transform~\cite{Lev}) the predicted values are
\bqa
{\overline{\mrx}}_4 &=& \frac{1}{3}\,\frac{\mrx_3}{\mrx_1\,\mrx_2}\,\lpar 4\,\mrx_1\,\mrx_3 - \mrx_2^2 \rpar
\nl
{\overline{\mrx}}_5 &=& \frac{1}{10}\,\frac{{\overline{\mrx}}_4}{\mrx_1\,\mrx_2\,\mrx_3}\,
\lpar \mrx_2^2\,\mrx_3 - 9\,\mrx_1\,\mrx_3^2 + 18\,\mrx_1\,\mrx_2\,{\overline{\mrx}}_4 \rpar \spp
\eqa
If we select $\mrx= 11/2$ and $\mrz_1 = - 1.25$, $\mrz_2 = 1.5 + 0.1\,\mri$ we obtain
\[
\begin{array}{rrrr}
&&& \\
\hline
&&& \\
\mrx_0 =    0.09868139     \quad&\quad 
\mrx_1 =  - 0.49260961     \quad&\quad
\mrx_2 =    1.97345217      \quad&\quad 
\mrx_3 =  - 6.94303796      \\
&&& \\
\hline
&&& \\
\mry_0 =  - 0.0008362581   \quad&\quad
\mry_1 =    0.0012450585   \quad&\quad
\mry_2 =  - 0.0042891402   \quad&\quad
\mry_3 =    0.0100186830    \\
&&& \\
\hline
\end{array}
\]
\paragraph{Truncation error in the approximation} \hspace{0pt} \\
For one{-}dimensional integrals the error in the truncated Sinc expansion can
be determined~\cite{Sinc,eSinc}. Much less is known in the case of multidimensional integrals; 
it seems natural to adopt variable precision control which 
is imposed through the computation of a sequence 
of approximants until the difference between two consecutive 
ones is less than some desired tolerance. 
Round{-}off and cancellation errors can be cured by a careful coding of the analytic results.

One significant test is given by the computation of
\bq
\mrH = \Bigl[ \prod_{\mrj=1}^{4}\,\int_{\mrL_mrj}\,\frac{\mrd \mrs_\mrj}{\tip}\,\mrz_{\mrj}^{\mrd_{\mrj}} \Bigr]\,
\frac{\eG{\mra + \mathbf{s}}}{\eG{\mrc + \mathbf{s}}}\,
\Bigl[ \prod_{\mrj=1}^{4}\,\eG{ - \mrs_\mrj}\,\eG{\mrb_{\mrj} + \mrs_{\mrj}} \Bigr] \spc
\eq
where we have selected
\bq
\mra = 0.6 \spc \quad
\mrc = 5 \spc \quad
\mrb_1 = 1.15 \spc \quad
\mrb_2 = 1.26 \spc \quad
\mrb_3 = 1.37 \spc \quad
\mrb_4 = 1 \spc
\eq
\bq
\mrx_1 = 1.11 \spc \quad
\mrx_2 = 0.22 \spc \quad
\mrx_3 = 1.33 \spc \quad
\mrx_4 = 2.04 \spc
\eq
with $\mry_\mrj = 0.01$ The contours $\mrL_\mrj$ are such that among all the poles of the integrand only the
poles of $\eG{ - \mrs_\mrj}$ lie to the right of $\mrL_\mrj$. In this case $\mrH$ is a Lauricella function,
$\mrF^{(4)}_{\mrD}$. Given $\mrs_\mrj = \sigma_\mrj + \mri\,\mrt_\mrj$ we select $\sigma_\mrj = - 0.1$.
As long as the conditions on $\mrL_\mrj$ are satisfied, the result for $\mrH$ does not depend on $\sigma_\mrj$.
From the point of view of numerical computation all sort of errors will be relevant. In the following
table we study $\mrH$ as a function of $\sigma_1$. The effect of the truncation error in the Sinc expansion
is easily seen by comparing the results, as a function of $\sigma_1$, for $\mrN = 30$ and $\mrN = 100$.
The differences for $\mrN = 100$ and different values of $\sigma_1$ are also due to round{-}off errors and
cancellations. The last two lines in the table reflect the fact that with $\sigma_1 > 0$ we are crossing
the pole at $\mrs_1 = 0$, \ie $\mrH$ is not anymore a Lauricella function. However, as long as we do
not cross another pole, there is a good agreement among different values of $\sigma_1$.
\[
\begin{array}{ccc}
&& \\
\hline
&& \\
\sigma_1\,/\,\mrN     \quad & \quad 30 \quad & \quad 100 \\    
&& \\
&& \\
\hline     
&&\\
- 0.1 \psp   0.471668235\tens - 0.133915676\,\mri \psp      0.471213764\tens - 0.132555537\,\mri  \\
&& \\    
- 0.2 \psp   0.471649625\tens - 0.133758349\,\mri \psp      0.471213720\tens - 0.132554966\,\mri  \\
&& \\
- 0.25 \psp  0.471745510\tens - 0.133537460\,\mri \psp      0.471213719\tens - 0.132553759\,\mri  \\
&& \\
+ 0.1 \psp  - 0.424528323\teno - 0.969441512\tenmt\,\mri \psp     - 0.421367595\teno - 0.990397834\tenmt\,\mri \\
&& \\
+ 0.2 \psp  - 0.421454870\teno - 0.975353506\tenmt\,\mri \psp     - 0.421360807\teno - 0.990414222\tenmt\,\mri \\
&& \\
\hline
\end{array}
\]
\section{Conclusions \label {conc}}
In this work we have considered multiple Mellin{-}Barnes integrals, akas Fox functions (akas Feynman integrals),
\ie we have studied Feynman integrals from the perspective of Fox functions, a generalization of
hypergeometric functions.
It is important to realize that singularities of these functions coincide with the Landau 
singularities~\cite{TR,Book}.

After determining the convergence of the integrals we are faced with the problem of their numerical
computation. Our strategy requires a numerical integration instead of closing the contour(s) and summing over
residues, bypassing the (multivariate) residue computation step~\cite{MBmr} that is typical of the MB{-}approach. 
The resulting integrals can be approximated, with high accuracy, by using Sinc numerical methods~\cite{Sinc}.

The convergence of MB integrals~\cite{Paris_Kaminski_2001} is controlled by the parameters 
$\alpha, \beta$ and $\lambda$ defined in
\eqn{cpar} and by the values of $\phi_\mrj = \marg (\mrz_\mrj)$. When the behavior of the integrand is
exponential the rate of convergence does not represent a problem. However, for one or more 
$\mid \phi_\mrj \mid = \alpha \pi/2$ there is a cancellation and the behavior of the integrand is power like,
requiring $\lambda < - 1$ for convergence. Even if $\lambda < - 1$ the rate of convergence can be rather slow;
this is a well{-}known problem (connected with Feynman integrals in the physical region) and several 
solutions have been introduced in the literature, mostly limited to one{-}dimensional integrals. Our
solution is based on the use of contiguity relations which allow to compute a slowly convergent MB integral
in terms of a combination of MB integrals where $\lambda \to \lambda - 1$.

\clearpage
\bibliographystyle{elsarticle-num}
\bibliography{NCFF}

\begin{thebibliography}{100}
\expandafter\ifx\csname url\endcsname\relax
  \def\url#1{\texttt{#1}}\fi
\expandafter\ifx\csname urlprefix\endcsname\relax\def\urlprefix{URL }\fi
\expandafter\ifx\csname href\endcsname\relax
  \def\href#1#2{#2} \def\path#1{#1}\fi

\bibitem{oFox}
C.~Fox, The {G} and {H} functions as symmetrical {F}ourier kernels, Trans.
  Amer. Math. Soc. 98 (1961) 395–429 (1961).

\bibitem{compH}
A.~Mathai, R.~Saxena, H.~Haubold, The {H}{-}function: Theory and applications,
  Springer New York, NY DOI: https://doi.org/10.1007/978-1-4419-0916-9 (2010).

\bibitem{HS}
N.~Hai, H.~Srivastava, The convergence problem of certain multiple
  {M}ellin-{B}arnes contour integrals representing {H}-functions in several
  variables, Computers Math. Applic. Vol. 29, No. 6, pp. 17-25, 1995 (1995).

\bibitem{Hus}
A.~Inayat-Hussain, New properties of hypergeometric series derivable from
  {F}eynman integrals {II}. a generalisation of the {H} function, J. Phys. A:
  Math. Gen. 20 4119 (1987).

\bibitem{BSid}
R.~Buschman, H.~Srivastava, The {H} function associated with a certain class of
  {F}eynman integrals, J. Phys. A: Math. Gen. 23 (1990) 4707-4710 (1990).

\bibitem{More}
H.~Srivastava, S.~Lin, P.~Wang, Some fractional-calculus results for the
  {H}-function associated with a class of {F}eynman integrals, Russian J. Math.
  Phys. 2006, 13, 94–100 (2006).

\bibitem{Passarino:2024ugq}
G.~Passarino, {Feynman integrals and Fox functions}\href
  {http://arxiv.org/abs/2405.18755} {\path{arXiv:2405.18755}}.

\bibitem{Boos:1990rg}
E.~E. Boos, A.~I. Davydychev, {A Method of evaluating massive Feynman
  integrals}, Theor. Math. Phys. 89 (1991) 1052--1063.
\newblock \href {http://dx.doi.org/10.1007/BF01016805}
  {\path{doi:10.1007/BF01016805}}.

\bibitem{Friot:2011ic}
S.~Friot, D.~Greynat, {On convergent series representations of Mellin-Barnes
  integrals}, J. Math. Phys. 53 (2012) 023508.
\newblock \href {http://arxiv.org/abs/1107.0328} {\path{arXiv:1107.0328}},
  \href {http://dx.doi.org/10.1063/1.3679686} {\path{doi:10.1063/1.3679686}}.

\bibitem{ZT}
O.~Zhdanov, A.~Tsikh, Investigation of multiple {M}ellin–{B}arnes integrals
  by means of multidimensional residues, Sibirsk. Mat. Zh., 39:2 (1998),
  281–298; Siberian Math. J., 39:2 (1998), 245–260 (1998).

\bibitem{banik2024geometrical}
S.~Banik, S.~Friot, Geometrical methods for the analytic evaluation of multiple
  {M}ellin-{B}arnes integrals (2024).
\newblock \href {http://arxiv.org/abs/2402.04174} {\path{arXiv:2402.04174}}.

\bibitem{Freitas:2010nx}
A.~Freitas, Y.-C. Huang, {On the Numerical Evaluation of Loop Integrals With
  Mellin-Barnes Representations}, JHEP 04 (2010) 074.
\newblock \href {http://arxiv.org/abs/1001.3243} {\path{arXiv:1001.3243}},
  \href {http://dx.doi.org/10.1007/JHEP04(2010)074}
  {\path{doi:10.1007/JHEP04(2010)074}}.

\bibitem{Kalmykov:2012rr}
M.~Y. Kalmykov, B.~A. Kniehl, {Mellin-Barnes representations of Feynman
  diagrams, linear systems of differential equations, and polynomial
  solutions}, Phys. Lett. B 714 (2012) 103--109.
\newblock \href {http://arxiv.org/abs/1205.1697} {\path{arXiv:1205.1697}},
  \href {http://dx.doi.org/10.1016/j.physletb.2012.06.045}
  {\path{doi:10.1016/j.physletb.2012.06.045}}.

\bibitem{Kalmykov:2016lxx}
M.~Y. Kalmykov, B.~A. Kniehl, {Counting the number of master integrals for
  sunrise diagrams via the Mellin-Barnes representation}, JHEP 07 (2017) 031.
\newblock \href {http://arxiv.org/abs/1612.06637} {\path{arXiv:1612.06637}},
  \href {http://dx.doi.org/10.1007/JHEP07(2017)031}
  {\path{doi:10.1007/JHEP07(2017)031}}.

\bibitem{HTF}
A.~Erd\`elyi, W.~Magnus, F.~Oberhettinger, F.~G. Tricomi, Higher Transcendental
  Functions, Vol.~1, McGraw-Hill, 1953, compiled by the staff of the Bateman
  Manuscript Project.

\bibitem{Sbook}
F.~Stenger, Numerical methods based on {S}inc and analytic functions, Springer
  New York 1993 (1993).
\newblock \href {http://dx.doi.org/10.1007/978-1-4612-2706-9}
  {\path{doi:10.1007/978-1-4612-2706-9}}.

\bibitem{Sinc}
F.~Stenger, Summary of {S}inc numerical methods, Journal of Computational and
  Applied Mathematics 121 (2000) 379 - 420 (2000).

\bibitem{SUGIHARA2004673}
M.~Sugihara, T.~Matsuo,
  \href{https://www.sciencedirect.com/science/article/pii/S0377042703008380}{Recent
  developments of the {S}inc numerical methods}, Journal of Computational and
  Applied Mathematics 164-165 (2004) 673--689, proceedings of the 10th
  International Congress on Computational and Applied Mathematics.
\newblock \href {http://dx.doi.org/https://doi.org/10.1016/j.cam.2003.09.016}
  {\path{doi:https://doi.org/10.1016/j.cam.2003.09.016}}.
\newline\urlprefix\url{https://www.sciencedirect.com/science/article/pii/S0377042703008380}

\bibitem{doi:10.1137/1.9781611971637}
J.~Lund, K.~L. Bowers,
  \href{https://epubs.siam.org/doi/abs/10.1137/1.9781611971637}{Sinc Methods
  for Quadrature and Differential Equations}, Society for Industrial and
  Applied Mathematics, 1992.
\newblock \href
  {http://arxiv.org/abs/https://epubs.siam.org/doi/pdf/10.1137/1.9781611971637}
  {\path{arXiv:https://epubs.siam.org/doi/pdf/10.1137/1.9781611971637}}, \href
  {http://dx.doi.org/10.1137/1.9781611971637}
  {\path{doi:10.1137/1.9781611971637}}.
\newline\urlprefix\url{https://epubs.siam.org/doi/abs/10.1137/1.9781611971637}

\bibitem{PhysRevD.61.125001}
R.~Easther, G.~Guralnik, S.~Hahn,
  \href{https://link.aps.org/doi/10.1103/PhysRevD.61.125001}{Fast evaluation of
  {F}eynman diagrams}, Phys. Rev. D 61 (2000) 125001.
\newblock \href {http://dx.doi.org/10.1103/PhysRevD.61.125001}
  {\path{doi:10.1103/PhysRevD.61.125001}}.
\newline\urlprefix\url{https://link.aps.org/doi/10.1103/PhysRevD.61.125001}

\bibitem{Petrov_2001}
D.~Petrov, R.~Easther, G.~Guralnik, S.~Hahn, W.-M. Wang,
  \href{http://dx.doi.org/10.1103/PhysRevD.63.105001}{Fermions, gauge theories,
  and the {S}inc function representation for {F}eynman diagrams}, Physical
  Review D 63~(10).
\newblock \href {http://dx.doi.org/10.1103/physrevd.63.105001}
  {\path{doi:10.1103/physrevd.63.105001}}.
\newline\urlprefix\url{http://dx.doi.org/10.1103/PhysRevD.63.105001}

\bibitem{osti_40205105}
R.~Easther, G.~Guralnik, S.~Hahn,
  \href{https://www.osti.gov/biblio/40205105}{Sinc function representation and
  three-loop master diagrams}, Physical Review D 63~(8).
\newblock \href {http://dx.doi.org/10.1103/PhysRevD.63.085017}
  {\path{doi:10.1103/PhysRevD.63.085017}}.
\newline\urlprefix\url{https://www.osti.gov/biblio/40205105}

\bibitem{SSinc}
G.~Baumann, N.~Sudland, \href{https://doi.org/10.3390/fractalfract6080449}{Sinc
  numeric methods for {F}ox-{H}, {A}leph, and {S}axena functions}, Fractal
  Fract. 2022, 6(8), 449 (2022).
\newline\urlprefix\url{https://doi.org/10.3390/fractalfract6080449}

\bibitem{HTgam}
A.~Erd\`elyi, F.~Tricomi, The asymptotic expansion of a ratio of {G}amma
  functions, {P}acific {J}ournal of {M}athematics, Vol.1 N. 1 (1951).

\bibitem{BRaa}
B.~Braaksma, Asymptotic expansions and analytic continuations for a class of
  {B}arnes-integrals, Compositio Mathematica, tome 15 (1962-1964), p. 239-341
  (1962).

\bibitem{Ext}
H.~Exton, Multiple hypergeometric functions and applications, John Wiley and
  Sons,Inc., New York–London–Sydney 1976, 312 pp (1976).

\bibitem{FDMB}
S.~Bezrodnykh, The {L}auricella hypergeometric function
  $\mathrm{F}^{\mrN}_{\sPD}$, Russian Mathematical Surveys, 2018, Volume 73,
  Issue 6, 941–1031 DOI: 10.1070/RM9841 (2083).

\bibitem{HY}
N.~Hai, Y.~S., The double {M}ellin-{B}arnes type integrals and their
  applications to convolution theory., Series on Soviet and East European
  Mathematics, 6. World Scientific Publishing, Singapore, New Jersey (1992).

\bibitem{SKL}
H.~M. Srivastava, P.~W. Karlsson, Multiple {G}aussian hypergeometric series,
  Mathematics and its applications. Chichester, UK: Halsted Press, Ellis
  Horwood Ltd. ISBN 0-470-20100-2. MR 0834385 (1985).

\bibitem{Asur}
M.~Schlosser, chapter in the book computer algebra in quantum field theory,
  Springer-Verlag Wien 2013 DOI: https://doi.org/10.1007/978-3-7091-1616-6
  (2013).

\bibitem{TBL}
V.~Tuan, R.~Buschman, Integral representations of generalized {L}auricella
  hypergeometric functions, International Journal of Mathematics and
  Mathematical Sciences 15(4) (1992).

\bibitem{Passarino:2006gv}
G.~Passarino, S.~Uccirati, {Two-loop vertices in quantum field theory: Infrared
  and collinear divergent configurations}, Nucl. Phys. B747 (2006) 113--189.
\newblock \href {http://arxiv.org/abs/hep-ph/0603121}
  {\path{arXiv:hep-ph/0603121}}, \href
  {http://dx.doi.org/10.1016/j.nuclphysb.2006.04.014}
  {\path{doi:10.1016/j.nuclphysb.2006.04.014}}.

\bibitem{Masa}
K.~Kono,
  \href{https://fractional-calculus.com/series_expansion_gamma_reciprocal.pdf}{Series
  expansion of {G}amma function and the reciprocal} (2017).
\newline\urlprefix\url{https://fractional-calculus.com/series_expansion_gamma_reciprocal.pdf}

\bibitem{Bell}
E.~Bell, Exponential polynomials, Ann. Math. 35, 258-277, 1934 (1934).

\bibitem{Weinzierl:2004bn}
S.~Weinzierl, {Expansion around half integer values, binomial sums and inverse
  binomial sums}, J. Math. Phys. 45 (2004) 2656--2673.
\newblock \href {http://arxiv.org/abs/hep-ph/0402131}
  {\path{arXiv:hep-ph/0402131}}, \href {http://dx.doi.org/10.1063/1.1758319}
  {\path{doi:10.1063/1.1758319}}.

\bibitem{Puhlfurst:2015vqx}
G.~Puhlf\"urst, S.~Stieberger, {A Feynman Integral and its Recurrences and
  Associators}, Nucl. Phys. B 906 (2016) 168--193.
\newblock \href {http://arxiv.org/abs/1511.03630} {\path{arXiv:1511.03630}},
  \href {http://dx.doi.org/10.1016/j.nuclphysb.2016.03.008}
  {\path{doi:10.1016/j.nuclphysb.2016.03.008}}.

\bibitem{psis}
J.~Gonz\'alez-Santander, S.~Lasheras, Finite and infinite hypergeometric sums
  involving the digamma function, Mathematics 2022, 10, 2990 (2022).

\bibitem{Passarino:2001wy}
G.~Passarino, {A Practical approach for exponentiation of QED corrections in
  arbitrary processes}, Nucl. Phys. B 619 (2001) 313--358.
\newblock \href {http://arxiv.org/abs/hep-ph/0108255}
  {\path{arXiv:hep-ph/0108255}}, \href
  {http://dx.doi.org/10.1016/S0550-3213(01)00542-9}
  {\path{doi:10.1016/S0550-3213(01)00542-9}}.

\bibitem{Kolbig:1983qt}
K.~S. Kolbig, {Nielsen's generalized polylogarithms}, SIAM J. Math. Anal. 17
  (1986) 1232--1258.
\newblock \href {http://dx.doi.org/10.1137/0517086}
  {\path{doi:10.1137/0517086}}.

\bibitem{Devoto:1983tc}
A.~Devoto, D.~W. Duke, {Table of Integrals and Formulae for Feynman Diagram
  Calculations}, Riv. Nuovo Cim. 7N6 (1984) 1--39.
\newblock \href {http://dx.doi.org/10.1007/BF02724330}
  {\path{doi:10.1007/BF02724330}}.

\bibitem{Sthe}
Y.~Sokhotskii, On definite integrals and functions used in series expansions,
  St. Petersburg (1873) (In Russian) (Dissertation) (1873).

\bibitem{SPF}
J.~Plemelj, Problems in the sense of {R}iemann and {K}lein, Differential
  equations, Functional analysis, New York, Interscience Publishers (1964).

\bibitem{FPI}
J.~Hadamard, Lectures on {C}auchy's problem in linear partial differential
  equations, Dover Phoenix editions, Dover Publications, New York, p. 316, ISBN
  978-0-486-49549-1 (1923).

\bibitem{GLerch}
V.~Kumar, On the generalized {H}urwitz–{L}erch zeta function and generalized
  {L}ambert transform, Journal of Classical Analysis, Vol. 17, N. 1(2021),55-67
  (2021).

\bibitem{FFPR}
G.~Passarino, Feynman fox integrals in the physical region, To be submitted.

\bibitem{Hreg}
L.~Blanchet, G.~Faye, Hadamard regularization, Journal of Mathematical Physics,
  41 (11): 7675–7714, arXiv:gr-qc/0004008 (2000).

\bibitem{Bardin:1999ak}
D.~Y. Bardin, G.~Passarino, {{The standard model in the making: Precision study
  of the electroweak interactions}}, {Oxford University Press, International
  series of monographs on physics. 104} (1999).

\bibitem{Kershaw:1973km}
D.~S. Kershaw, {Feynman amplitudes as power series}, Phys. Rev. D 8 (1973)
  2708--2713.
\newblock \href {http://dx.doi.org/10.1103/PhysRevD.8.2708}
  {\path{doi:10.1103/PhysRevD.8.2708}}.

\bibitem{PhysRevD.9.370}
A.~C.~T. Wu, \href{https://link.aps.org/doi/10.1103/PhysRevD.9.370}{Generalized
  {E}uler-{P}ochhammer integral representation for single-loop {F}eynman
  amplitudes}, Phys. Rev. D 9 (1974) 370--373.
\newblock \href {http://dx.doi.org/10.1103/PhysRevD.9.370}
  {\path{doi:10.1103/PhysRevD.9.370}}.
\newline\urlprefix\url{https://link.aps.org/doi/10.1103/PhysRevD.9.370}

\bibitem{PhysRevD.11.452}
K.~Mano, \href{https://link.aps.org/doi/10.1103/PhysRevD.11.452}{Comment on
  generalized {E}uler-{P}ochhammer integral representation for single-loop
  {F}eynman amplitudes}, Phys. Rev. D 11 (1975) 452--454.
\newblock \href {http://dx.doi.org/10.1103/PhysRevD.11.452}
  {\path{doi:10.1103/PhysRevD.11.452}}.
\newline\urlprefix\url{https://link.aps.org/doi/10.1103/PhysRevD.11.452}

\bibitem{Ghs}
I.~Gel'fand, R.~V. Graev M.I.~and, General hypergeometric systems of equations
  and series of hypergeometric type, Uspekhi Mat. Nauk 47:4 (1992),3-82,
  Russian Math. Surveys 47:4 (1992), 1-88 (1992).

\bibitem{Moriello:2019yhu}
F.~Moriello, {Generalised power series expansions for the elliptic planar
  families of Higgs + jet production at two loops}, JHEP 01 (2020) 150.
\newblock \href {http://arxiv.org/abs/1907.13234} {\path{arXiv:1907.13234}},
  \href {http://dx.doi.org/10.1007/JHEP01(2020)150}
  {\path{doi:10.1007/JHEP01(2020)150}}.

\bibitem{Armadillo:2022ugh}
T.~Armadillo, R.~Bonciani, S.~Devoto, N.~Rana, A.~Vicini, {Evaluation of
  Feynman integrals with arbitrary complex masses via series expansions},
  Comput. Phys. Commun. 282 (2023) 108545.
\newblock \href {http://arxiv.org/abs/2205.03345} {\path{arXiv:2205.03345}},
  \href {http://dx.doi.org/10.1016/j.cpc.2022.108545}
  {\path{doi:10.1016/j.cpc.2022.108545}}.

\bibitem{Kershaw:1971rc}
D.~Kershaw, {Algebraic factorization of scattering amplitudes at physical
  Landau singularities}, Phys. Rev. D5 (1972) 1976--1982.
\newblock \href {http://dx.doi.org/10.1103/PhysRevD.5.1976}
  {\path{doi:10.1103/PhysRevD.5.1976}}.

\bibitem{Ferroglia:2002mz}
A.~Ferroglia, M.~Passera, G.~Passarino, S.~Uccirati, {All purpose numerical
  evaluation of one loop multileg Feynman diagrams}, Nucl. Phys. B650 (2003)
  162--228.
\newblock \href {http://arxiv.org/abs/hep-ph/0209219}
  {\path{arXiv:hep-ph/0209219}}, \href
  {http://dx.doi.org/10.1016/S0550-3213(02)01070-2}
  {\path{doi:10.1016/S0550-3213(02)01070-2}}.

\bibitem{Passarino:2018wix}
G.~Passarino, {Peaks and cusps: anomalous thresholds and LHC physics}\href
  {http://arxiv.org/abs/1807.00503} {\path{arXiv:1807.00503}}.

\bibitem{mSinc}
A.~Asharabi, F.~Al-Haddad, On multidimensional {S}inc-{G}auss sampling formulas
  for analytic functions, Electronic Transactions on Numerical Analysis. Volume
  55, pp. 242–262, 2022 (2022).

\bibitem{GSinc}
W.~Ye, A.~Entezari, IEEE Transactions on Image 1 June 2012 (2012).

\bibitem{eSinc}
T.~Okayama, T.~Matsuo, M.~Sugihara,
  \href{http://www.keisu.t.u-tokyo.ac.jp/research/techrep/index.html}{Error
  estimates with explicit constants for {S}inc approximation, {S}inc quadrature
  and {S}inc indefinite integration} (2009).
\newline\urlprefix\url{http://www.keisu.t.u-tokyo.ac.jp/research/techrep/index.html}

\bibitem{Lip}
M.~Searc/'oid, Lipschitz functions, Metric Spaces, Springer undergraduate
  mathematics series, Berlin, New York: Springer-Verlag, ISBN
  978-1-84628-369-7.

\bibitem{Sth}
M.~Sugihara, Near optimality of the {S}inc approximation, Mathematics of
  Computation, Volume 72, Number 242, Pages 767–786 (2002).

\bibitem{multiSinc}
W.~Hackbusch, B.~Khoromskij, Tensor-product approximation to operators and
  functions in high dimensions, Journal of Complexity 23 (2007) 697–714
  (2007).

\bibitem{Koro}
N.~Korobov, Doklady Akademii Nauk SSSR 124 1207–1210 (Russian) (1959).

\bibitem{Kort}
N.~Korobov, Doklady Akademii Nauk SSSR 132 1009–1012 (Russ.). Eng. trans.
  Soviet Math. Doklady, 1, 696-700 (1960).

\bibitem{PK}
P.~Keast, Optimal parameters for multidimensional integration, SIAM J Numer
  Anal, 10, pp. 831-838.

\bibitem{Rand}
R.~Cranley, T.~Patterson, Randomization of number theoretic methods for
  multiple integration, SIAM Journal on Numerical Analysis, Vol. 13, No. 6
  (Dec., 1976), pp. 904-914 (1976).

\bibitem{pillichshammer2020notekorobovlatticerules}
F.~Pillichshammer, \href{https://arxiv.org/abs/2010.03286}{A note on {K}orobov
  lattice rules for integration of analytic functions} (2020).
\newblock \href {http://arxiv.org/abs/2010.03286} {\path{arXiv:2010.03286}}.
\newline\urlprefix\url{https://arxiv.org/abs/2010.03286}

\bibitem{CRo}
H.~Niederreiter, Random number generation and quasi-{M}onte {C}arlo methods,
  volume 63 of CBMS-NSF Regional Conference Series in Applied Mathematics.
  Society for Industrial and Applied Mathematics (SIAM), Philadelphia, PA, 1992
  (1992).

\bibitem{CRoo}
H.~Niederreiter, On a number-theoretical integration method, Aequationes
  Mathematicae, 8(1972), pp. 304-11 (1972).

\bibitem{CRt}
I.~Sloan, S.~Joe, Lattice methods for multiple integration, Oxford Science
  Publications. The Clarendon Press, Oxford University Press, New York, 1994
  (1994).

\bibitem{KRd}
D.~T.~P. Nguyen, D.~Nuyens, Multivariate integration over $\mathbb{R}^{\mrs}$
  with exponential rate of convergence, Journal of Computational and Applied
  Mathematics, 315:327–342, 2017 (2017).

\bibitem{Per}
A.~Sidi, A new variable transformation for numerical integration, Numerical
  Integration, IV (Oberwolfach, 1992), volume 112 of Internat. Ser. Numer.
  Math., pages 359–373. Birkhäuser, Basel, 1993 (1993).

\bibitem{BP}
S.~Zaremba, La m\'ethode des ``bons treillis'' pour le calcul des int\'egrales
  multiples, Applications of Number Theory to Numerical Analysis, pages
  39–119. Academic Press, New York, 1972 (1972).

\bibitem{RPBo}
R.~P. Brent, Fast multiple-precision evaluation of elementary functions,
  Journal of the ACM 23 (1976), no. 2, 242–251 (1976).

\bibitem{RPBt}
R.~P. Brent, On the accuracy of asymptotic approximations to the log-{G}amma
  and {R}iemann-{S}iegel theta functions, Journal of the Australian
  Mathematical Society 107 (2018), no. 3, 319–337 (2018).

\bibitem{HAL}
F.~Johansson, Arbitrary-precision computation of the {G}amma function, Maple
  Transactions, 2023, 3 (1), 10.5206/mt.v3i1.14591. hal-03346642 (2023).

\bibitem{Hornc}
H.~M. Srivastava, P.~W. Karlsson, Multiple {G}aussian hypergeometric series,
  Ellis Horwood Series: Mathematics and its Applications, Chichester: Ellis
  Horwood Ltd (1985).

\bibitem{DEC}
J.~Borwein, A.~Lewis, Decomposition of multivariate functions, Can. J.
  Math.Vol. 44 (3), 1992 pp. 463-482 (1992).

\bibitem{ADEC}
F.~Kuo, I.~Sloan, G.~Wasilkowski, H.~Wozniakowski, On decompositions of
  multivariate functions, MATHEMATICS OF COMPUTATION Volume 79, Number 270,
  April 2010, Pages 953–966 (2010).

\bibitem{diaz2025simplewayreducenumber}
M.~Diaz, I.~Gonzalez, I.~Kondrashuk, E.~A. Notte-Cuello,
  \href{https://arxiv.org/abs/2412.13512}{A simple way to reduce the number of
  contours in the multi-fold mellin-barnes integrals} (2025).
\newblock \href {http://arxiv.org/abs/2412.13512} {\path{arXiv:2412.13512}}.
\newline\urlprefix\url{https://arxiv.org/abs/2412.13512}

\bibitem{rBeta}
E.~Stade, The reciprocal of the beta function, Annales de l’institut
  {F}ourier, tome 44, n. 1 (1994), p. 93-108 (1994).

\bibitem{weniger}
E.~J. Weniger, \href{https://arxiv.org/abs/math/0306302}{Nonlinear sequence
  transformations for the acceleration of convergence and the summation of
  divergent series} (2003).
\newblock \href {http://arxiv.org/abs/math/0306302}
  {\path{arXiv:math/0306302}}.
\newline\urlprefix\url{https://arxiv.org/abs/math/0306302}

\bibitem{Ames}
L.~Ames, Evaluation of slowly convergent series, Annals of Mathematics, Vol. 3,
  No. 1/4 (1901 - 1902), pp. 185-192 (1901).

\bibitem{Shanks}
Non-linear transformation of divergent and slowly convergent sequences, Journal
  of Mathematics and Physics, 34 (1–4): 1-42 (1955).

\bibitem{Spouge}
J.~Spouge, Computation of the gamma, digamma, and trigamma functions, Journal
  on Numerical Analysis 31 (1994), no. 3, 931–944 (1994).

\bibitem{Binet}
J.~Binet, M\'emoire sur les int\'egrales d\'efinies eul\'eriennes et sur leur
  application \'a la th\'eorie des suites ainsi qu’\'a l’\'evaluation des
  fonctions des grands nombres, Journal de l’\'Ecole Polytechnique (1839),
  no. XVI, 123–343 (1839).

\bibitem{Burn}
F.~Wilton, A proof of {B}urnside’s formula for $\ln{\Gamma}(1 + \mrx)$ and
  certain allied properties of the {R}iemann $\zeta$-function, Messenger of
  Mathematics 52, 90-93 (1922).

\bibitem{Barnes}
E.~Barnes, Messenger of Mathematics 29, 64-128 (1899).

\bibitem{Tmem}
E.~W. NG,
  \href{https://ntrs.nasa.gov/api/citations/19740015043/downloads/19740015043.pdf}{A
  comparison of computational methods and algorithms for the complex {G}amma
  function}, Technical Memorandum 33-686 (1974).
\newline\urlprefix\url{https://ntrs.nasa.gov/api/citations/19740015043/downloads/19740015043.pdf}

\bibitem{Spira}
R.~Spira, Calculation of the {G}amma function by {S}tirling's formula, Math.
  Comp. Vol. 25, Pp. 317-322 (1971).

\bibitem{tBurn}
W.~Burnside, A rapidly convergent series for $\ln\,\mrn\,!$, Messenger Math. 46
  (1917), 157–159 (1917).

\bibitem{abramowitz+stegun}
M.~Abramowitz, I.~A. Stegun, Handbook of Mathematical Functions with Formulas,
  Graphs, and Mathematical Tables, ninth dover printing, tenth gpo printing
  Edition, Dover, New York, 1964.

\bibitem{comb}
P.~Olver, Classical invariant theory, Cambridge University Press. p. 101. ISBN
  0-521-55821-2. MR 1694364 (1999).

\bibitem{Carlson}
B.~Carlson, Numerical computation of real or complex elliptic integrals,
  Numerical Algorithms 10, no. 1 (1995) 13--26.

\bibitem{Gquot}
M.~Chamberland, A.~Straub,
  \href{http://45.76.13.230/downloads/pub/gammaquotients.pdf}{On gamma
  quotients and infinite products} (2010).
\newline\urlprefix\url{http://45.76.13.230/downloads/pub/gammaquotients.pdf}

\bibitem{Horn35}
J.~Horn, {Hypergeometrische Funktionen zweier Ver\"nderlichen}, Mathematische
  Annalen 105.
\newblock \href {http://dx.doi.org/10.1007/BF01455825}
  {\path{doi:10.1007/BF01455825}}.

\bibitem{deDoncker:2018nqe}
E.~de~Doncker, A.~Almulihi, F.~Yuasa, {High-speed evaluation of loop integrals
  using lattice rules}, J. Phys. Conf. Ser. 1085~(5) (2018) 052005.
\newblock \href {http://dx.doi.org/10.1088/1742-6596/1085/5/052005}
  {\path{doi:10.1088/1742-6596/1085/5/052005}}.

\bibitem{deDoncker:2020gub}
E.~de~Doncker, F.~Yuasa, A.~Almulihi, N.~Nakasato, H.~Daisaka, T.~Ishikawa,
  {Numerical multi-loop integration on heterogeneous many-core processors}, J.
  Phys. Conf. Ser. 1525 (2020) 012002.
\newblock \href {http://dx.doi.org/10.1088/1742-6596/1525/1/012002}
  {\path{doi:10.1088/1742-6596/1525/1/012002}}.

\bibitem{Borowka:2018goh}
S.~Borowka, G.~Heinrich, S.~Jahn, S.~P. Jones, M.~Kerner, J.~Schlenk, {A GPU
  compatible quasi-Monte Carlo integrator interfaced to pySecDec}, Comput.
  Phys. Commun. 240 (2019) 120--137.
\newblock \href {http://arxiv.org/abs/1811.11720} {\path{arXiv:1811.11720}},
  \href {http://dx.doi.org/10.1016/j.cpc.2019.02.015}
  {\path{doi:10.1016/j.cpc.2019.02.015}}.

\bibitem{dagum1998openmp}
L.~Dagum, R.~Menon, Openmp: an industry standard api for shared-memory
  programming, Computational Science \& Engineering, IEEE 5~(1) (1998) 46--55.

\bibitem{David:2013gaa}
A.~David, G.~Passarino, {How well can we guess theoretical uncertainties?},
  Phys. Lett. B 726 (2013) 266--272.
\newblock \href {http://arxiv.org/abs/1307.1843} {\path{arXiv:1307.1843}},
  \href {http://dx.doi.org/10.1016/j.physletb.2013.08.025}
  {\path{doi:10.1016/j.physletb.2013.08.025}}.

\bibitem{Weniger:1997zz}
E.~J. Weniger, {Performance of superconvergent perturbation theory}, Phys.Rev.
  A56 (1997) 5165--5168.
\newblock \href {http://dx.doi.org/10.1103/PhysRevA.56.5165}
  {\path{doi:10.1103/PhysRevA.56.5165}}.

\bibitem{Lev}
D.~Levin, {}, Int. J. Comput. Math. 3 (1973) 371.

\bibitem{TR}
T.~Regge, Algebraic topology methods in the theory of {F}eynman relativistic
  amplitudes, In: Battelle Rencontres - 1967 Lectures in Mathematics and
  Physics. Ed. by C. M. DeWitt and J. A. Wheeler. 1967, pp. 433–458 (1967).

\bibitem{Book}
G.~Passarino, \href{https://doi.org/10.1142/11643}{S matrix and {F}eynman
  amplitudes}, in Tullio {R}egge: An Eclectic Genius (2019).
\newblock \href {http://dx.doi.org/10.1142/11643} {\path{doi:10.1142/11643}}.
\newline\urlprefix\url{https://doi.org/10.1142/11643}

\bibitem{MBmr}
O.~Zhdanov, A.~Tsikh, Investigation of multiple {M}ellin-{B}arnes integrals by
  means of multidimensional residues, Siberian Math. J. 39 (1998) 245 (1998).

\bibitem{Paris_Kaminski_2001}
R.~B. Paris, D.~Kaminski, Asymptotics and Mellin-Barnes Integrals, Encyclopedia
  of Mathematics and its Applications, Cambridge University Press, 2001.

\end{thebibliography}
\end{document}